\documentclass[twocolumn]{aastex631}
\usepackage{tablefootnote}
\usepackage{hyperref}
\usepackage{amsmath}

\begin{document}

\title{Polarization properties of 128 non-repeating fast radio bursts from the first CHIME/FRB baseband catalog}
\shorttitle{Polarization properties of CHIME/FRB non-repeaters}

\author[0000-0002-8897-1973]{Ayush Pandhi}
\affiliation{David A. Dunlap Department of Astronomy and Astrophysics, University of Toronto, 50 St. George Street, Toronto, ON M5S 3H4, Canada}
\affiliation{Dunlap Institute for Astronomy and Astrophysics, University of Toronto, 50 St. George Street, Toronto, ON M5S 3H4, Canada}

\author[0000-0002-4795-697X]{Ziggy Pleunis}
\affiliation{Dunlap Institute for Astronomy and Astrophysics, University of Toronto, 50 St. George Street, Toronto, ON M5S 3H4, Canada}
\affiliation{Anton Pannekoek Institute for Astronomy, University of Amsterdam, Science Park 904, 1098 XH Amsterdam, The Netherlands}
\affiliation{ASTRON, Netherlands Institute for Radio Astronomy, Oude Hoogeveensedijk 4, 7991 PD Dwingeloo, The Netherlands}

\author[0000-0001-7348-6900]{Ryan Mckinven}
\affiliation{Department of Physics, McGill University, 3600 rue University, Montreal, QC H3A 2T8, Canada}
\affiliation{Trottier Space Institute, McGill University, 3550 rue University, Montreal, QC H3A 2A7, Canada}

\author[0000-0002-3382-9558]{B. M. Gaensler}
\affiliation{David A. Dunlap Department of Astronomy and Astrophysics, University of Toronto, 50 St. George Street, Toronto, ON M5S 3H4, Canada}
\affiliation{Dunlap Institute for Astronomy and Astrophysics, University of Toronto, 50 St. George Street, Toronto, ON M5S 3H4, Canada}
\affiliation{Department of Astronomy and Astrophysics, 1156 High Street, University of California Santa Cruz, Santa Cruz, CA 95064, USA}

\author[0000-0003-0607-8194]{Jianing Su}
\affiliation{David A. Dunlap Department of Astronomy and Astrophysics, University of Toronto, 50 St. George Street, Toronto, ON M5S 3H4, Canada}

\author[0000-0002-3616-5160]{Cherry Ng}
\affiliation{Laboratoire de Physique et Chimie de l'Environnement et de l'Espace - Université d'Orléans/CNRS, 45071, Orléans Cedex 02, France}

\author[0000-0002-3615-3514]{Mohit Bhardwaj}
\affiliation{Department of Physics, Carnegie Mellon University, 5000 Forbes Avenue, Pittsburgh, 15213, PA, USA}

\author[0000-0002-1800-8233]{Charanjot Brar}
\affiliation{Department of Physics, McGill University, 3600 rue University, Montr\'eal, QC H3A 2T8, Canada}
\affiliation{Trottier Space Institute, McGill University, 3550 rue University, Montr\'eal, QC H3A 2A7, Canada}

\author[0000-0003-2047-5276]{Tomas Cassanelli}
\affiliation{Department of Electrical Engineering, Universidad de Chile, Av. Tupper 2007, Santiago 8370451, Chile}

\author[0000-0001-6422-8125]{Amanda Cook}
\affiliation{David A. Dunlap Department of Astronomy and Astrophysics, University of Toronto, 50 St. George Street, Toronto, ON M5S 3H4, Canada}
\affiliation{Dunlap Institute for Astronomy and Astrophysics, University of Toronto, 50 St. George Street, Toronto, ON M5S 3H4, Canada}

\author[0000-0002-8376-1563]{Alice P. Curtin}
\affiliation{Department of Physics, McGill University, 3600 rue University, Montr\'eal, QC H3A 2T8, Canada}
\affiliation{Trottier Space Institute, McGill University, 3550 rue University, Montr\'eal, QC H3A 2A7, Canada}

\author[0000-0001-9345-0307]{Victoria M. Kaspi}
\affiliation{Department of Physics, McGill University, 3600 rue University, Montr\'eal, QC H3A 2T8, Canada}
\affiliation{Trottier Space Institute, McGill University, 3550 rue University, Montr\'eal, QC H3A 2A7, Canada}

\author[0000-0002-5857-4264]{Mattias Lazda}
\affiliation{David A. Dunlap Department of Astronomy and Astrophysics, University of Toronto, 50 St. George Street, Toronto, ON M5S 3H4, Canada}
\affiliation{Dunlap Institute for Astronomy and Astrophysics, University of Toronto, 50 St. George Street, Toronto, ON M5S 3H4, Canada}

\author[0000-0002-4209-7408]{Calvin Leung}
\affiliation{Department of Astronomy, University of California Berkeley, Berkeley, CA 94720, USA}
\affiliation{NHFP Einstein Fellow}

\author[0000-0001-7931-0607]{Dongzi Li}
\affiliation{Department of Astrophysical Sciences, Princeton University, Princeton, NJ 08544, USA}

\author[0000-0002-4279-6946]{Kiyoshi W. Masui}
\affiliation{MIT Kavli Institute for Astrophysics and Space Research, Massachusetts Institute of Technology, 77 Massachusetts Ave, Cambridge, MA 02139, USA}
\affiliation{Department of Physics, Massachusetts Institute of Technology, 77 Massachusetts Ave, Cambridge, MA 02139, USA}

\author[0000-0002-2551-7554]{Daniele Michilli}
\affiliation{MIT Kavli Institute for Astrophysics and Space Research, Massachusetts Institute of Technology, 77 Massachusetts Ave, Cambridge, MA 02139, USA}
\affiliation{Department of Physics, Massachusetts Institute of Technology, 77 Massachusetts Ave, Cambridge, MA 02139, USA}

\author[0000-0003-0510-0740]{Kenzie Nimmo}
\affiliation{MIT Kavli Institute for Astrophysics and Space Research, Massachusetts Institute of Technology, 77 Massachusetts Ave, Cambridge, MA 02139, USA}

\author[0000-0002-8912-0732]{Aaron~B.~Pearlman}
\affiliation{Department of Physics, McGill University, 3600 rue University, Montr\'eal, QC H3A 2T8, Canada}
\affiliation{Trottier Space Institute, McGill University, 3550 rue University, Montr\'eal, QC H3A 2A7, Canada}
\affiliation{Banting Fellow}
\affiliation{McGill Space Institute Fellow}
\affiliation{FRQNT Postdoctoral Fellow}

\author[0000-0002-9822-8008]{Emily Petroff}
\affiliation{Perimeter Institute for Theoretical Physics, 31 Caroline Street N, Waterloo, ON N25 2YL, Canada}

\author[0000-0001-7694-6650]{Masoud Rafiei-Ravandi}
\affiliation{Department of Physics, McGill University, 3600 rue University, Montr\'eal, QC H3A 2T8, Canada}
\affiliation{Trottier Space Institute, McGill University, 3550 rue University, Montr\'eal, QC H3A 2A7, Canada}

\author[0000-0003-3154-3676]{Ketan R. Sand}
\affiliation{Department of Physics, McGill University, 3600 rue University, Montr\'eal, QC H3A 2T8, Canada}
\affiliation{Trottier Space Institute, McGill University, 3550 rue University, Montr\'eal, QC H3A 2A7, Canada}

\author[0000-0002-7374-7119]{Paul Scholz}
\affiliation{Department of Physics and Astronomy, York University, 4700 Keele Street, Toronto, ON MJ3 1P3, Canada}
\affiliation{Dunlap Institute for Astronomy and Astrophysics, University of Toronto, 50 St. George Street, Toronto, ON M5S 3H4, Canada}

\author[0000-0002-6823-2073]{Kaitlyn Shin}
\affiliation{MIT Kavli Institute for Astrophysics and Space Research, Massachusetts Institute of Technology, 77 Massachusetts Ave, Cambridge, MA 02139, USA}
\affiliation{Department of Physics, Massachusetts Institute of Technology, 77 Massachusetts Ave, Cambridge, MA 02139, USA}

\author[0000-0002-2088-3125]{Kendrick Smith}
\affiliation{Perimeter Institute for Theoretical Physics, 31 Caroline Street N, Waterloo, ON N25 2YL, Canada}

\author[0000-0001-9784-8670]{Ingrid Stairs}
\affiliation{Department of Physics and Astronomy, University of British Columbia, 6224 Agricultural Road, Vancouver, BC V6T 1Z1 Canada}

\correspondingauthor{Ayush Pandhi}
\email{ayush.pandhi@astro.utoronto.ca}

\begin{abstract}
We present a $400-800$~MHz polarimetric analysis of 128 non-repeating fast radio bursts (FRBs) from the first CHIME/FRB baseband catalog, increasing the total number of FRB sources with polarization properties by a factor of $\sim 3$. $89$ FRBs have $> 6\sigma$ linearly polarized detections, 29 FRBs fall below this significance threshold and are deemed linearly unpolarized, and for 10 FRBs the polarization data are contaminated by instrumental polarization. For the 89 polarized FRBs, we find Faraday rotation measure (RM) amplitudes, after subtracting approximate Milky Way contributions, in the range $0.5 - 1160~\mathrm{rad}~\mathrm{m}^{-2}$ with a median of $53.8~\mathrm{rad}~\mathrm{m}^{-2}$. Most non-repeating FRBs in our sample have RMs consistent with Milky Way-like host galaxies and their linear polarization fractions range from $\leq 10$\% to $100$\% with a median of $63$\%. We see marginal evidence that non-repeating FRBs have more constraining lower limits than repeating FRBs for the host electron-density-weighted line-of-sight magnetic field strength. We classify the non-repeating FRB polarization position angle (PA) profiles into four archetypes: (i) single component with constant PA (57\% of the sample), (ii) single component with variable PA (10\%), (iii) multiple components with a single constant PA (22\%), and (iv) multiple components with different or variable PAs (11\%). We see no evidence for population-wide frequency-dependent depolarization and, therefore, the spread in the distribution of fractional linear polarization is likely intrinsic to the FRB emission mechanism. Finally, we present a novel method to derive redshift lower limits for polarized FRBs without host galaxy identification and test this method on 20 FRBs with independently measured redshifts.
\end{abstract}


\section{Introduction} \label{sec:intro}
Fast radio bursts (FRBs), discovered by \cite{2007Sci...318..777L}, are millisecond-duration extragalactic radio transients. So far, 759 FRB sources have been published\footnote{Based on the \href{https://doi.org/10.7298/X3BX-0E49}{FRB Newsletter Volume 04, Issue 12} published in December 2023.}  \citep[the largest sample being the first Canadian Hydrogen Intensity Mapping Experiment FRB (CHIME/FRB) catalog;][]{2021ApJS..257...59C} and $\sim$3\% of the FRB sources are known to repeat \citep{2023ApJ...947...83C}. Repeating FRBs have been observed over a wide range of frequencies, from as low as 110~MHz \citep[][]{2021ApJ...911L...3P} up to 8~GHz \citep[][]{2018ApJ...863....2G,  2023MNRAS.524.3303B}. Apparent non-repeating FRBs have been detected from 300~MHz \citep{2020ApJ...904...92P} to 1.53~GHz \citep[e.g.,][]{2023arXiv230703344L}. Over $40$ FRBs have been localized to external galaxies \citep[e.g.,][]{2017Natur.541...58C, 2017ApJ...834L...7T, 2022ApJ...927L...3N, 2023ApJ...950..134M, 2023arXiv231010018B} with diverse host galaxy properties and local environments \citep[e.g.,][]{2021ApJ...917...75M, 2022AJ....163...69B, 2023ApJ...954...80G, 2023arXiv230402638I}. However, the specific origins and emission mechanisms of FRBs remain elusive.  Furthermore, it is not clear whether repeating and non-repeating FRBs have distinct origins or, even if all FRBs repeat. 

\subsection{Polarization information from FRB voltage data}
FRBs are typically highly linearly polarized \citep[e.g.,][]{2015MNRAS.447..246P, 2015Natur.528..523M}. As such, FRB polarimetry can help to elucidate properties of both their emission mechanism and local magneto-ionic environment in the immediate vicinity of the source. The polarization properties of FRBs are fully encapsulated in the Stokes $I$, $Q$, $U$, and $V$ dynamic spectra. Here, $I$ refers to the total intensity of the emission, $Q$ and $U$ are the linearly polarized components, and $V$ is the circularly polarized component. The Stokes parameters are a function of both time ($t$) and frequency ($\nu$). By integrating the $Q$ and $U$ dynamic spectra over the FRB burst envelope and then averaging over the emitting band of the FRB, we can derive the average linear and circular polarization fraction for the FRB as $L/I \equiv \left[ Q^2 + U^2 \right]^{1/2} / I$ and $|V|/I$, respectively.\footnote{For CHIME/FRB data, a debiasing is applied following Equation 11 of \cite{2001ApJ...553..341E}, which becomes important in the low signal-to-noise regime.} 

As the linearly polarized radio emission, with an intrinsic position angle $\psi_0$, propagates through magneto-ionic media, it undergoes Faraday rotation. This causes a wavelength squared $\left(\lambda^2\right)$-dependent rotation of the observed linear polarization position angle, $\psi(\lambda^2)$:
\begin{equation}
\psi(\lambda^2) = \psi_0 + {\rm{RM}}\lambda^{2}\,, \label{eq:pa}
\end{equation}
where RM is the rotation measure, which is related to the integrated number density of electrons and parallel component of the magnetic field along the line of sight (LoS):
\begin{equation}
{\rm{RM}} = 0.812 \int_{z}^{0} \frac{n_e(z) B_\parallel(z)}{(1 + z)^2}\frac{\mathrm{d}l(z)}{\mathrm{d}z}\mathrm{d}z~\mathrm{rad}~\mathrm{m}^{-2} \label{eq:RM} \,.
\end{equation}
Here, $z$ is the redshift of the FRB, $n_e(z)$ is the electron density in units of cm$^{-3}$, $B_\parallel(z)$ is the LoS magnetic field strength in units of $\mu$G, $\mathrm{d}l(z)$ is the LoS line element at $z$ in units of kpc, and we integrate from the FRB source (at $z$) to the observer (at $z=0$). A positive (negative) RM implies that the average LoS magnetic field is pointed towards (away from) the observer. Simultaneously, the radio emission experiences a $\lambda^2$ dispersion by the same free electrons, which is characterized by the dispersion measure (DM):
\begin{equation}
{\rm{DM}}= \int_{0}^{z} \frac{n_e(z)}{(1+z)} \frac{\mathrm{d}l(z)}{\mathrm{d}z}\mathrm{d}z~\mathrm{pc}~\mathrm{cm}^{-3} \label{eq:DM} \,.
\end{equation}

\subsection{Polarization as a probe of FRB emission mechanisms and local environments}
Coherent emission models for FRBs often invoke either a neutron star magnetosphere \citep[e.g.,][]{2017MNRAS.468.2726K, 2018ApJ...868...31Y, 2021ApJ...922..166L} or a synchrotron maser origin \citep[e.g.,][]{2014MNRAS.442L...9L, 2019MNRAS.485.4091M, 2020ApJ...896..142B}. Both sets of models explain the typically high linear polarization fraction seen for FRBs. Magnetospheric models are able to reproduce both constant and varying polarization angle (PA) profiles across the burst based on their magnetospheric configuration or LoS geometry. Synchrotron maser models, however, predict a constant PA across bursts \citep[][]{2019MNRAS.485.4091M, 2020ApJ...896..142B} and are thus disfavored for bursts that show significant PA swings \citep[e.g., as is the case for some FRB 20180301A bursts and for FRB 20221022A;][]{2020Natur.586..693L, 2024arXiv240209304M}. 

We expect different levels of typical $\left| \mathrm{RM} \right|$ contributions from different local environments. For instance, we might expect FRBs located in clean environments, such as globular clusters, to have a local environment $\left| \mathrm{RM} \right|$ contributions of $\sim 0~\mathrm{rad}~\mathrm{m}^{-2}$. Meanwhile, \cite{2019MNRAS.488.4220H} predict a median $\left| \mathrm{RM} \right| \sim 10~\mathrm{rad}~\mathrm{m}^{-2}$ contribution from a Milky Way (MW)-like FRB host galaxy and $\sim 10^2~\mathrm{rad}~\mathrm{m}^{-2}$ for FRBs located near star forming regions (based on models of the MW electron density and Galactic magnetic field). For an FRB embedded within a young supernova remnant (with a typical age of $\sim 10^2 - 10^3$~years), we might expect $\left| \mathrm{RM} \right|$s of up to $\sim 10^3~\mathrm{rad}~\mathrm{m}^{-2}$ contributed from the local environment \citep{2018ApJ...861..150P}.

\subsection{Current observations and understanding of FRB polarimetry}
Currently, there exist polarimetric observations for $36$ non-repeating FRBs \citep[e.g.,][]{2015MNRAS.447..246P, 2015Natur.528..523M, 2023arXiv230806813S} and for over $2000$ bursts from $18$ repeating FRB sources \citep[e.g.,][]{2022Natur.609..685X, 2023ApJ...950...12M, 2023ApJ...951...82M, 2023Sci...380..599A}. For a summary of the published polarization properties of FRBs from the literature, see Table \ref{tb:lit_pol} in Appendix \ref{app:A}. The primary focus of FRB polarimetry studies thus far has been on the temporal evolution of polarization properties of prolific repeating sources. The general picture of FRB polarization thus far has been that:
\begin{enumerate}
    \item FRBs are typically highly linearly polarized (the median $L/I$ of currently published FRBs with polarization properties is $\sim 75$\%). Many FRBs are also consistent with being $100$\% linearly polarized.
    \item Approximately $13$\% of the currently published FRB sources with polarimetry show at least one burst with substantial circular polarization (i.e., $\gtrsim 20$\%) over their burst profile \citep[e.g.,][]{2020ApJ...891L..38C, 2023arXiv230806813S}.
    \item Typically FRBs have $|\mathrm{RM}|$s on the order of $\sim 10^2~\mathrm{rad}~\mathrm{m}^{-2}$.
    \item A few prolific repeaters originate from extremely dense and/or highly magnetic environments \citep[e.g., FRB 20121102A and FRB 20190520B, which show $|\mathrm{RM}|$s up to $\sim 10^4 - 10^5~\mathrm{rad}~\mathrm{m}^{-2}$;][]{2018Natur.553..182M, 2023Sci...380..599A}.
    \item At least some of the known repeaters for which we have detected polarized bursts reside in dynamic magneto-ionic environments such that they undergo RM variations on days to years timescales and, in some cases, exhibit changes in the sign of their RM \citep[][]{2021ApJ...908L..10H, 2023Sci...380..599A, 2023ApJ...950...12M}.
\end{enumerate}

Multi-band polarimetry (between $\sim 0.1 - 5~\mathrm{GHz}$) by \cite{2022Sci...375.1266F} found that five repeating FRBs (FRB 20121102A, FRB 20190520B, FRB 20190303A, FRB 20190417A, and FRB 20201124A) show decreasing $L/I$ with decreasing observing frequency. \cite{2022Sci...375.1266F} suggest that the observed frequency dependence in $L/I$ is an imprint of depolarization caused by multi-path propagation of the FRB emission through an inhomogeneous magneto-ionic medium, parameterized by
\begin{equation}
f_\mathrm{RM}(\lambda) = 1 - \mathrm{exp}\left[ -2 \lambda^4 \sigma_\mathrm{RM}^2 \right]\,, \label{eq:depol}
\end{equation}
originally formulated by \cite{1966MNRAS.133...67B} for incoherent synchrotron radiation\footnote{\cite{2022MNRAS.510.4654B}, on the other hand, argue that depolarization from propagation of \emph{coherent} radiation through a magnetized scattering screen instead follows a power-law dependence on frequency.}. Here $f_\mathrm{RM}(\lambda)$ is the fractional decrease in the observed linear polarization from its intrinsic value, $\lambda$ is the wavelength in the observer frame, and $\sigma_\mathrm{RM}$ is a measure of the dispersion about the mean observed RM. Note that the exponent in Equation \ref{eq:depol} is independent of redshift as the $(1+z)$ scaling on $\sigma_\mathrm{RM}$ and $\lambda$ cancel out. However, in order to derive the typical depolarization wavelength, $\lambda_\mathrm{depol}$, we would need a redshift-dependent coordinate transform in $\lambda$ to the rest frame of the FRB. \cite{2022Sci...375.1266F} find a positive correlation between $\sigma_\mathrm{RM}$ and temporal scattering, suggesting that both $\sigma_\mathrm{RM}$ and the temporal scattering originate from the same region near the FRB source. Furthermore, the FRBs with the largest $\sigma_\mathrm{RM}$ (FRB 20121102A and FRB 20190520B) are associated with persistent radio sources \citep{2017ApJ...834L...8M, 2022Natur.606..873N}, consistent with a supernova remnant or pulsar wind nebula origin, which is in agreement with their progenitors being associated with dense local environments.

\subsection{Dichotomy between repeating and non-repeating FRBs}
There is some evidence for differences between repeating and non-repeating FRBs. \cite{2021ApJ...923....1P} find a dichotomy in burst properties of repeating and non-repeating FRBs, with repeating FRBs having larger average burst widths and narrower average emitting bandwidths than non-repeating FRBs in the first CHIME/FRB catalog \citep[][]{2021ApJS..257...59C}. Whether repeating and non-repeating FRBs originate from (i) the same source population, (ii) the same source population with different environments or propagation effects, or (iii) different source populations remains largely uncertain. Polarimetry provides insight into both the emission mechanism and local magneto-ionic environment of FRBs and is therefore a powerful tool to distinguish between these scenarios. \cite{2022Sci...375.1266F} show that the $|$RM$|$ distributions of the repeating and non-repeating sources differ (with a $p$-value of $0.02$ using the Kolmogorov-Smirnov test), possibly indicating a divergence in local environments of the two populations. Note, however, that this difference in $|$RM$|$ was only seen in a small sample (9 repeating and 12 non-repeating FRBs), and is potentially subject to selection bias due to intra-channel depolarization in the coarser filter bank data for non-repeating FRBs. This is because discovery searches often use coarser filter banks for computational efficiency, which is not necessary for follow-up observations of known repeating sources. Hence, the difference in $|$RM$|$ between repeating and non-repeating FRBs may not be representative of the full observed FRB sample and would be greatly benefited by polarimetric studies on a larger sample of FRBs.

\subsection{Polarization properties of CHIME/FRB non-repeaters}
In this paper, we perform a systematic analysis of the polarization properties for 128 non-repeating FRBs from the first CHIME/FRB baseband catalog \citep{2023arXiv231100111T}. The $L/I$, RM, DM, and lower limits on the LoS-averaged magnetic field strength in the FRB host galaxy of our polarized sample are compared to the polarization properties of the 13 polarized, repeating FRB sources observed by CHIME/FRB \citep{2021ApJ...910L..18B, 2023ApJ...950...12M, 2023ApJ...951...82M}. Furthermore, we determine the extent to which the PA varies across each burst in our sample of non-repeating FRBs and what fraction of these FRBs exhibit PA variations. For a subset of broadband emitting FRBs, we derive a model-agnostic depolarization ratio across the CHIME/FRB band and test the prevalence of frequency-dependent depolarization in our FRB sample, comparing to the spectral depolarization seen in some repeating FRBs. Finally, the polarization properties of the repeating and non-repeating sources are compared with their burst rates as estimated by \cite{2023ApJ...947...83C}.

This paper is structured as follows. In Section \ref{sec:data}, we provide an overview of the non-repeating FRBs from the first CHIME/FRB baseband catalog, a comparison sample of repeating FRBs observed by CHIME/FRB, the derived polarization properties of the CHIME/FRB FRBs, and describe the statistical methods used throughout this paper. The results of the non-repeater polarimetry, including $L/I$, RM, lower limits on the LoS-averaged magnetic field strength in the FRB host galaxy, PA variability, depolarization, and comparisons to the polarized, repeating CHIME/FRB sample are presented in Section \ref{sec:results}. Interpretation of our results is provided in Section \ref{sec:discussion}. We conclude by summarizing our findings and identifying avenues for future work.

\section{Voltage data, derived polarization properties, and statistical methods} \label{sec:data}
In this Section, we define the repeating and non-repeating FRB population for which we conduct our polarization analysis, and then summarize the CHIME/FRB polarization pipeline and additional analyses that we perform to produce our polarimetric results.

\subsection{CHIME/FRB voltage data} \label{sec:rep_nonrep_data}
\subsubsection{Non-repeating FRBs}
\cite{2023arXiv231100111T} present the first CHIME/FRB baseband catalog, providing the full voltage data for all 140 FRBs recorded by the triggered baseband system between 2018 December 9 and 2019 July 1 \citep[for a detailed description of the baseband analysis system on CHIME/FRB, see][]{2021ApJ...910..147M}. Baseband data are only stored for those events exceeding a signal-to-noise (S/N) threshold of $\text{S/N}>10-12$ (the threshold has changed over the course of operation). These data are saved with a time and frequency resolution of $2.56~\mu\mathrm{s}$ and $390~\mathrm{kHz}$, respectively, with full polarization information (i.e., Stokes $I$, $Q$, $U$, and $V$). \cite{2023arXiv231100111T} describe the processing of baseband data in detail. To summarize, the initial estimate of the FRB sky position is refined by forming in software a large array of beams on the sky, and then is precisely localizing by maximizing the S/N with a more densely packed array of beams formed around the initial position. This process typically returns sky localization regions with sub-arcminute precision \citep{2023arXiv231100111T}; a single beam is then formed at the best estimated FRB sky position and those data are recorded as a ``single-beam'' file. These single-beam files serve as inputs to the CHIME/FRB polarization pipeline, described by \cite{2021ApJ...920..138M} and summarized in Section \ref{sec:pipeline_products} below. The 128 non-repeating FRBs with single-beam files presented by \cite{2023arXiv231100111T} form the sample of sources that are the focus of this paper.

\subsubsection{Repeating FRBs}
One of the goals of this paper is to compare the polarization properties of our non-repeating sample to that of known repeating FRBs. To avoid confounding effects in the polarization properties arising from different observing frequencies, selection effects, and processing stages, we opt to compare our CHIME/FRB non-repeaters with a sample of repeating FRBs with polarization measurements also obtained by CHIME/FRB. The total comparison sample consists of 13 repeating sources with a collective 82 bursts detected between December 2018 and December 2021 that are collated from results by \cite{2021ApJ...910L..18B}, \cite{2022ATel15679....1M}, \cite{2023ApJ...950...12M}, and \cite{2023ApJ...951...82M}, which make use of the CHIME/FRB polarization pipeline described by \cite{2021ApJ...920..138M}.

\subsection{Polarization pipeline products} \label{sec:pipeline_products}
For each burst, we begin with the Stokes $I$, $Q$, $U$, and $V$ dynamic spectra channelized to a time resolution of $2.56~\mu\mathrm{s}$ and a frequency resolution of $390~\mathrm{kHz}$. 
We obtain the best fit structure optimizing DM, $\mathrm{DM}_\mathrm{obs,struct}$, based on the {\tt DM\_phase} algorithm \citep{2019ascl.soft10004S}. The Stokes data are coherently de-dispersed to their respective $\mathrm{DM}_\mathrm{obs,struct}$ and a radio frequency interference (RFI) mask is applied in the baseband pipeline processing stage \citep{2021ApJ...910..147M}. To derive the polarization properties, we process each burst using the CHIME/FRB polarization pipeline \citep[for a full description of the pipeline, see][]{2021ApJ...920..138M}. Below, we provide a summary of the pipeline and highlight any changes from the version presented by \cite{2021ApJ...920..138M}.

The burst envelope limits in time are identified as the points at which the total intensity drops to $20\%$ of the peak burst intensity. A Gaussian function is fit to the intensity spectrum as a function of frequency and the burst spectral limits are set at the $3\sigma$ level of the Gaussian fit (with a minimum of $400$~MHz and maximum of $800$~MHz). Stokes $I$, $Q$, $U$, and $V$ spectra are generated by integrating the signal between the burst envelope time limits with uniform weights in time. Then the $L/I$ spectrum is computed and averaged over the burst spectral limits to provide the $L/I$ of the FRB, averaged over both time and frequency.

The observing frequencies, and Stokes spectra with associated standard deviation uncertainties, form an input to RM-synthesis \citep{2005A&A...441.1217B}, which outputs an observed RM estimate, $\mathrm{RM}_\mathrm{obs}$, with measurement uncertainties $\mathrm{FWHM}/(2 \times \mathrm{S/N})$. Here, the FWHM refers to the full-width at half maximum of the cleaned Faraday dispersion function (FDF) peak \citep[the cleaning is done using the {\tt RM-CLEAN} framework;][]{2009A&A...503..409H} and the S/N is that of the peak polarized intensity in Faraday depth space. A minimum threshold of $\mathrm{S/N} = 6$ in $L$ is required to warrant a linearly polarized detection. This threshold was chosen after internal testing on many FRBs over the course of CHIME/FRB operations. Setting it lower results in an increase of false detections caused by instrumental polarization (i.e., at $\mathrm{RM}_\mathrm{obs} \sim 0~\mathrm{rad}~\mathrm{m}^{-2}$), while increasing the threshold causes the pipeline to miss some marginal linearly polarized detections that would otherwise be well-fit by the pipeline.

Another independent $\mathrm{RM}_\mathrm{obs}$ is estimated by the QU-fitting algorithm \citep[implemented with a modified version of V1.2 of the {\tt RM-tools} package;][]{2020ascl.soft05003P}. We use the same default model as \cite{2021ApJ...920..138M} with $\mathrm{RM}_\mathrm{obs}$, PA, $L/I$, and the physical (cable) delay between the two linear polarizations, $\tau_\mathrm{delay}$, as fitted parameters and assume uniform priors on each of them. In this model, $L/I$ is assumed to be constant over the burst spectral range (though we also explore $L/I$ variations as a function of frequency using a slightly different method in Section \ref{sec:results_beam_depol}). The Stokes $I$ spectrum is fit using a univariate spline fit, $I_\mathrm{mod}$, from the {\tt scipy.interpolate.UnivariateSpline} module, and is also used as an input in the QU-fitting step. We set the Stokes $V$ model to $V_\mathrm{mod} = 0$ to limit ambiguity of unmodelled instrumental sources of circular polarization (which can contribute up to $|V|/I \sim 20\%$) with intrinsic signal. Models of the Stokes $Q$ and $U$ ($Q_\mathrm{mod}$ and $U_\mathrm{mod}$, respectively) spectra are then expressed as,
\begin{equation}
Q_\mathrm{mod} = I_\mathrm{mod} (L/I) \mathrm{cos}(2(\mathrm{RM}_\mathrm{obs}\lambda^2 + \psi_0))\,, \label{eq:Q_mod}
\end{equation}
\begin{equation}
U_\mathrm{mod} = I_\mathrm{mod} (L/I) \mathrm{sin}(2(\mathrm{RM}_\mathrm{obs}\lambda^2 + \psi_0))\,. \label{eq:U_mod}
\end{equation}
Maximum likelihood estimates for the fitted parameters are derived via Nested Sampling \citep{2004AIPC..735..395S} to best fit $Q_\mathrm{mod}$ and $U_\mathrm{mod}$ to the observed spectra, $Q_\mathrm{obs}$ and $U_\mathrm{obs}$. Examples illustrating the fitted Stokes spectra and posterior distributions for the polarization properties incorporated in the QU-fitting model are presented for a simulated FRB and a CHIME/FRB detection, respectively, by \cite{2021ApJ...920..138M} in their Figures 1 (panels c and d) and 6. A phase shift is introduced due to the time delay, $\tau_\mathrm{delay}$, between the signals from the two polarized feeds due to their differential path lengths through the system. This causes a mixing between Stokes $U$ and $V$ that is characterized by the matrix
\begin{equation}
\begin{pmatrix}
U_\mathrm{obs}\\
V_\mathrm{obs}
\end{pmatrix}
=
\begin{pmatrix}
\mathrm{cos}(2\pi\nu\tau_\mathrm{delay}) & -\mathrm{sin}(2\pi\nu\tau_\mathrm{delay})\\
\mathrm{sin}(2\pi\nu\tau_\mathrm{delay}) & \mathrm{cos}(2\pi\nu\tau_\mathrm{delay})
\end{pmatrix}
\begin{pmatrix}
U\\
V
\end{pmatrix}
\,. \label{eq:delay_matrix}
\end{equation}
We thus update the Stokes $Q$ and $U$ models (now $Q'_\mathrm{mod}$ and $U'_\mathrm{mod}$) to account for $\tau_\mathrm{delay}$, while still assuming $V_\mathrm{mod} = 0$, as follows,
\begin{equation}
Q'_\mathrm{mod} = Q_\mathrm{mod}\,, \label{eq:Q_mod_2}
\end{equation}
\begin{equation}
U'_\mathrm{mod} = U_\mathrm{mod} \left[ \mathrm{cos}(2\pi\nu\tau_\mathrm{delay}) \right]\,. \label{eq:U_mod_2}
\end{equation}
For a non-zero $\tau_\mathrm{delay}$, the QU-fitting routine may incorrectly converge on the wrong sign for the $\mathrm{RM}_\mathrm{obs}$ estimate. In these cases, however, the FDF appears to have two mirrored peaks about $\mathrm{RM}_\mathrm{obs} = 0~\mathrm{rad}~\mathrm{m}^{-2}$, but the higher of the two peaks is always the correct sign of the $\mathrm{RM}_\mathrm{obs}$ for the phase offsets regularly observed in CHIME/FRB data \citep[for more details, see Appendix A by][]{2021ApJ...920..138M}. Throughout this paper, we will use the FDF derived $\mathrm{RM}_\mathrm{obs}$ for all analyses.

For finite frequency resolution, the linear polarization angle changes within one frequency channel. The intra-channel change in polarization angle, $\delta \psi$, for an event with $\mathrm{RM}_\mathrm{obs}$ observed at a central frequency $\nu_\mathrm{c}$ with frequency channel width $\delta \nu$ follows
\begin{equation}
\delta\psi = \frac{-2 c^2 \mathrm{RM}_\mathrm{obs} \delta\nu}{\nu_c^3} \,. \label{eq:d_pa}
\end{equation}
Thus, for high $\mathrm{RM}_\mathrm{obs}$ magnitude events, the intra-channel change in polarization angle can be significant and cause depolarization of the observed signal. The decrease in $L/I$ 
due to intra-channel depolarization, $f_\mathrm{channel}$, follows
\begin{equation}
f_\mathrm{channel} = \frac{\mathrm{sin}(\delta\psi)}{\delta\psi}\, \label{eq:depol_intra_channel}
\end{equation}
\citep{1966ARA&A...4..245G}.

For $\nu_\mathrm{c} = 600$~MHz and $\delta\nu = 390$~kHz, the sensitivity to polarized emission is halved at $\mathrm{RM}_\mathrm{obs} \sim 5000~\mathrm{rad}~\mathrm{m}^{-2}$, meaning that the range of $\mathrm{RM}_\mathrm{obs}$ that can be deduced at the native resolution of these data is limited to a few thousand~rad~m$^{-2}$ \citep[][]{2021ApJ...920..138M}. In some cases, instrumental polarization may cause a peak in the FDF centered at $0~\mathrm{rad}~\mathrm{m}^{-2}$ with a full width at half maximum equal to the theoretical RM spread function of $\sim 9~\mathrm{rad}~\mathrm{m}^{-2}$ for the full CHIME/FRB frequency band \citep{2020MNRAS.496.2836N}. The instrumental polarization induced artefact can sometimes exceed the polarized intensity S/N of the true $\mathrm{RM}_\mathrm{obs}$ peak. In cases where this occurs and a $> 6 \sigma$ secondary peak in the FDF can be clearly identified, the native RM search is re-run with the peak at $0~\mathrm{rad}~\mathrm{m}^{-2}$ masked out.

\subsection{Temporal PA variations} \label{sec:pa_var}
From the Stokes $Q$ and $U$ dynamic spectra, we can derive the observed linear polarization angle $\psi$ of the FRB emission as a function of time $t$ and observing wavelength $\lambda$, such that,
\begin{equation}
\psi(t,\lambda) = \frac{1}{2} \mathrm{arctan}\left( \frac{U_\mathrm{obs}(t,\lambda)}{Q_\mathrm{obs}(t,\lambda)} \right)\,. \label{eq:pa_def}
\end{equation}
With the observed $Q_\mathrm{obs}$ and $U_\mathrm{obs}$ spectra, estimated $\mathrm{RM}_\mathrm{obs}$, and $\psi_0$ in hand, we can de-rotate the linear polarization vector to remove the Faraday rotation effect,
\begin{equation}
\left[ Q + iU \right]_\mathrm{obs,derot} = \left[ Q + iU \right]_\mathrm{obs} \times \mathrm{exp}\left[ 2i \left( \mathrm{RM}_\mathrm{obs} \left( \lambda^2 - \lambda_0^2 \right) + \psi_0 \right) \right]\,. \label{eq:qu_derot}
\end{equation}
In the equation above, the subscript ``obs'' refers to the observed linear polarization vector, and $\lambda_0$ is the reference wavelength at which $\psi_0$ is measured (here, we set $\lambda_0 = 0$~m such that $\psi_0$ is referenced to zero wavelength). From Equation \ref{eq:qu_derot}, we can extract the de-rotated $Q_\mathrm{obs,derot}$ and $U_\mathrm{obs,derot}$ dynamic spectra as the real and imaginary components of $\left[ Q + iU \right]_\mathrm{obs,derot}$, respectively. Using the same procedure as Equation \ref{eq:pa_def} and integrating over the burst spectral range, we can then compute the temporal PA variation across the burst width as
\begin{equation}
\psi_0(t) = \frac{1}{2} \mathrm{arctan}\left( \frac{U_\mathrm{obs,derot}(t)}{Q_\mathrm{obs,derot}(t)} \right)\,. \label{eq:pa_derot}
\end{equation}
Note that we do not calibrate for true value of the PA due to unknown beam phase effects off of CHIME's meridian axis. Hence, $\psi_0(t)$ is centered such that it has a median of $0~\mathrm{deg}$ and is only used to characterize the relative time evolution in PA across the burst. By propagating the uncertainties in the $Q_\mathrm{obs,derot}$ and $U_\mathrm{obs,derot}$ spectra, we estimate the PA measurement uncertainties, $\sigma_\psi$, following the process described by \cite{2019ApJ...878...92V}. These measurement uncertainties can become inaccurate for low S/N in the linear polarization $L \equiv \left[ Q^2 + U^2 \right]^{1/2}$. To ensure that our PA profiles are robust, we set a minimum linear polarization S/N threshold $\mathrm{S/N}(L)_\mathrm{thresh} = 5$, below which points on the $\psi_0(t)$ profile are masked out. This step is particularly important when characterizing any time variable PA behavior. For a detailed explanation on how this threshold was derived, see Appendix \ref{app:B}.

To quantify the magnitude of deviation from a constant $\psi_0(t)$, we conduct a reduced chi-squared $\chi_\nu^2$ test \citep[implemented using the package {\tt scipy.stats.chisquare} V1.10.0;][]{2020SciPy-NMeth}, evaluating the goodness of fit between the observed $\psi_0(t)$ and a constant $\psi_0(t) = 0$ model.

\subsection{Foreground DM correction} \label{sec:dm_corr}
For each polarized FRB in our sample, we have estimates for the observed DM and RM ($\mathrm{DM}_\mathrm{obs}$ and $\mathrm{RM}_\mathrm{obs}$). These quantities, however, are integrated over the full path length between the emitting source and the observer and, therefore, likely encompass distinct contributing media along the LoS. We can express $\mathrm{DM}_\mathrm{obs}$ in terms of its individual contributing components,
\begin{equation}
\mathrm{DM}_{\mathrm{obs}} = \mathrm{DM}_{\mathrm{disk}} + \mathrm{DM}_{\mathrm{halo}} + \mathrm{DM}_{\mathrm{IGM}}(z) + \frac{\mathrm{DM}_{\mathrm{host}}}{(1 + z)}\,, \label{eq:dm_comps}
\end{equation}
where $\rm{DM}_{\rm{disk}}$ is contribution from warm ionized gas in the MW disk ($T \lesssim 10^4$ K) and $\rm{DM}_{\rm{halo}}$ is from the extended hot Galactic halo ($T \sim 10^6 - 10^7$ K). $\rm{DM}_{\rm{IGM}}(z)$ is the collective contributions from the intergalactic medium (IGM) and intervening systems. Each contributing system along the LoS in the IGM has a different (unknown) redshift which is not corrected for in the $\rm{DM}_{\rm{IGM}}(z)$ term. $\mathrm{DM}_{\mathrm{host}}/(1 + z)$ is from the host galaxy of the source at redshift $z$ and its local environment \citep[and references therein]{2020ApJ...888..105Y}. Here, we ignore the ionospheric contribution to the FRB DM as it typically adds on the order of $\sim 10^{-5}~\mathrm{pc}~\mathrm{cm}^{-3}$ \citep{2016ApJ...821...66L}.

The $\mathrm{DM}_{\mathrm{disk}}$ contribution is estimated by utilizing the thermal electron density $n_\mathrm{e}$ model of \citet[][hereafter YMW16]{2017ApJ...835...29Y}. For a given sky position, the {\tt PyGEDM} package \citep[V3.1.1;][]{2021PASA...38...38P} allows us to integrate over the full extent of the MW along that LoS and obtain the value of $\mathrm{DM}_{\mathrm{disk}}$ according to YMW16. For each FRB, we take the $\mathrm{DM}_{\mathrm{disk}}$ contribution at the best-fit position using the baseband localizations of \cite{2023arXiv231100111T}, which have a typical $1\sigma$ uncertainty region of $\lesssim 1~\mathrm{arcmin}$. For the MW halo, we assume a fiducial $\mathrm{DM}_{\mathrm{halo}} = 30~\mathrm{pc}~\mathrm{cm}^{-3}$ contribution based on estimates by \cite{2015MNRAS.451.4277D}, \cite{2020ApJ...888..105Y}, and \cite{2023ApJ...946...58C}. Our results are not sensitive to the exact value of $\mathrm{DM}_{\mathrm{halo}}$ as long as it is assumed to be uniform across the sky. In general, FRBs have significant observed excess DM contributions after subtracting the MW DM which, for a subset of arcsecond and sub-arcsecond localized FRBs, scales roughly linearly with the host galaxy redshift \citep[i.e., the ``Macquart relation'';][]{2020Natur.581..391M}. Unfortunately, the FRBs in our sample are yet to be associated with host galaxies and, thus, do not have reliable redshift estimates. As such, we do not attempt to estimate and subtract $\mathrm{DM}_{\mathrm{IGM}}(z)$ from $\mathrm{DM}_{\mathrm{obs}}$.

Removing the MW disk and halo DM contributions gives us a foreground corrected DM estimate that encapsulates the totality of the extragalactic dispersion incurred by the FRB emission,
\begin{equation}
\mathrm{DM}_\mathrm{EG} = \mathrm{DM}_{\mathrm{IGM}}(z) + \frac{\mathrm{DM}_{\mathrm{host}}}{(1+z)}\,, \label{eq:dm_eg}
\end{equation}
and is an upper limit on the host galaxy/local environment DM. We highlight here that $\mathrm{DM}_\mathrm{EG}$ is reported in our observer frame, not the rest frame of the FRB.

\subsection{Foreground RM correction} \label{sec:rm_corr}
Analogously, we can perform the same decomposition as in Section \ref{sec:dm_corr} for $\mathrm{RM}_\mathrm{obs}$ and write,
\begin{equation}
\mathrm{RM}_{\mathrm{obs}} = \mathrm{RM}_{\mathrm{ion}} + \mathrm{RM}_{\mathrm{disk}} + \mathrm{RM}_{\mathrm{halo}} + \mathrm{RM}_{\mathrm{IGM}}(z) + \frac{\mathrm{RM}_{\mathrm{host}}}{(1+z)^2}\,, \label{eq:rm_comps}
\end{equation}
where $\mathrm{RM}_{\mathrm{ion}}$ is the ionospheric contribution from the Earth's atmosphere at the time of observation. We opt to not correct for the ionospheric contribution to the RM since it only adds on the order of $\sim \pm 1~\mathrm{rad}~\mathrm{m}^{-2}$ \citep{2019MNRAS.484.3646S}. Stochastic RM variations of this amplitude do not affect our results (presented in Section \ref{sec:results}) in a meaningful way. Hence, we move forward with the assertion that $\mathrm{RM}_{\mathrm{ion}} = 0~\mathrm{rad}~\mathrm{m}^{-2}$.

\cite{2022A&A...657A..43H} construct an all-sky interpolated map of the foreground MW RM contribution using a Bayesian inference scheme applied to all Faraday rotation data \citep[a total of $55,190$ individual RMs, mostly from radio galaxies;][]{2023ApJS..267...28V} available by the end of 2020. This map has a pixel scale of $\sim 1.3 \times 10^{-2}~\mathrm{deg}^2$ and provides the best estimate of the Galactic RM (i.e., $\mathrm{RM}_{\mathrm{disk}} + \mathrm{RM}_{\mathrm{halo}}$) sky to date. We estimate the $\mathrm{RM}_{\mathrm{disk}} + \mathrm{RM}_{\mathrm{halo}}$ towards each FRB by taking the RM from the \cite{2022A&A...657A..43H} map at the best-fit sky position of the FRB.

For most extragalactic, polarized radio sources, namely radio galaxies, the foremost contributor to $\mathrm{RM}_{\mathrm{obs}}$ is propagation through the MW \citep[][]{2010MNRAS.409L..99S, 2015A&A...575A.118O}. In the IGM, $B_{\parallel}$ undergoes multiple field reversals along the full path length, which is much larger than the magnetic field scales and $B_{\parallel}$ in the IGM is fairly weak \citep[$\leq 21$~nG;][]{2016Sci...354.1249R}. Therefore, we expect the amplitude of $\mathrm{RM}_\mathrm{IGM}$ to be small. From observations, we see that the residual extragalactic RM distribution (after subtracting an estimate of the Galactic RM contribution) for a set of polarized radio galaxies is centered around $0~\mathrm{rad}~\mathrm{m}^{-2}$ \citep[e.g., see][]{2022MNRAS.512..945C} with some finite width that can be quantified by measuring the rms of the distribution. This residual extragalactic RM rms has been measured at $144$~MHz \citep[$0.52~\mathrm{rad}~\mathrm{m}^{-2}$;][]{2022MNRAS.512..945C} and at $1.4$~GHz \citep[$14.9~\mathrm{rad}~\mathrm{m}^{-2}$;][]{2019ApJ...878...92V}. Since we expect the amplitude of $\mathrm{RM}_{\mathrm{IGM}}$ to be small and centered on $0~\mathrm{rad}~\mathrm{m}^{-2}$, while the typical $|\mathrm{RM}_{\mathrm{obs}}|$ of FRBs to date is $\gtrsim 10^2~\mathrm{m}^{-2}$, we adopt the assumption that $\mathrm{RM}_{\mathrm{IGM}} = 0~\mathrm{rad}~\mathrm{m}^{-2}$ (i.e., $|\mathrm{RM}_{\mathrm{IGM}}| \ll |\mathrm{RM}_{\mathrm{obs}}|$).

After subtracting the foreground MW RM and asserting that $\mathrm{RM}_{\mathrm{IGM}} = \mathrm{RM}_{\mathrm{ion}} = 0~\mathrm{rad}~\mathrm{m}^{-2}$, the redshift-corrected RM serves as a rough estimate for the total RM contributed by the FRB host galaxy and local environment, 
\begin{equation}
\mathrm{RM}_\mathrm{EG} = \frac{\mathrm{RM}_{\mathrm{host}}}{(1+z)^2} \,. \label{eq:rm_eg}
\end{equation} 
As with $\mathrm{DM}_\mathrm{EG}$, we do not have redshift information available for our FRBs and therefore $\mathrm{RM}_\mathrm{EG}$ is left in our observer frame and is not converted to the rest frame of the FRB.

\subsection{Parallel magnetic field lower limit} \label{sec:bpar_lim}
The ratio between RM and DM, along the same LoS, provides an estimate of the electron density weighted average magnetic field strength parallel to the LoS and has often been used to study Galactic pulsars \citep[e.g.,][]{2019MNRAS.484.3646S, 2020MNRAS.496.2836N}. In the context of FRBs, we would like to study the rest frame electron density weighted average magnetic field strength parallel to the LoS in the FRB host galaxy and local environment,
\begin{equation}
\left<B_{\parallel,\mathrm{host}}\right> = 1.232 \frac{\left|\mathrm{RM}_\mathrm{EG}\right|}{\mathrm{DM}_\mathrm{EG}} (1+z)\,. \label{eq:b_host}
\end{equation}
Lacking redshift information, however, we use the $\mathrm{DM}_\mathrm{EG}$ and $\mathrm{RM}_\mathrm{EG}$ to calculate a lower limit on $\left|\left<B_{\parallel,\mathrm{host}}\right>\right|$ as measured in our observer frame, which we define as
\begin{equation}
\left|\beta\right| = 1.232 \frac{\left|\mathrm{RM}_\mathrm{EG}\right|}{\mathrm{DM}_\mathrm{EG}}~\mu\text{G} \leq \left|\left<B_{\parallel,\mathrm{host}}\right>\right|\label{eq:B} \,.
\end{equation}
The reason $\left|\beta\right|$ is a lower limit on $\left|\left<B_{\parallel,\mathrm{host}}\right>\right|$ is multifaceted: (i) $\mathrm{DM}_\mathrm{EG}$ will almost always have a significant $\mathrm{DM}_{\mathrm{IGM}}(z)$ contribution compared to the negligible $\mathrm{RM}_{\mathrm{IGM}} = 0~\mathrm{rad}~\mathrm{m}^{-2}$, (ii) $\left|\mathrm{RM}_\mathrm{EG}\right| \leq \left|\mathrm{RM}_\mathrm{host}\right|$ from Equation \ref{eq:rm_eg}, and (iii) $\left|\mathrm{RM}_\mathrm{EG}\right|$ scales as $(1 + z)^{-2}$ (see Equation \ref{eq:rm_eg}) while $\mathrm{DM}_\mathrm{EG}$ scales as $(1 + z)^{-1}$ (see Equation \ref{eq:dm_eg}) meaning that a redshift correction on $\left|\mathrm{RM}_\mathrm{EG}\right| / \mathrm{DM}_\mathrm{EG}$ for $z \geq 0$ would lead to a larger ratio. It is possible that peculiar LoS exists in which one of our assumptions breaks down and $\left|\beta\right| \leq \left|\left<B_{\parallel,\mathrm{host}}\right>\right|$ no longer holds (e.g., if a strongly magnetized intervening medium, which imparts a large RM, is unaccounted for). However, we argue that Equation \ref{eq:B} is applicable to the vast majority of FRBs and is particularly useful when applied to large populations of FRBs, as we do in this study. 

We emphasize again that $L/I$, $\psi$, $\mathrm{DM}_\mathrm{EG}$, $\mathrm{RM}_\mathrm{EG}$, and $\left|\beta\right|$ are observer frame quantities while $\mathrm{DM}_\mathrm{host}$, $\mathrm{RM}_\mathrm{host}$, and $\left<B_{\parallel,\mathrm{host}}\right>$ are in the FRB rest frame. Any results and/or discussion related to $\mathrm{DM}_\mathrm{EG}$, $\mathrm{RM}_\mathrm{EG}$, and $\left|\beta\right|$ throughout this work will be in the observer frame unless otherwise specified. 

\subsection{Statistical tests} \label{sec:stats_tests}
\subsubsection{Comparing distributions}
One of the primary objectives of this work is to rigorously examine whether polarization properties of repeating and non-repeating FRBs arise from the same or distinct distributions. To this end, we employ the Anderson-Darling \citep[AD;][]{ADtest} and Kolmogorov-Smirnov \citep[KS;][]{smirnov1948, kolmogorov1956} tests. We utilize the Python implementation of these tests ({\tt anderson\_ksamp} and {\tt ks\_2samp}, respectively) from the {\tt scipy.stats} V1.9.3 package \citep{2020SciPy-NMeth} and apply them to the $L/I$, $\mathrm{DM}_\mathrm{EG}$, $|\mathrm{RM}_\mathrm{EG}|$, and $\left|\beta\right|$ distributions of our repeating and non-repeating FRB samples. The AD test returns a test statistic $S_\mathrm{AD}$ and a $p$-value ($p_\mathrm{AD}$; floored at $0.001$ and capped at $0.25$), while the KS test returns test statistic $S_\mathrm{KS}$ and $p$-value $p_\mathrm{KS}$, assuming a two-sided null hypothesis. The $p$-values from both of these tests signify the level at which we can reject the null hypothesis that the repeating and non-repeating polarized properties originate from the same underlying population (i.e., we reject the null hypothesis at a $1 - p$ level).

Here, the KS test is measuring the supremum between the observed repeating and non-repeating FRB cumulative distribution functions (CDFs) for a given parameter and is, therefore, sensitive to sharp differences between the distributions. On the other hand, the AD test statistic is more sensitive to smaller but more persistent differences, even in the tails of the distributions. While we present the results of both tests, we note that the KS test generally tends to produce more conservative significance levels across our sample when compared to the AD test.

When parts of data sets (but not all of the data set) are left-censored (i.e., contain upper limits, which is the case for some values of $L/I$ here and for all burst rates of non-repeating FRBs), we additionally employ statistical methods that take censoring into account \citep[see, e.g.,][]{2012msma.book.....F}. For this, we use methods implemented in R's NADA V1.6-1.1 package\footnote{\url{https://rdrr.io/cran/NADA/}} \citep{helsel2005nondetects, NADApackage, Rsoftware}. To begin with, we additionally derive statistics from empirical cumulative distribution functions calculated using the Kaplan-Meier method \citep{kaplan-meier1958}, as implemented in the \texttt{cenfit} function of NADA. When left-censored data are present, we also perform the Peto \& Peto \citep{peto-peto1972} and log-rank tests \citep{mantel1966}, also as implemented in the \texttt{cendiff} function of NADA.

For all four tests, we set a significance level at $\alpha = 2.7 \times 10^{-3}$ (i.e., at the $3\sigma$ level), such that when the $p$-value from either test is less than $\alpha$, we consider the result statistically significant. We will refer to a result as being ``marginally significant'' when it satisfies $2.7 \times 10^{-3} < \alpha < 0.05$ and not significant if $\alpha \geq 0.05$.

\subsubsection{Correlations between parameters}
In some cases we would like to evaluate whether a positive or negative correlation exists between two parameters in our data. For this purpose, we use Spearman's rank correlation coefficient \citep{spearman1904}, which measures the preponderance of a monotonic relationship between two input parameters. This test is implemented with the {\tt scipy.stats.spearmanr} V1.9.3 Python module \citep{2020SciPy-NMeth}. The test returns a Spearman rank correlation coefficient $S_\mathrm{SR}$ which is $0$ if there is no correlation and $+1$ ($-1$) if there is a perfectly positive (negative) monotonic correlation and also an associated $p$-value, $p_\mathrm{SR}$. This is used throughout Section \ref{sec:results} to determine the significance of any possible correlations between $\mathrm{DM}_\mathrm{EG}$ and $|\mathrm{RM}_\mathrm{EG}|$, $L/I$ and $|\mathrm{RM}_\mathrm{EG}|$, and burst rate with various polarization properties. When testing for correlations between singly or multiply left-censored data sets, we also compute Kendall's $\tau$ correlation coefficient \citep{kendall1938}, as implemented in the \texttt{cenken} function of NADA.  We use the same significance levels for these tests as we defined for the AD and KS tests above.

\section{Results} \label{sec:results}
\subsection{Polarization properties of the non-repeating FRBs}
In this section, we provide the derived polarization results for the non-repeating FRBs in the first CHIME/FRB baseband catalog. 

Of the 128 non-repeating FRBs, 89 produce a $> 6\sigma$ linearly polarized detection and are well characterized by the CHIME/FRB polarization pipeline. Of these, five are contaminated by instrumental polarization (i.e., the FDF peak from the Faraday rotation was smaller than the peak caused by instrumental polarization), but their values of $\mathrm{RM}_\mathrm{obs}$ are corrected by re-fitting to a secondary peak in the FDF. In all five cases, the polarized fit was significantly improved after adjusting for the instrumental polarization (see Section \ref{sec:pipeline_products}). However, note that we cannot derive a robust $L/I$ measurement for these five FRBs and, as such, they are excluded from any statistical tests applied to the $L/I$ data. 

There are 29 FRBs for which no significant linear polarization was detected. We consider these bursts unpolarized and derive upper limits for their linear polarization fraction $L/I$. First, we compute the S/N in Stokes $I$, $\mathrm{S/N}(I)$ by averaging over the emitting band of the FRB, subtracting the mean $I$ from an off-pulse (i.e., noise) region from the $I$ time profile across the FRB burst duration, and dividing by the standard deviation in the off-pulse region. Given the S/N in Stokes $I$, $\mathrm{S/N}(I)$, and our linear polarization detection threshold on the S/N in $L$, $\mathrm{S/N}(L) > 6$, we can find the requisite minimum $L/I$ needed to satisfy $\mathrm{S/N}(L) = 6$ as $L/I = 6/(\mathrm{S/N}(I))$. This serves as the upper limit for the time and frequency averaged $L/I$ of a given unpolarized FRB. For another 10 FRBs, we detect a significant linear polarization but are uncertain about the accuracy of the polarimetric results due to low S/N in Stokes $I$ and/or strong instrumental polarization contamination and, therefore, do not report their polarization properties.

\begin{figure*}[ht!]
\begin{center}
    \includegraphics[width=0.205\textwidth]{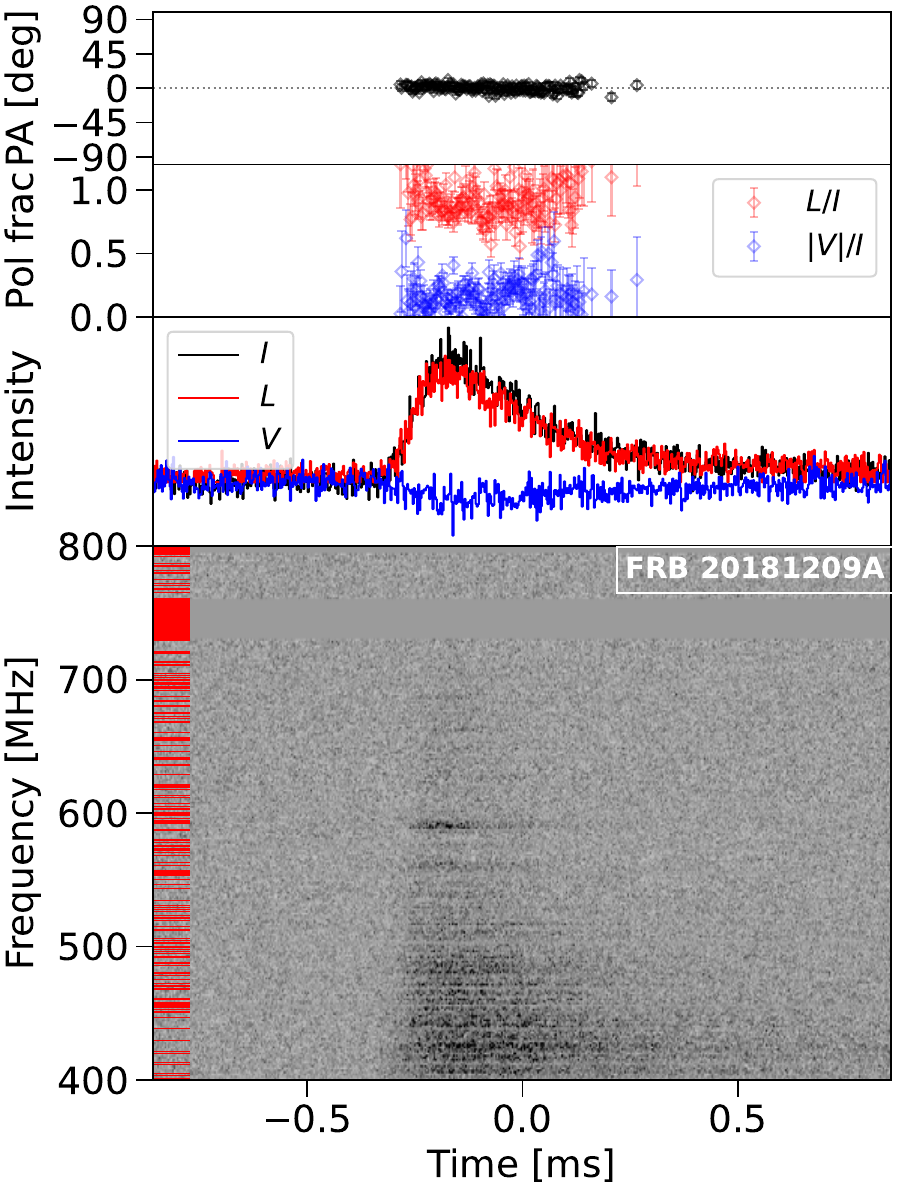}
    \includegraphics[width=0.205\textwidth]{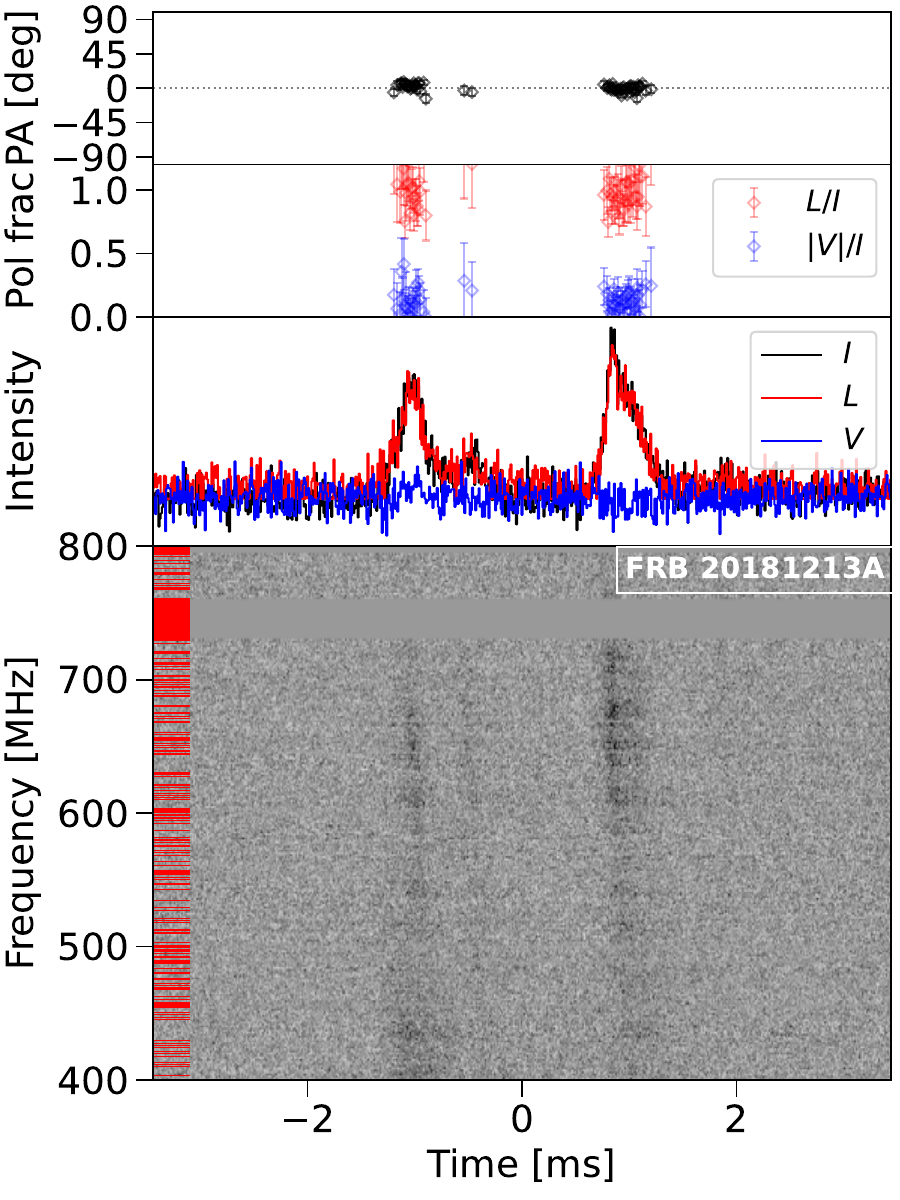}
    \includegraphics[width=0.205\textwidth]{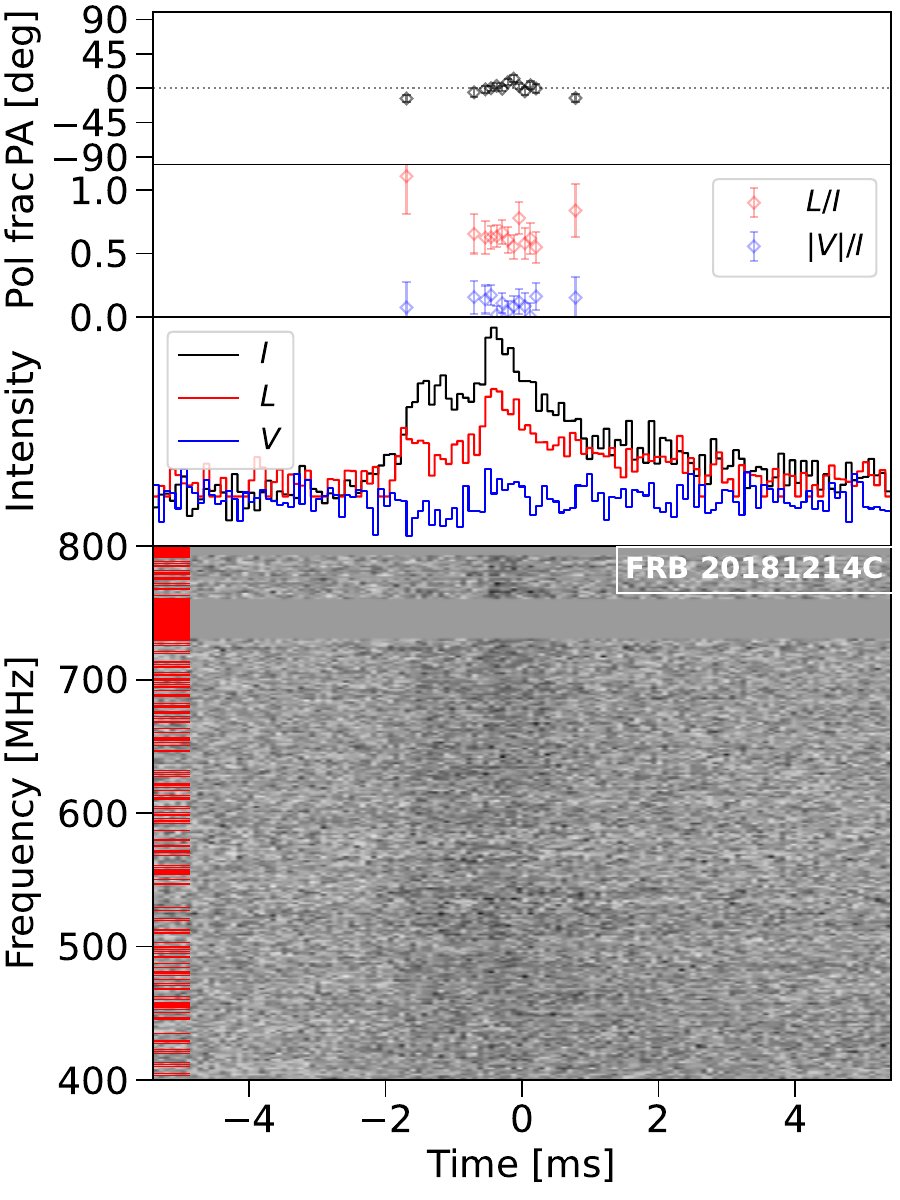}
    \includegraphics[width=0.205\textwidth]{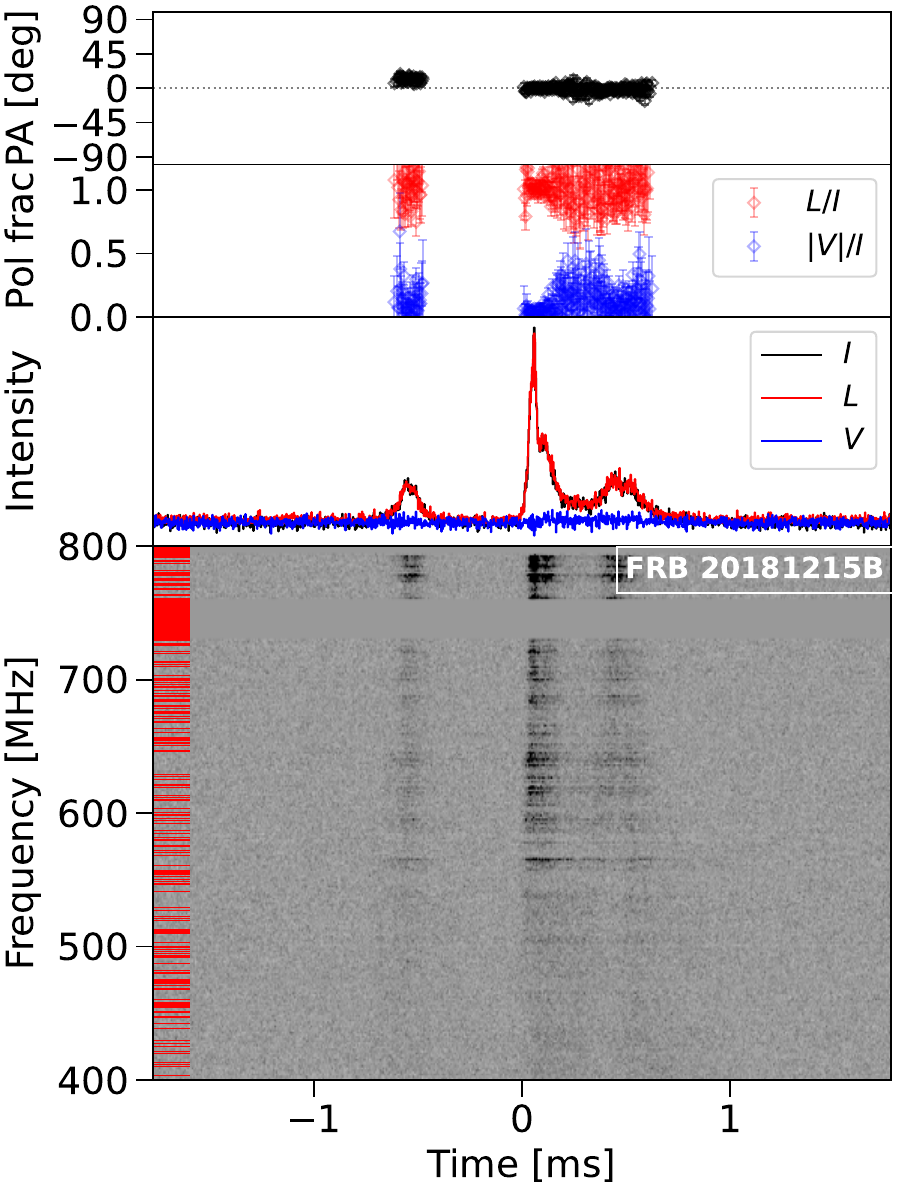}

    \includegraphics[width=0.205\textwidth]{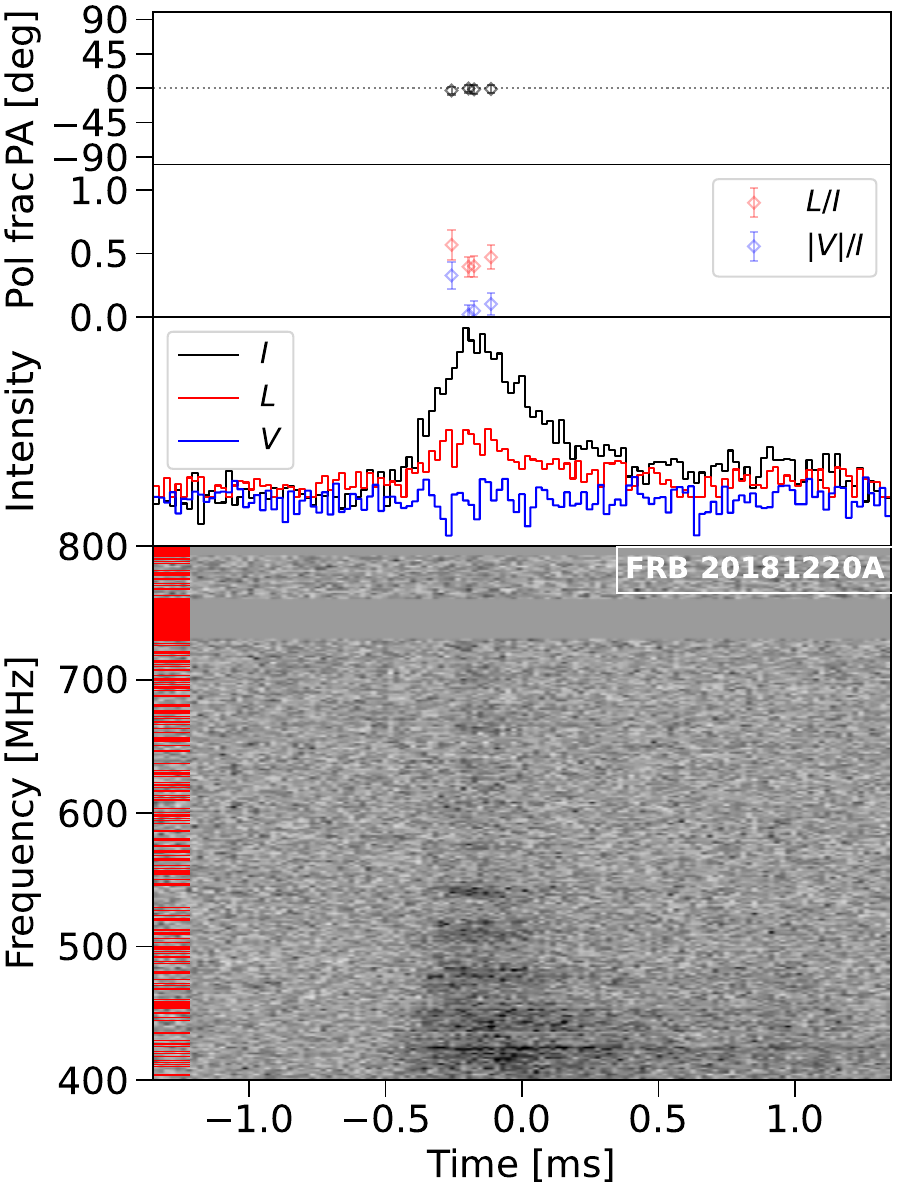}
    \includegraphics[width=0.205\textwidth]{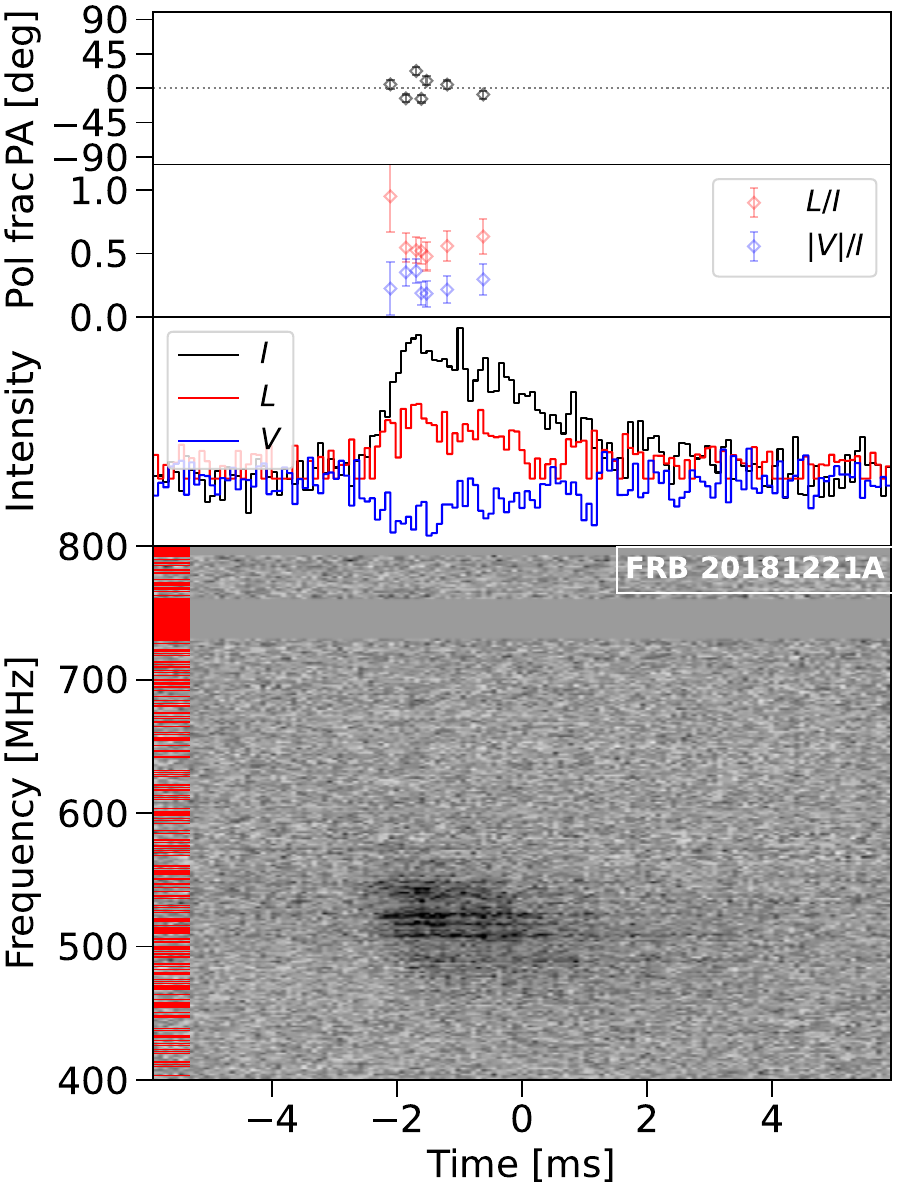}
    \includegraphics[width=0.205\textwidth]{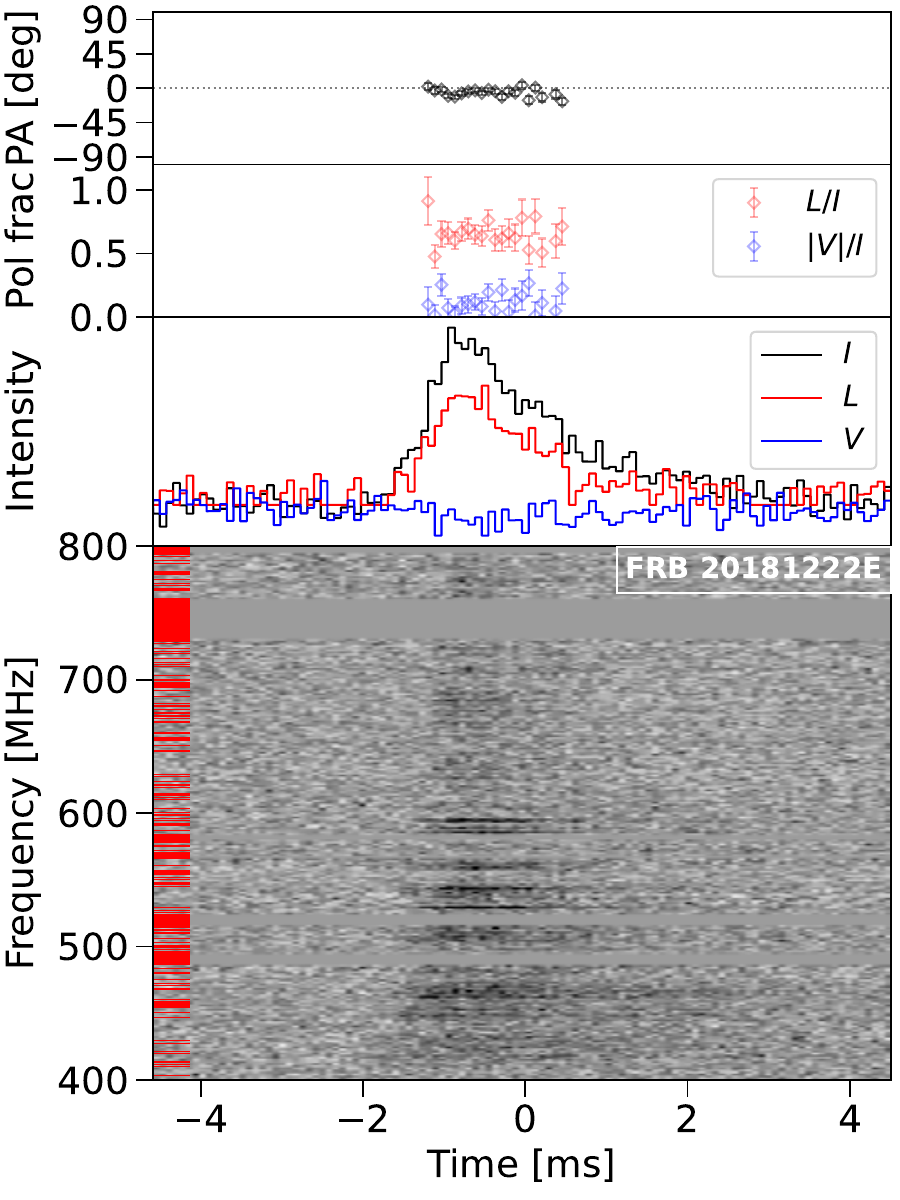}
    \includegraphics[width=0.205\textwidth]{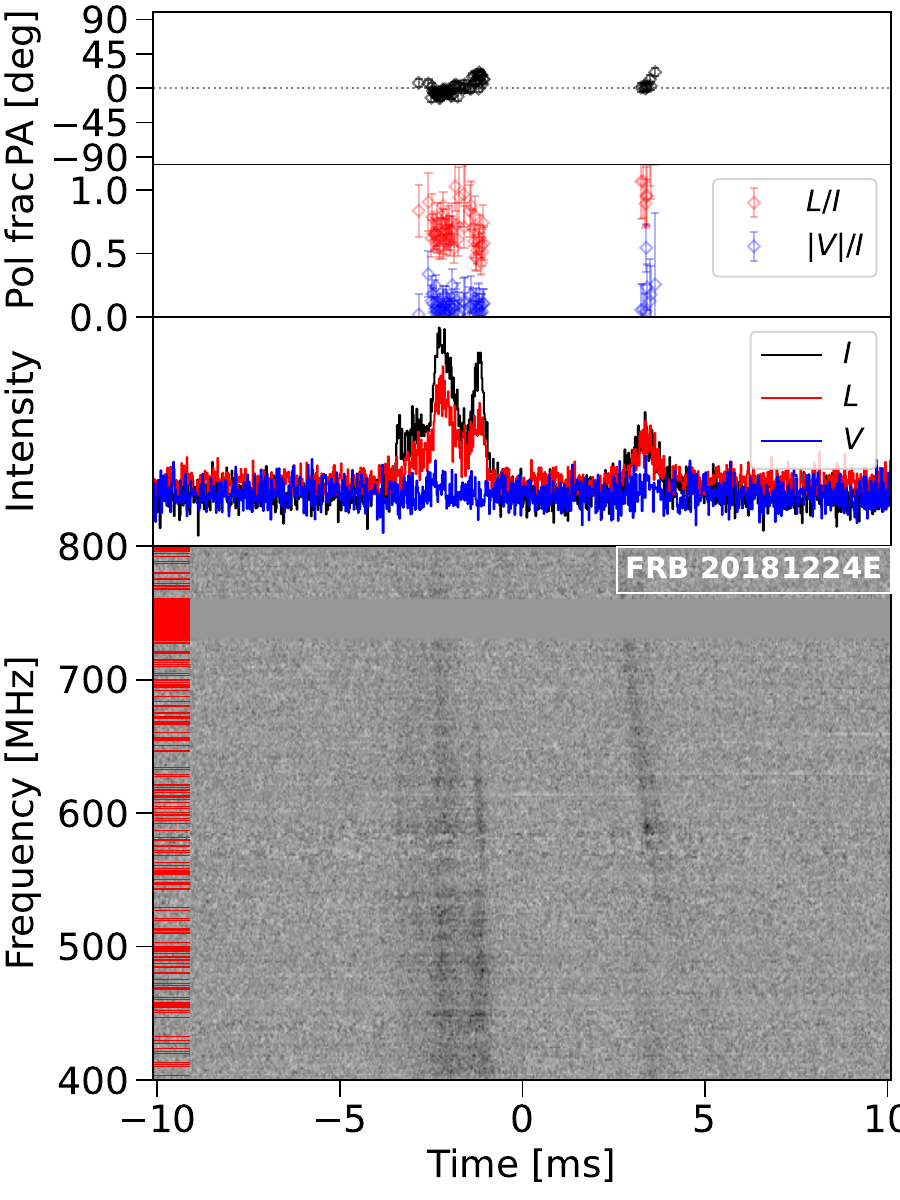}
    
    \includegraphics[width=0.205\textwidth]{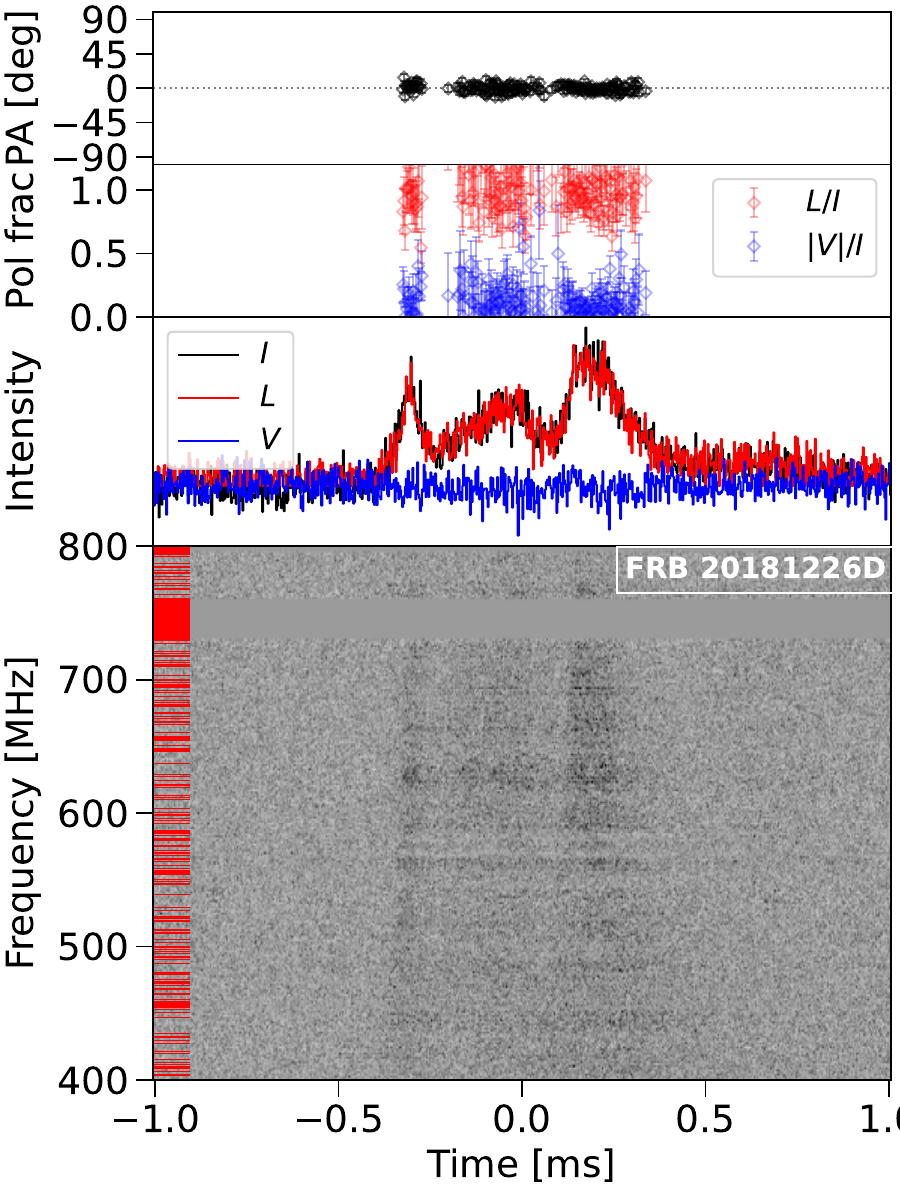}
    \includegraphics[width=0.205\textwidth]{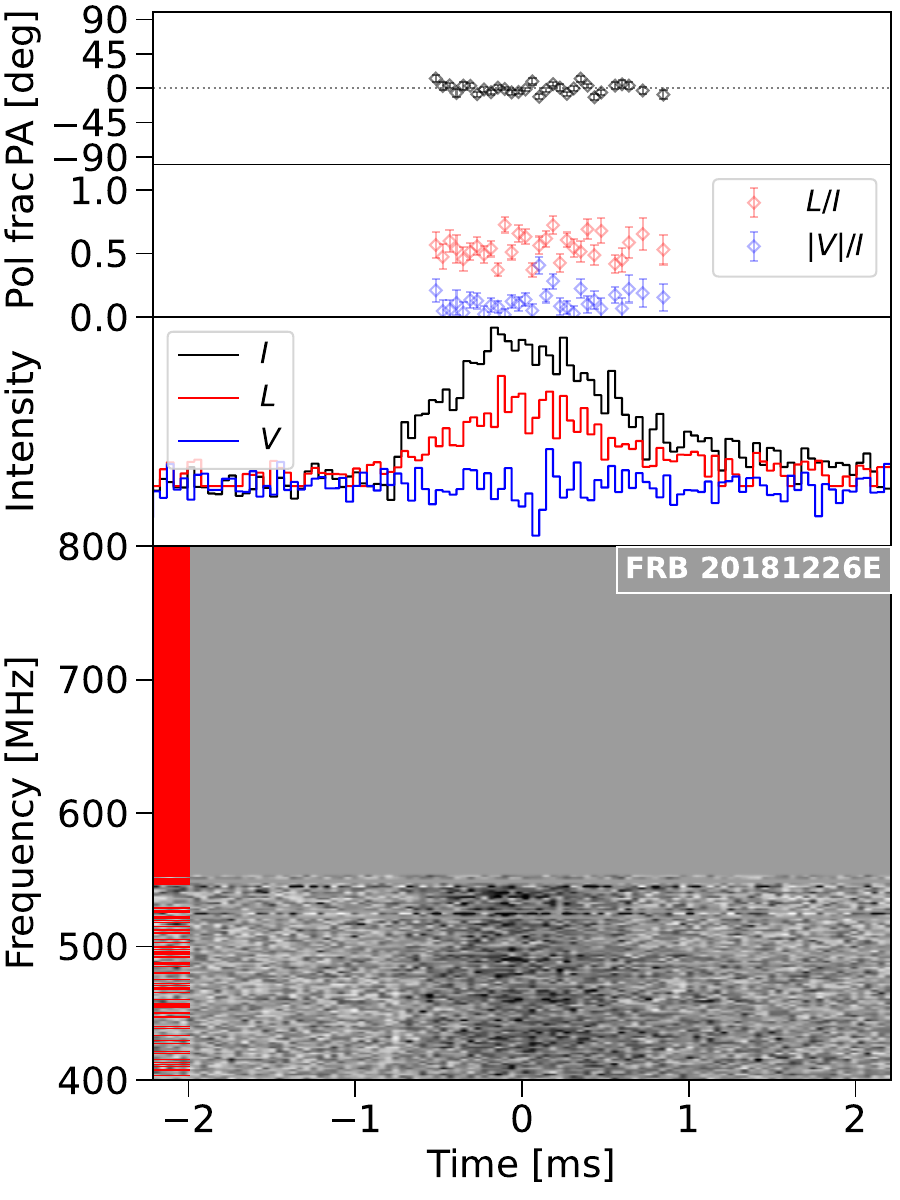}
    \includegraphics[width=0.205\textwidth]{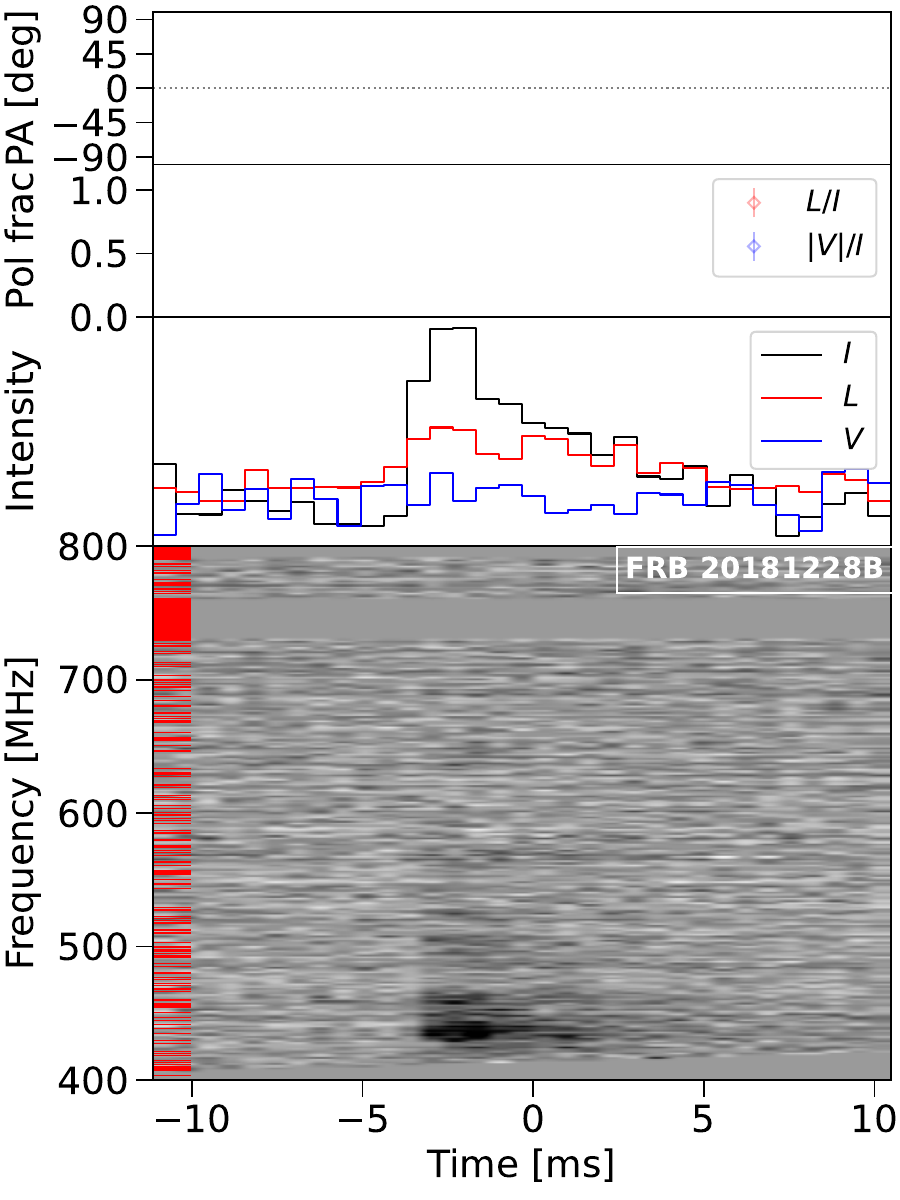}
    \includegraphics[width=0.205\textwidth]{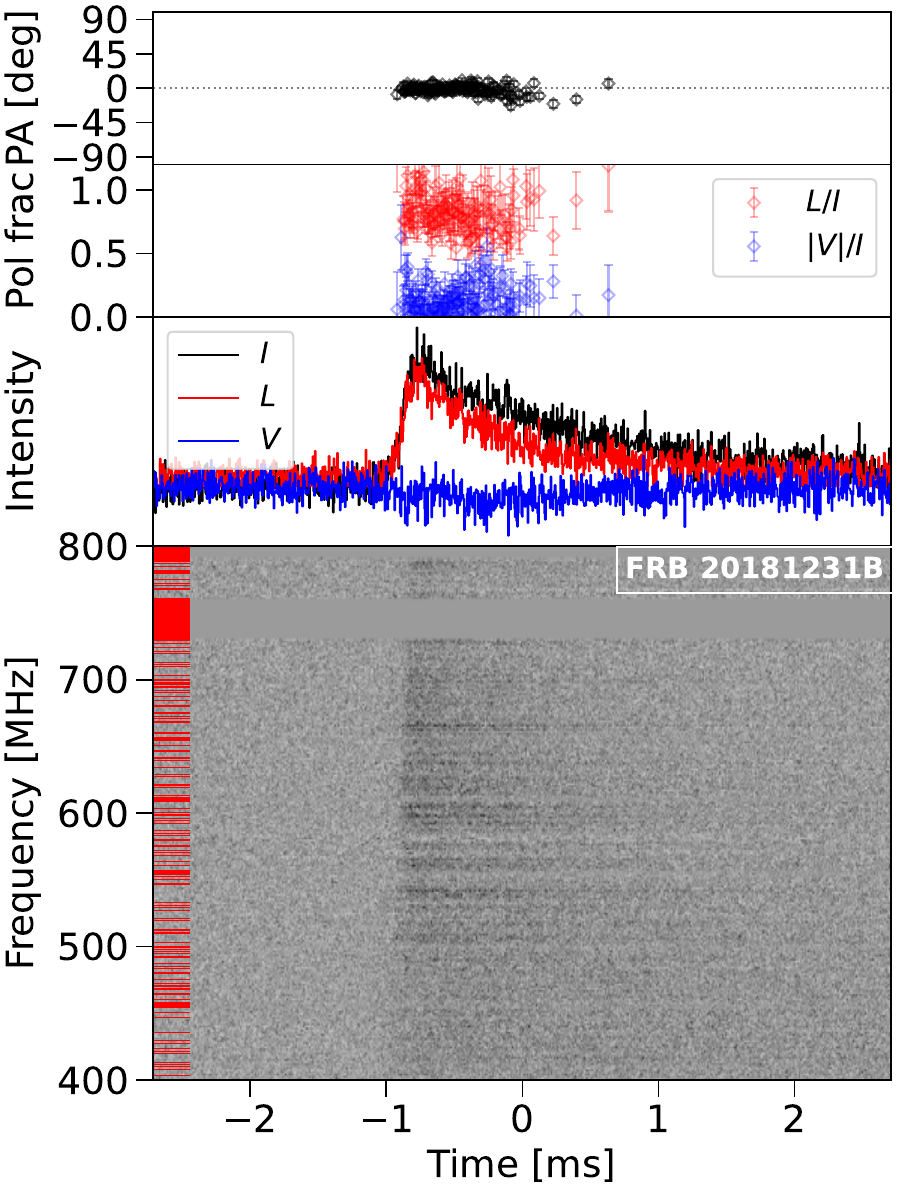}
    
    \includegraphics[width=0.205\textwidth]{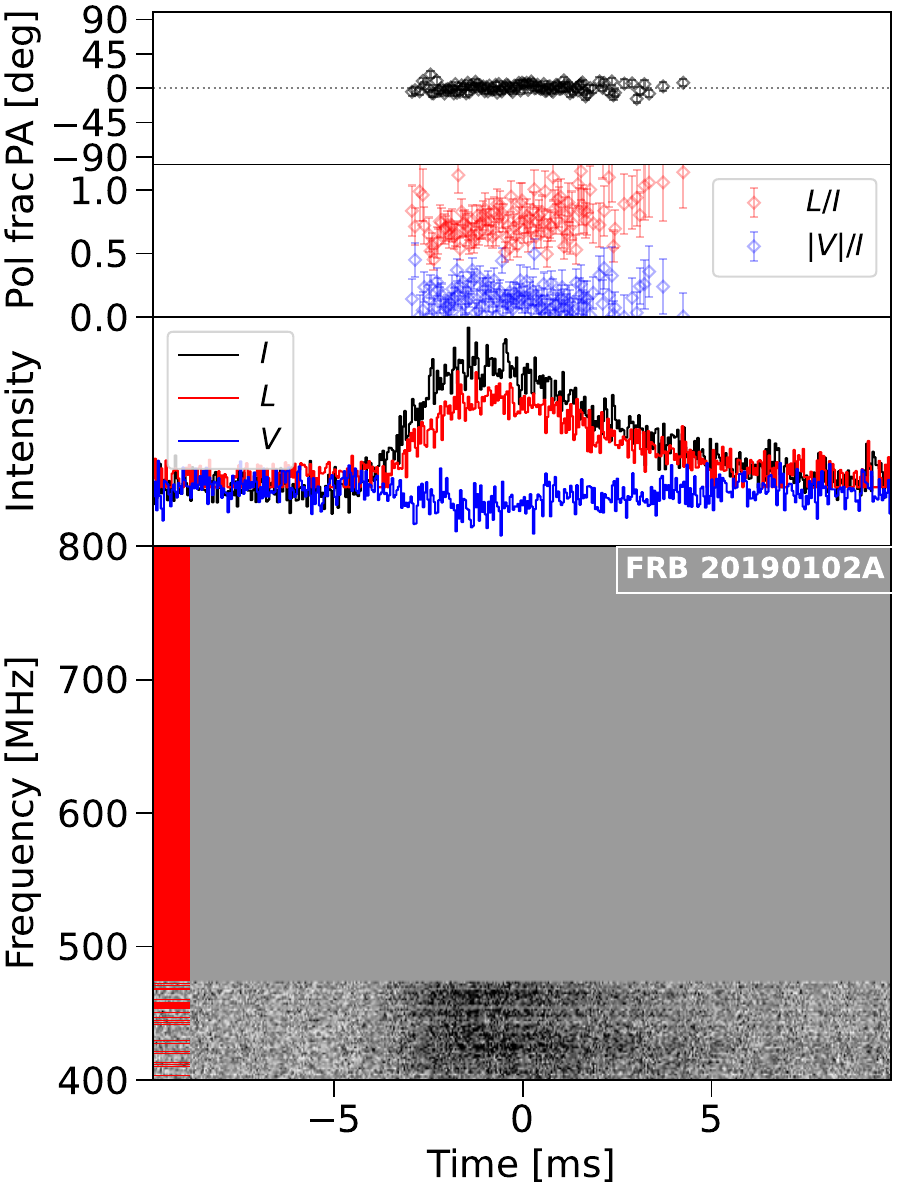}
    \includegraphics[width=0.205\textwidth]{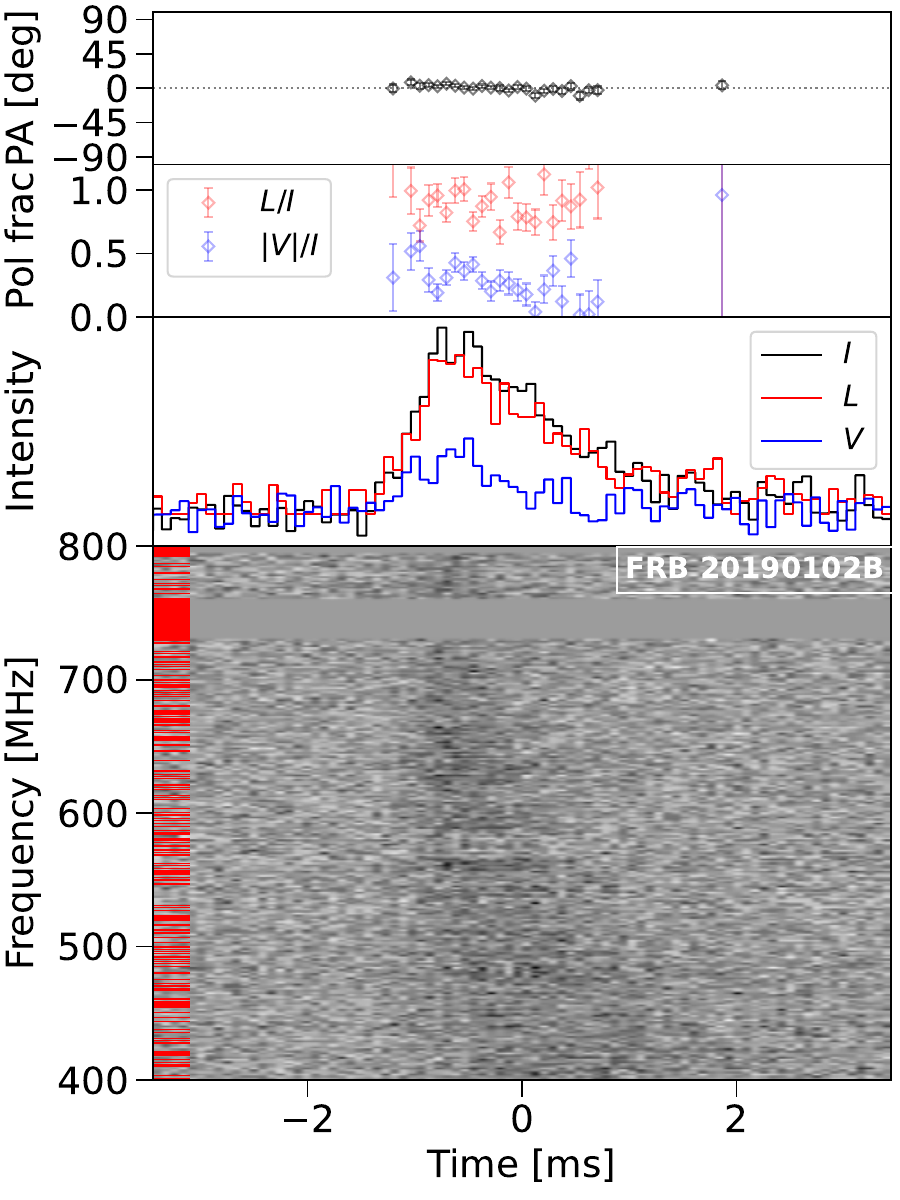}
    \includegraphics[width=0.205\textwidth]{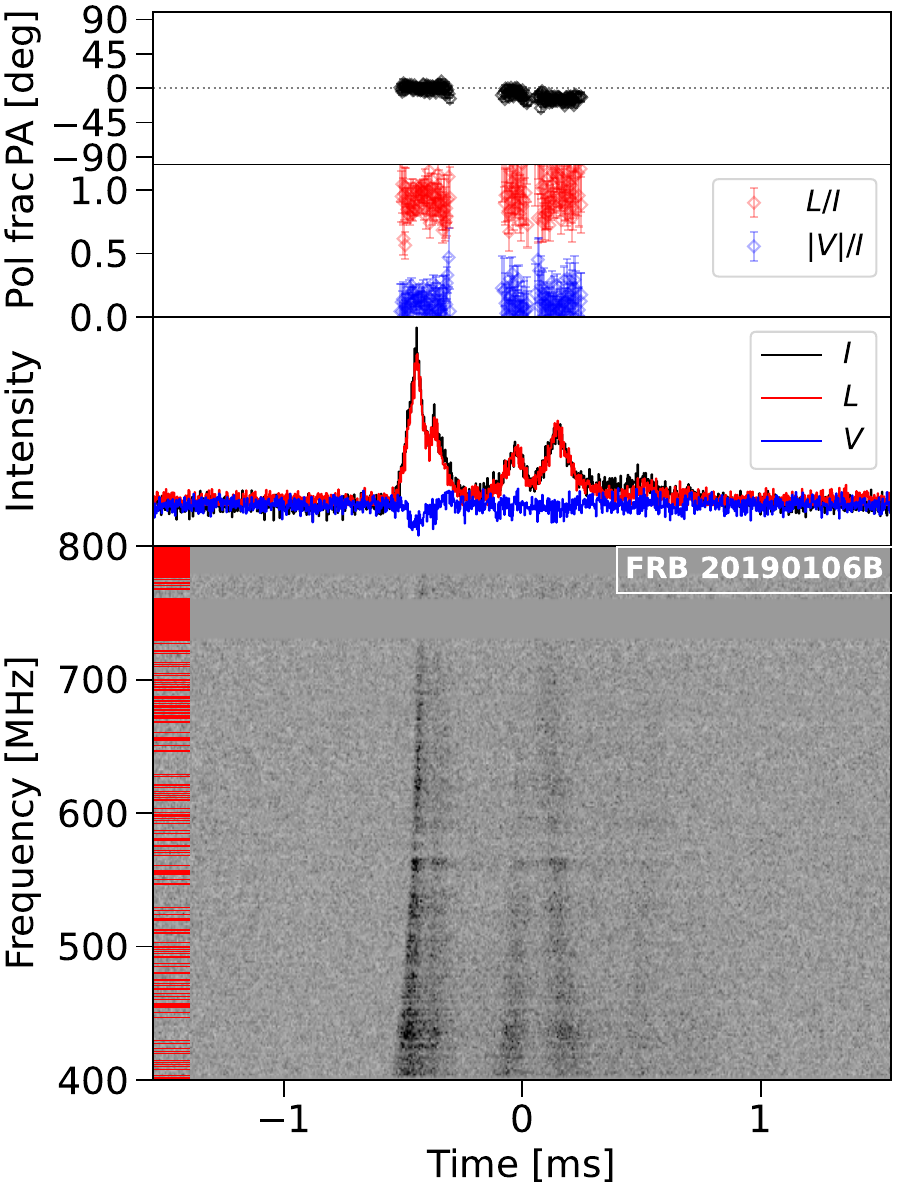}
    \includegraphics[width=0.205\textwidth]{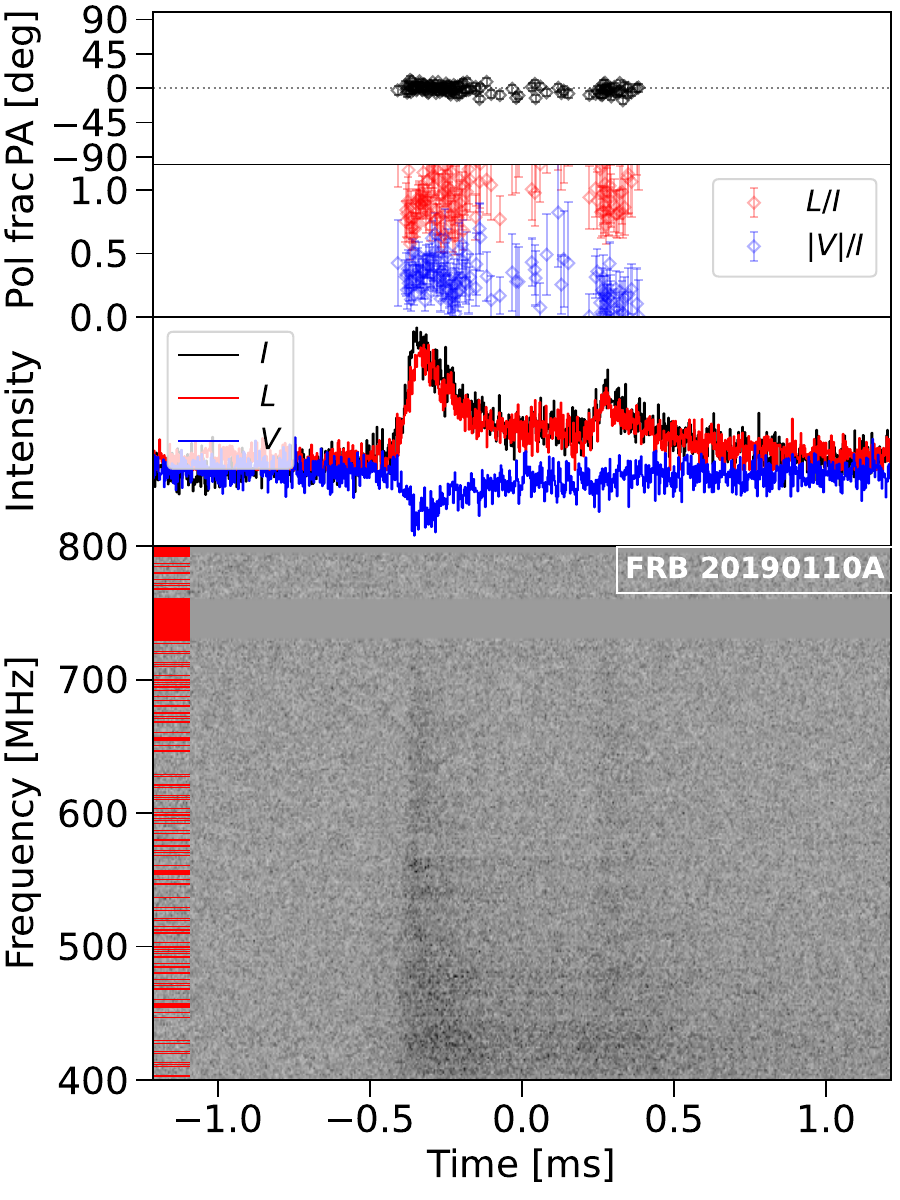}
    \caption{Total intensity waterfalls (de-dispersed to the associated $\mathrm{DM}_\mathrm{obs,struct}$) and temporal profiles of Stokes $I$ (black line), $L = \sqrt{Q^2 + U^2}$ (red line), $V$ (blue line), linear polarization fraction ($L/I$; red circles), circular polarization fraction ($|V|/I$; blue circles), and linear polarization position angle (PA; black circles) for the first 16 FRBs from Table \ref{tb:pol_results}. All points plotted on the $L/I$, $|V|/I$, and PA profiles exceed a $L$ S/N limit of 5 (see Appendix \ref{app:B} for details). Some faint, low linear polarization FRBs, therefore, may have very few (or no) points on their profiles, for example FRB 20181228B. Channels masked out due to radio frequency interference are highlighted by red streaks on the left-hand side of the total intensity waterfall plots. As some of the data have been downsampled by a factor of $n_\mathrm{down} > 1$, the resolution of the flagged radio frequency interference channels and Stokes images do not always match exactly.}
    \label{fig:waterfalls}
\end{center}
\end{figure*}

\begin{figure*}[ht!]
\begin{center}
    \includegraphics[width=0.246\textwidth]{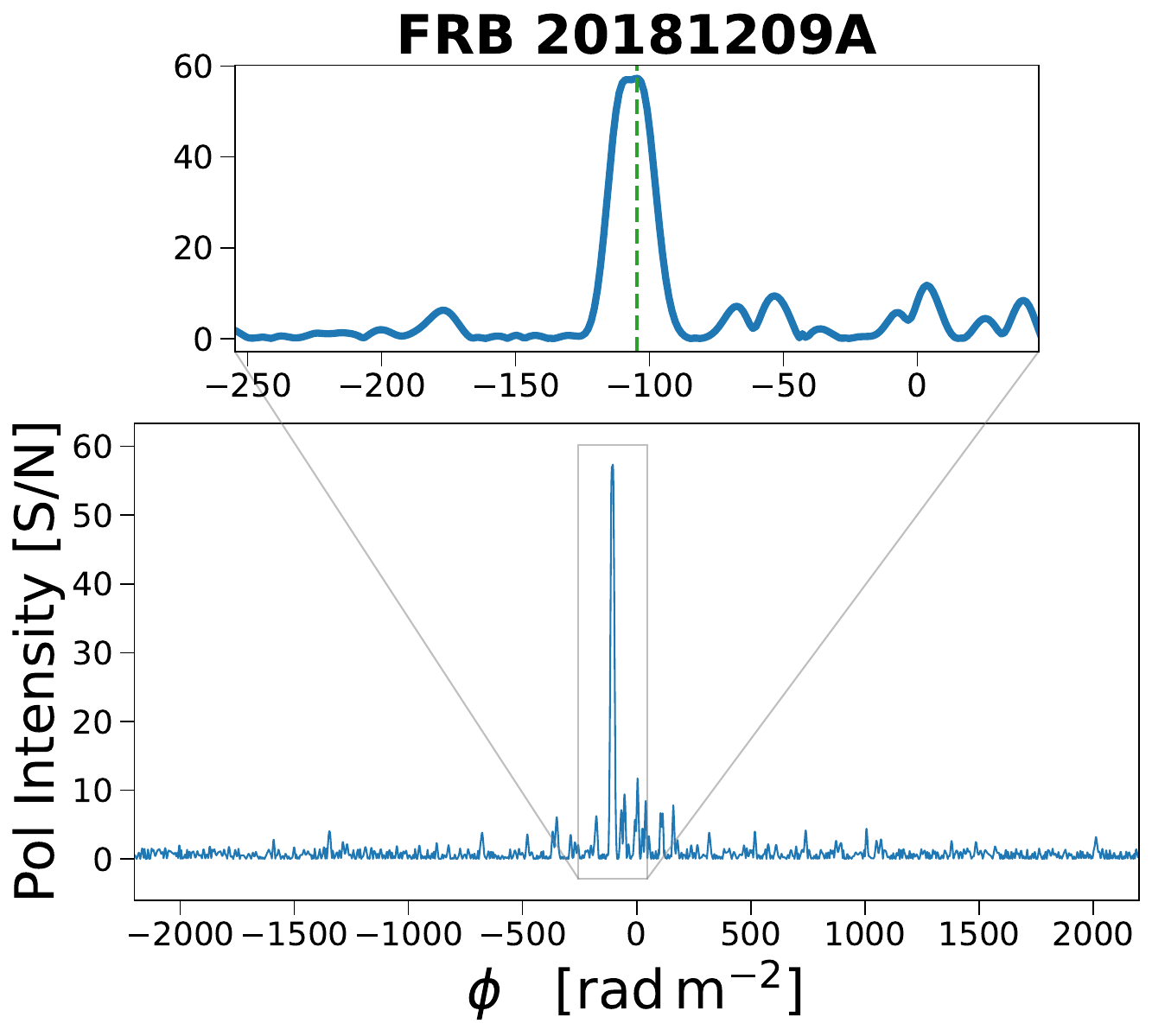}
    \includegraphics[width=0.246\textwidth]{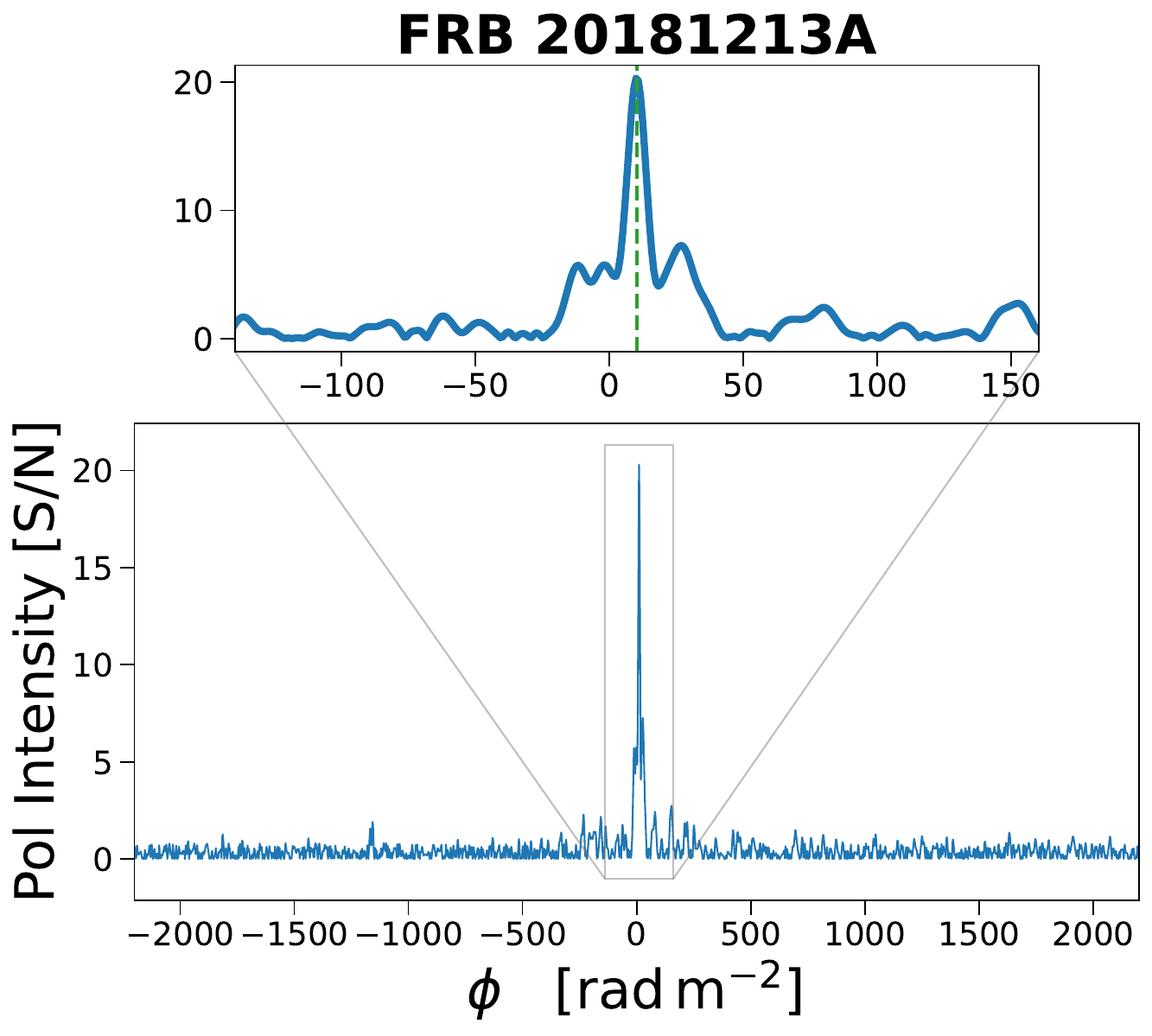}
    \includegraphics[width=0.246\textwidth]{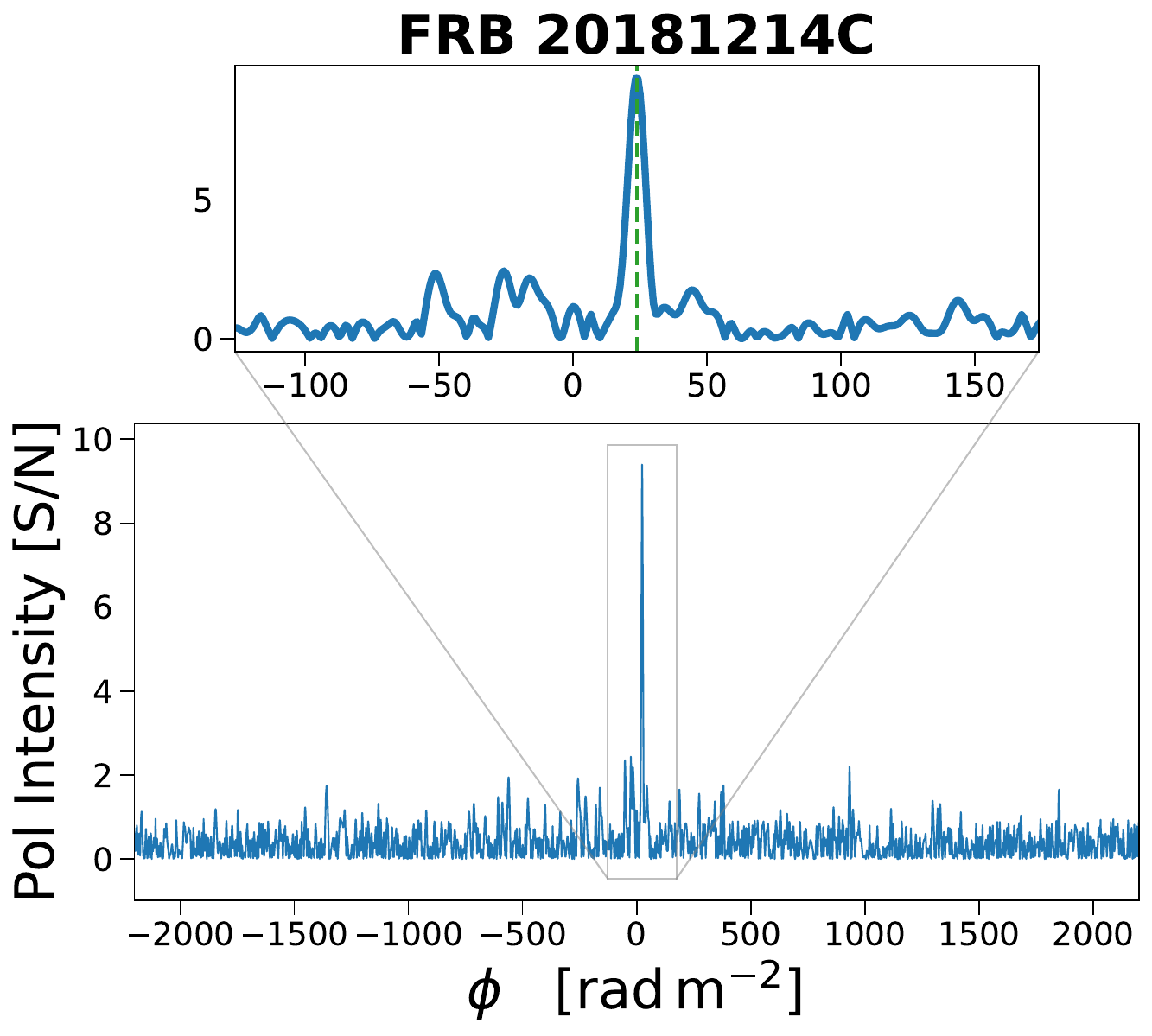}
    \includegraphics[width=0.246\textwidth]{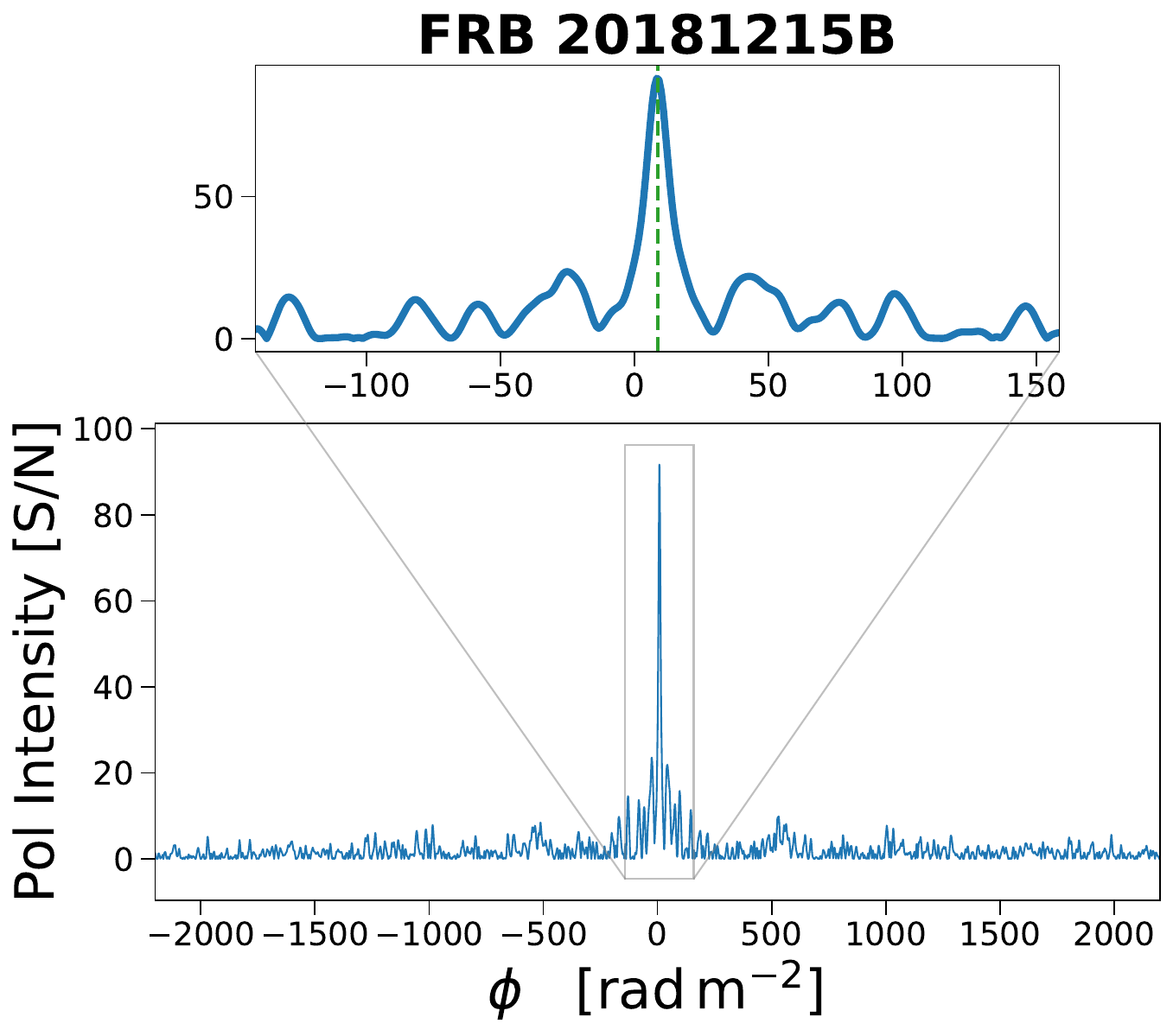}
    
    \includegraphics[width=0.246\textwidth]{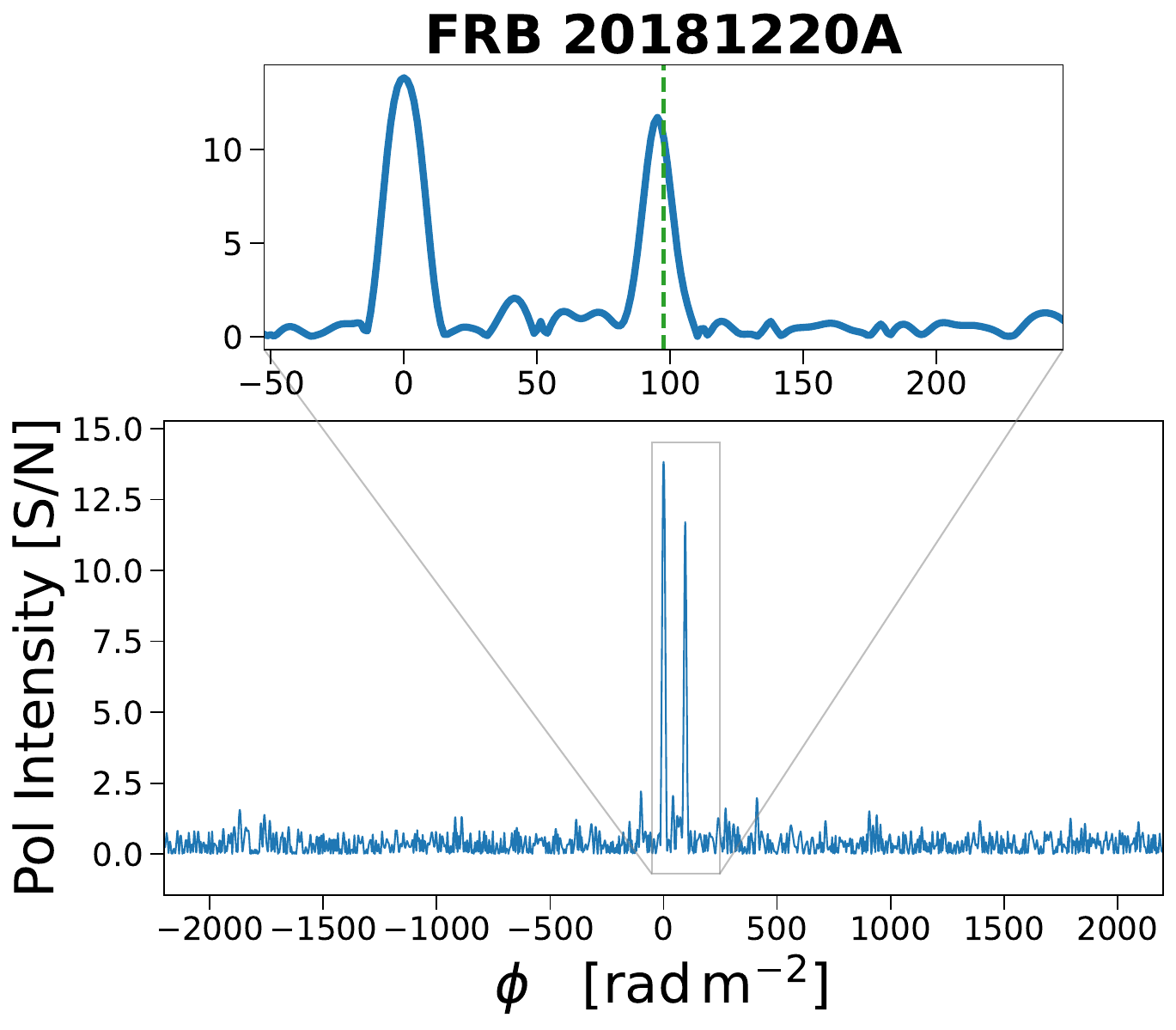}
    \includegraphics[width=0.246\textwidth]{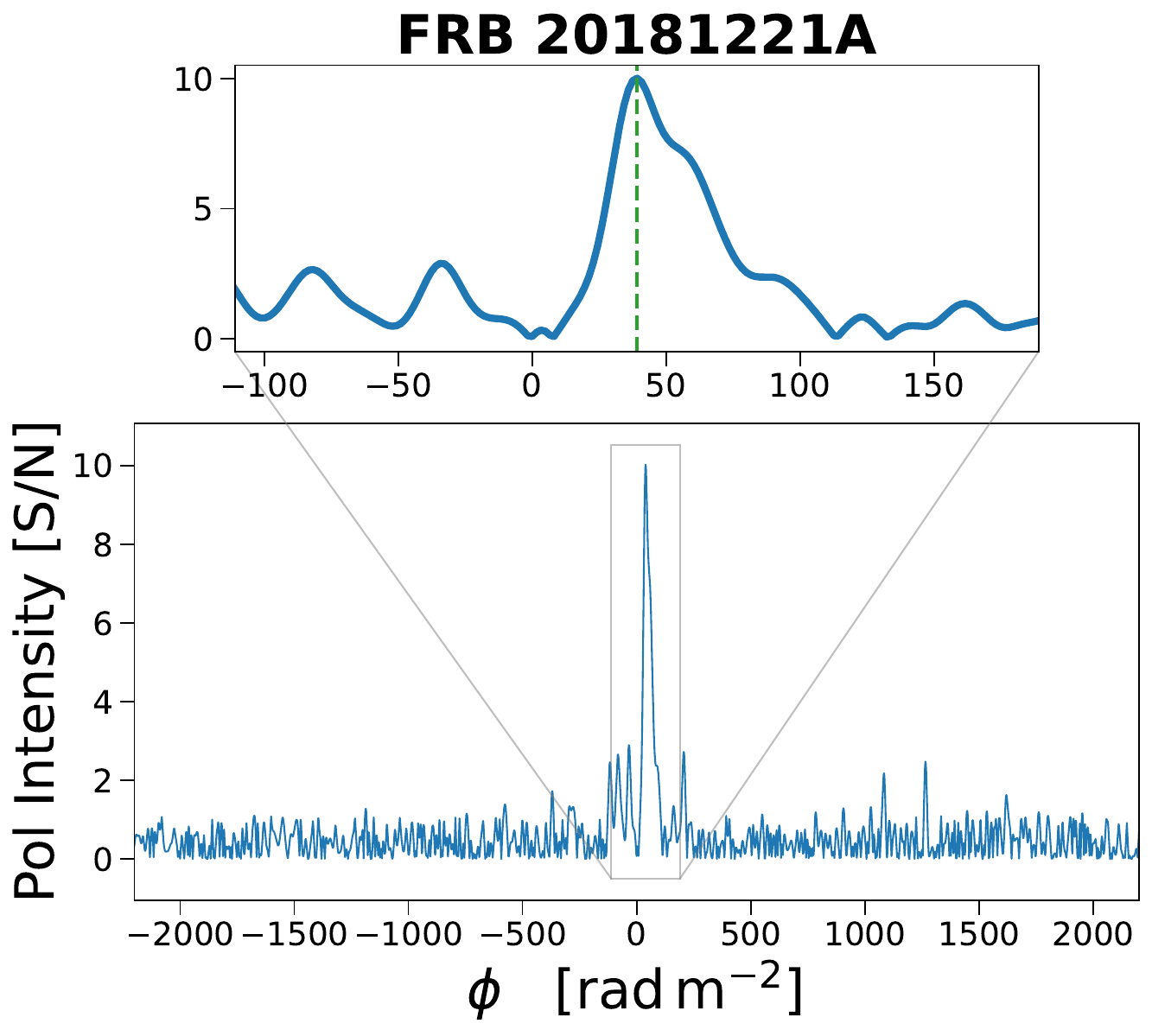}
    \includegraphics[width=0.246\textwidth]{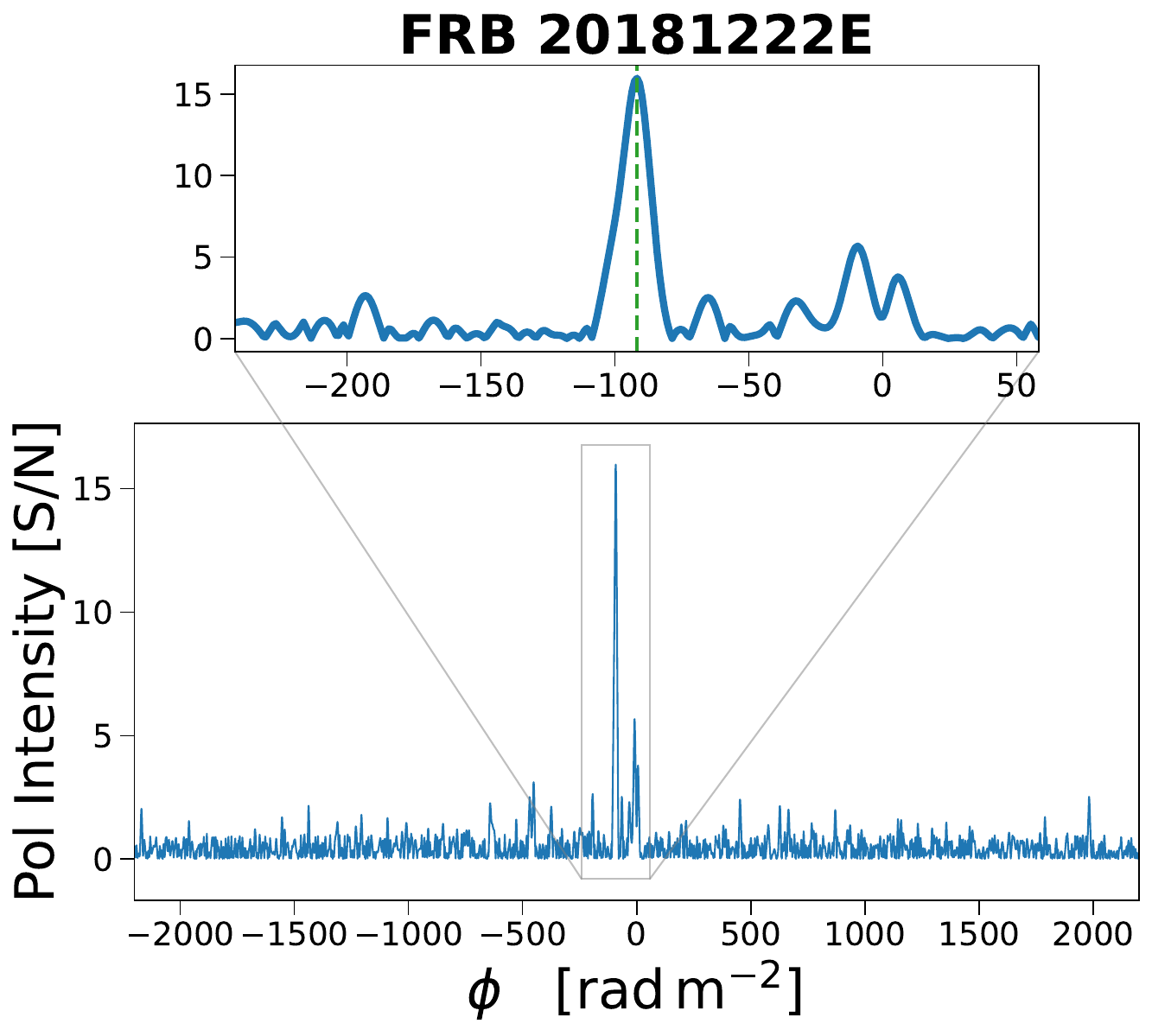}
    \includegraphics[width=0.246\textwidth]{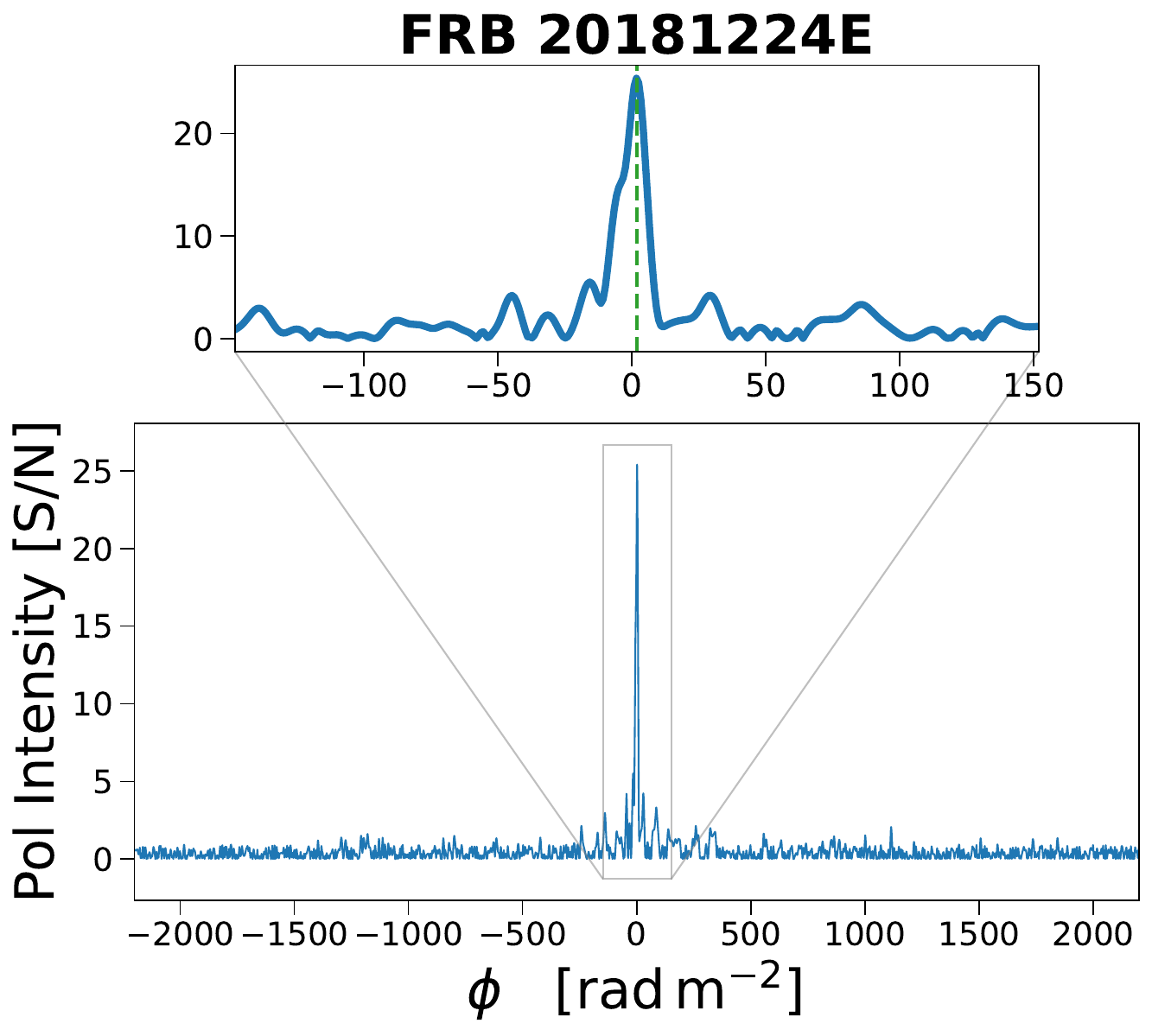}
    
    \includegraphics[width=0.246\textwidth]{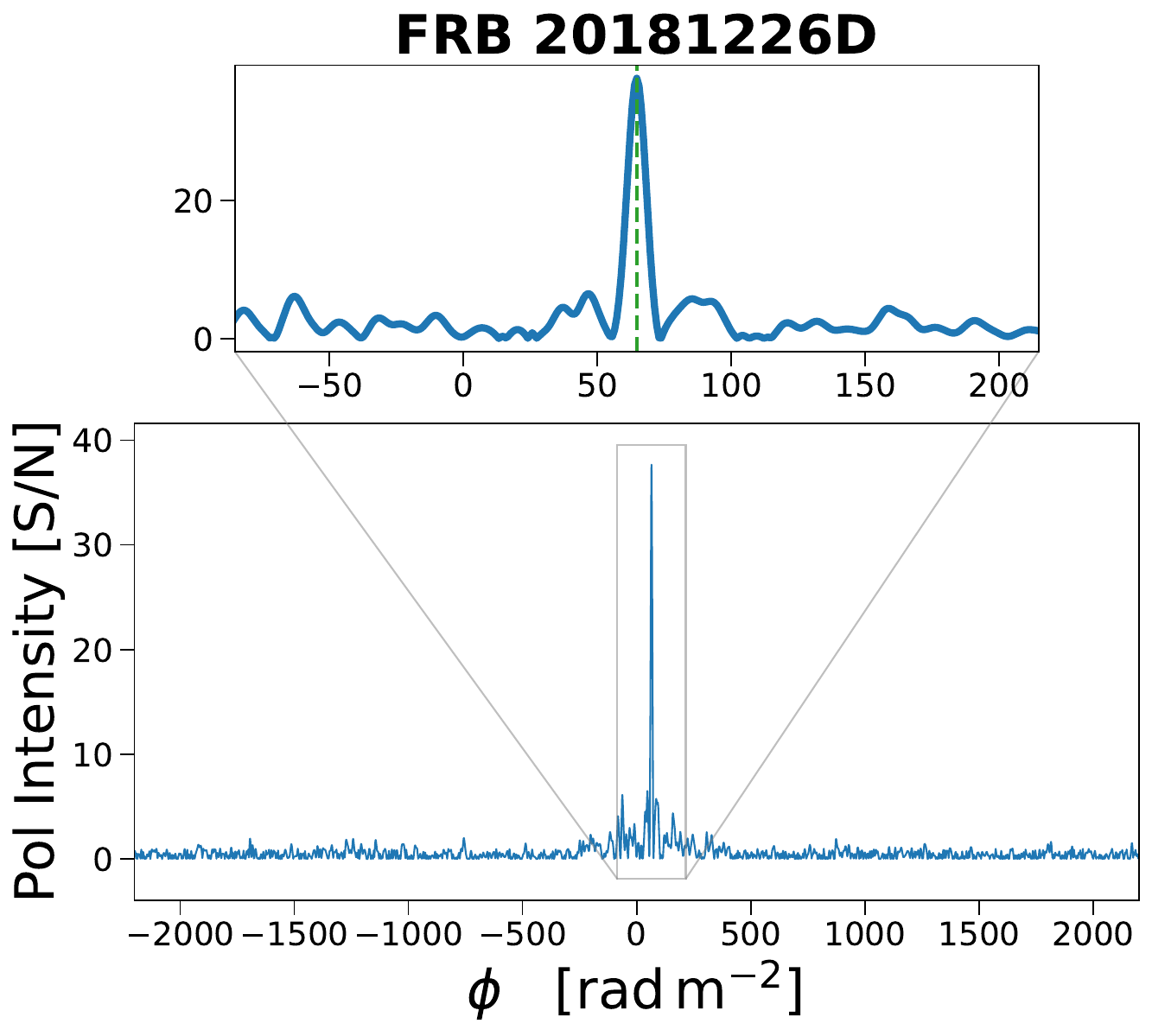}
    \includegraphics[width=0.246\textwidth]{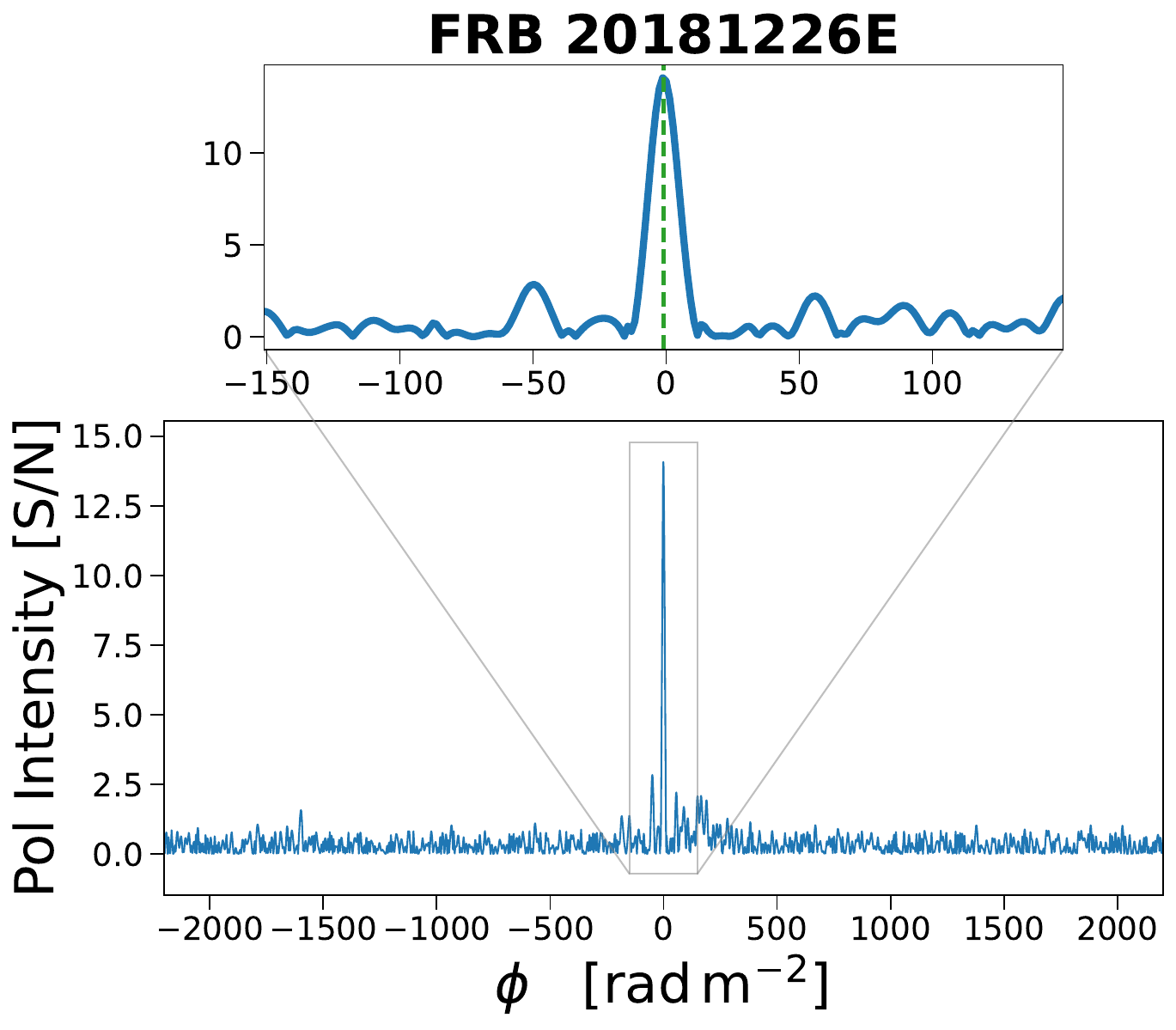}
    \includegraphics[width=0.246\textwidth]{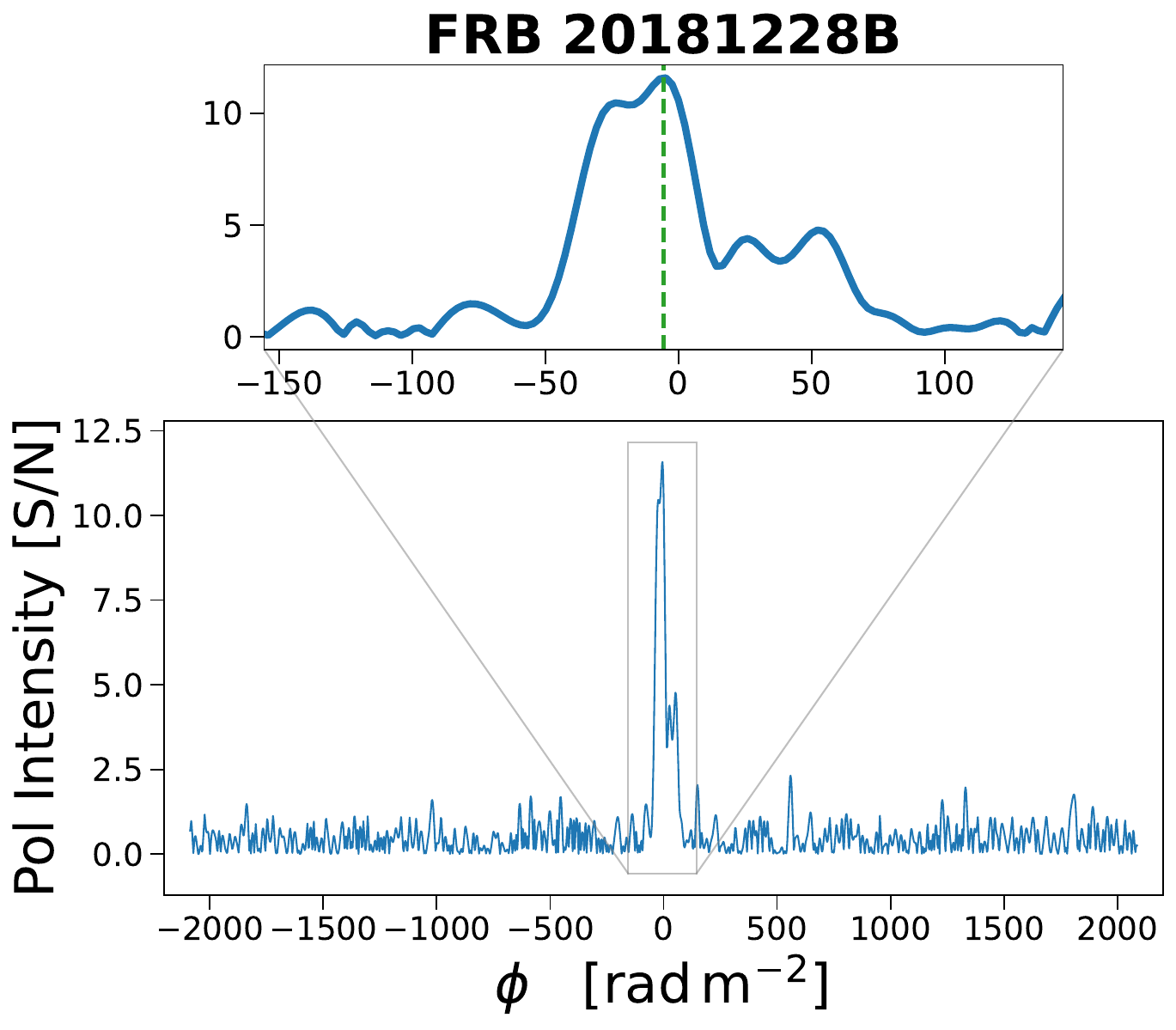}
    \includegraphics[width=0.246\textwidth]{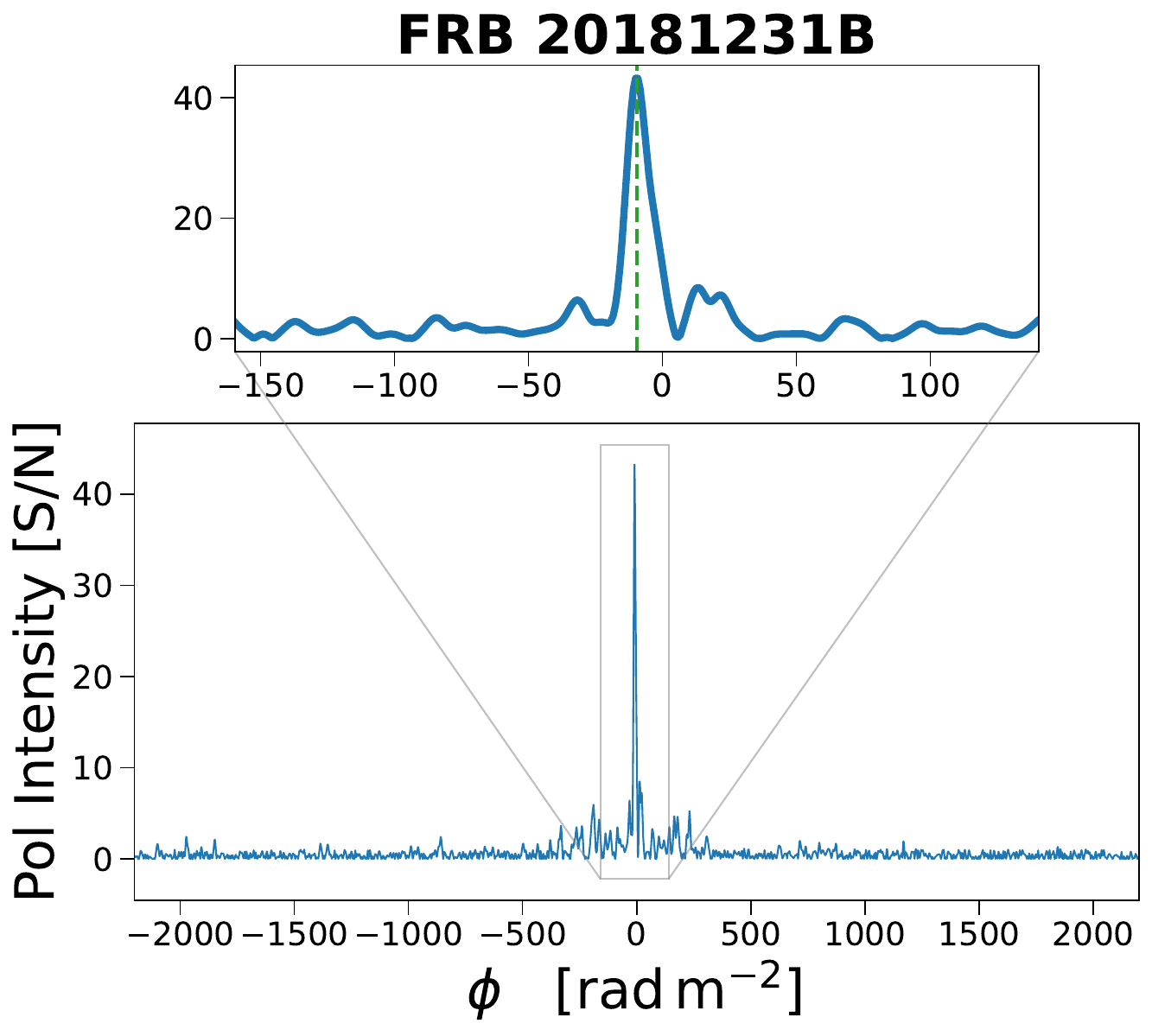}
    
    \includegraphics[width=0.246\textwidth]{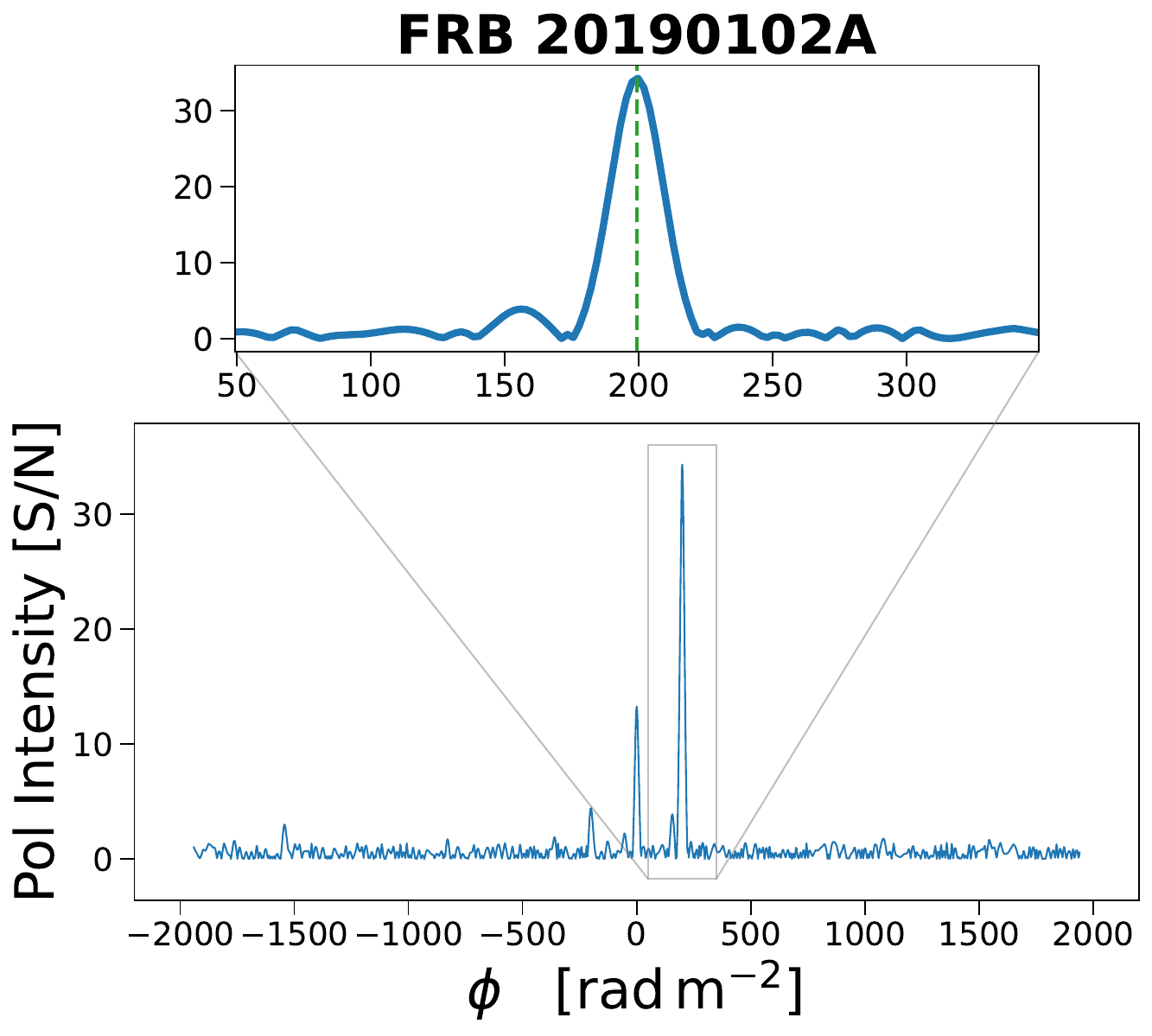}
    \includegraphics[width=0.246\textwidth]{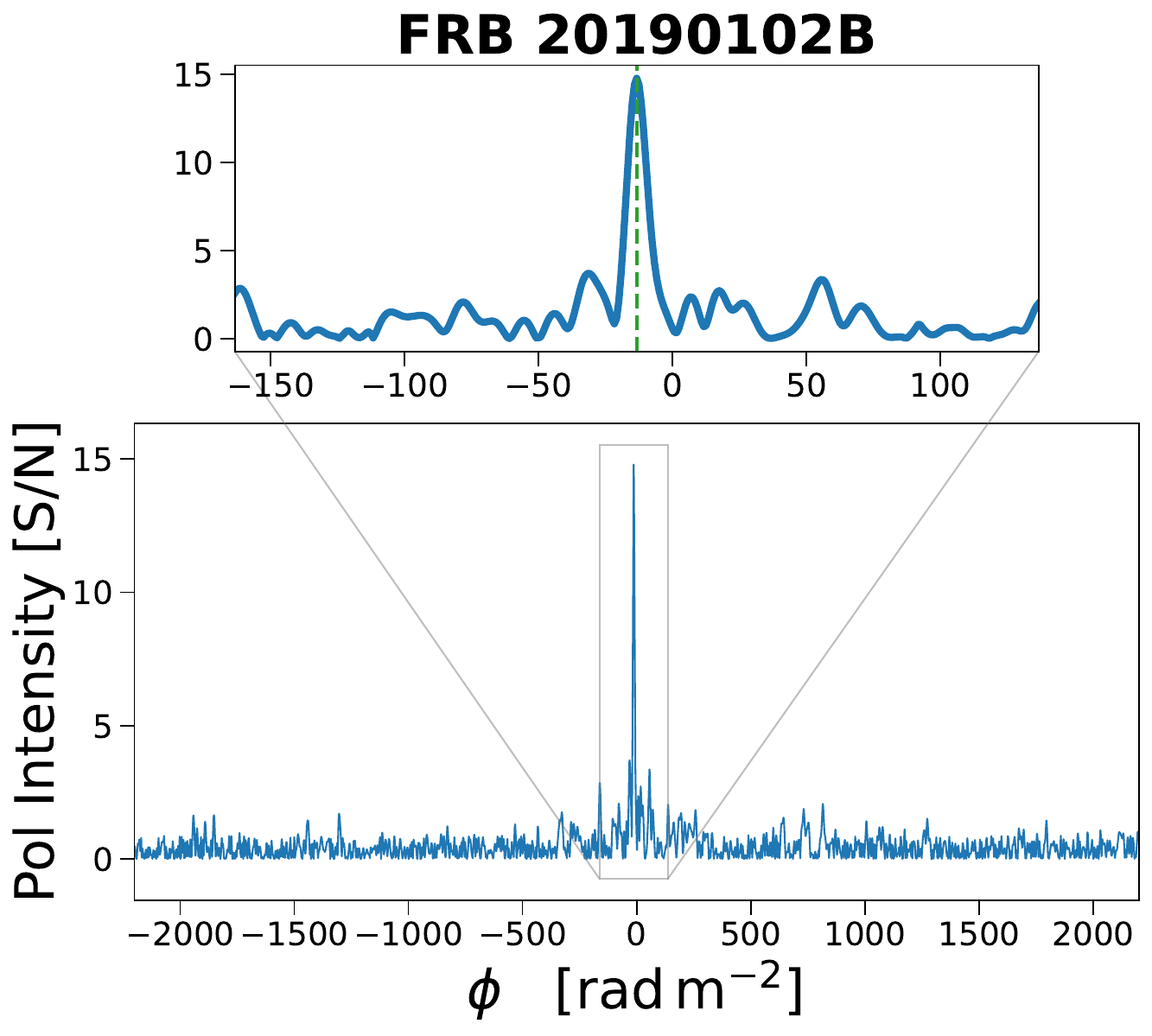}
    \includegraphics[width=0.246\textwidth]{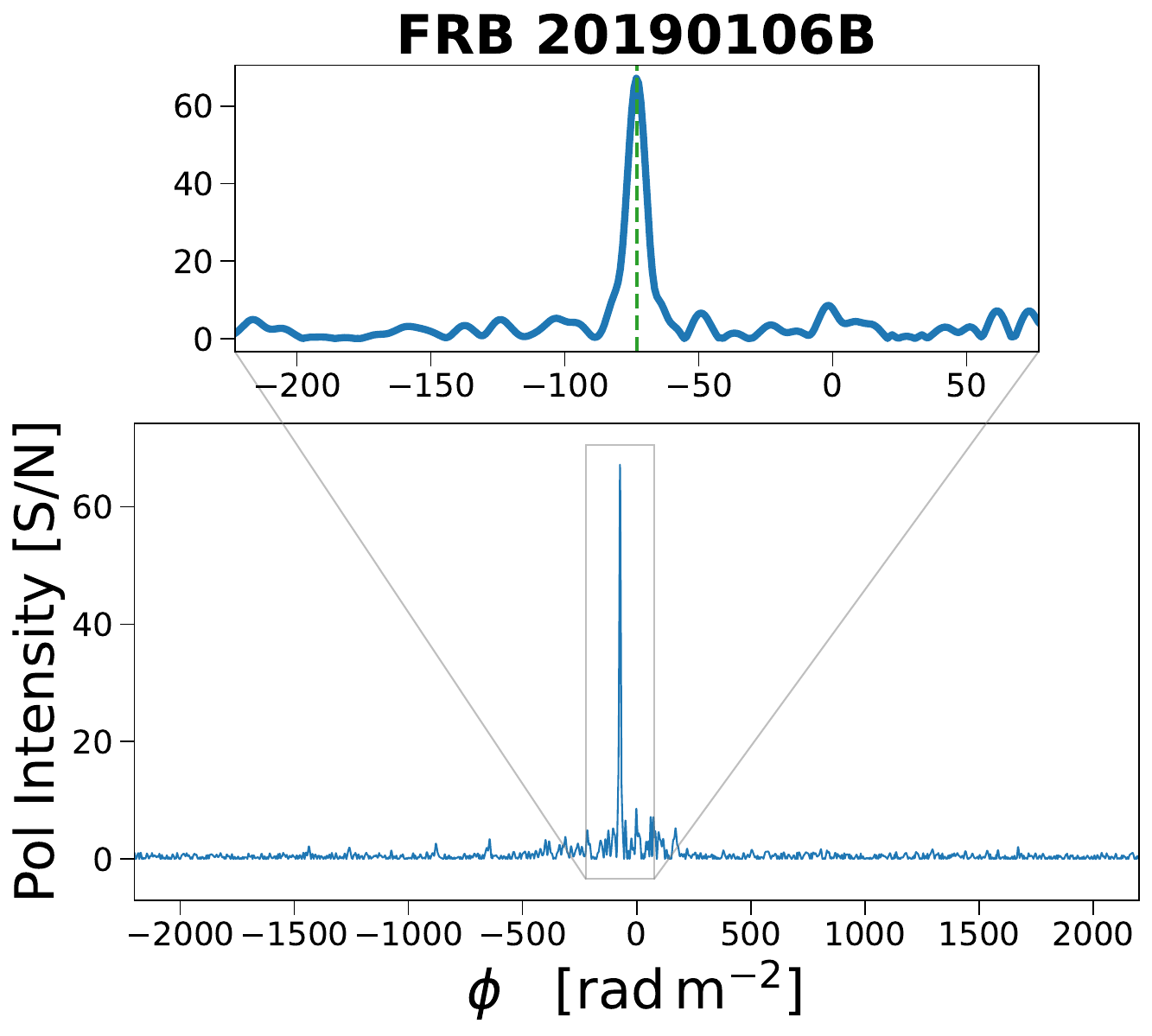}
    \includegraphics[width=0.246\textwidth]{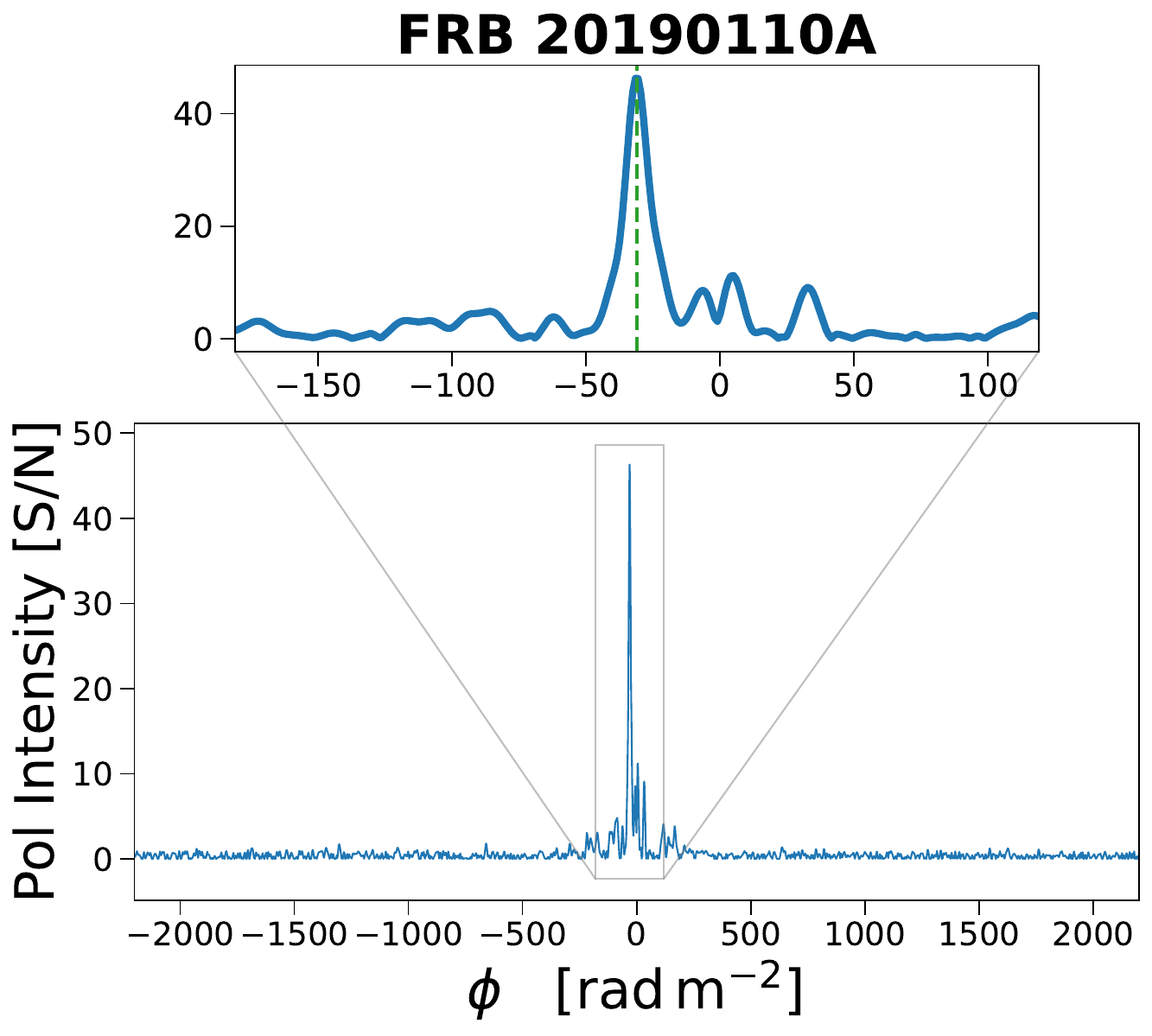}
    \caption{FDFs for the same sample of 16 FRBs presented in Figure \ref{fig:waterfalls} are shown as blue curves. The x-axis ($\phi$, in units of rad~m$^{-2}$) refers to the Faraday depth, which is an extension of $\mathrm{RM}_\mathrm{obs}$ in the scenario where we have multiple components of Faraday rotation in the same polarized signal. A zoomed in inset is presented centered at the position of the best fit $\mathrm{RM}_\mathrm{obs}$, which is marked as a green dashed line. In the case where the instrumental polarization is the dominant feature in the FDF, a comparable secondary peak is fit to derive the true $\mathrm{RM}_\mathrm{obs}$ (e.g., as seen in FRB 20181220A).}
    \label{fig:fdfs}
\end{center}
\end{figure*}

The polarization properties of the 89 polarized and 29 unpolarized FRBs are reported in Table \ref{tb:pol_results}.  Figure \ref{fig:waterfalls} displays the summary plots for the first 16 FRBs from Table \ref{tb:pol_results}. As described in Appendix \ref{app:B}, we only plot points on the $L/I$, $|V|/I$, and PA profiles if they pass the $\mathrm{S/N}(L)_\mathrm{thresh} \geq 5$ S/N limit. As a result, some faint and/or weakly polarized bursts may only have a few points or, in some cases, no points plotted in their respective profiles. Figure \ref{fig:fdfs} shows the FDFs for the same 16 FRBs as in Figure \ref{fig:waterfalls}. In the case of FRB 20181220A, we see two peaks in the FDF, with the best fit $\mathrm{RM}_\mathrm{obs}$ aligning with the lower peak. Here the peak at $0~\mathrm{rad}~\mathrm{m}^{-2}$ is due to instrumental polarization and, as mentioned in Section \ref{sec:pipeline_products}, the polarized fits for some events are improved by adopting the $\mathrm{RM}_\mathrm{obs}$ values of comparable secondary peaks in the FDF. The full set of summary plots, FDFs, Stokes $Q$, $U$, and $V$ waterfall data, and associated scripts describing how to use the data files are made available online.\footnote{\url{https://doi.org/10.11570/23.0029}}

\begin{figure*}[ht!]
\begin{center}
    \includegraphics[width=0.325\textwidth]{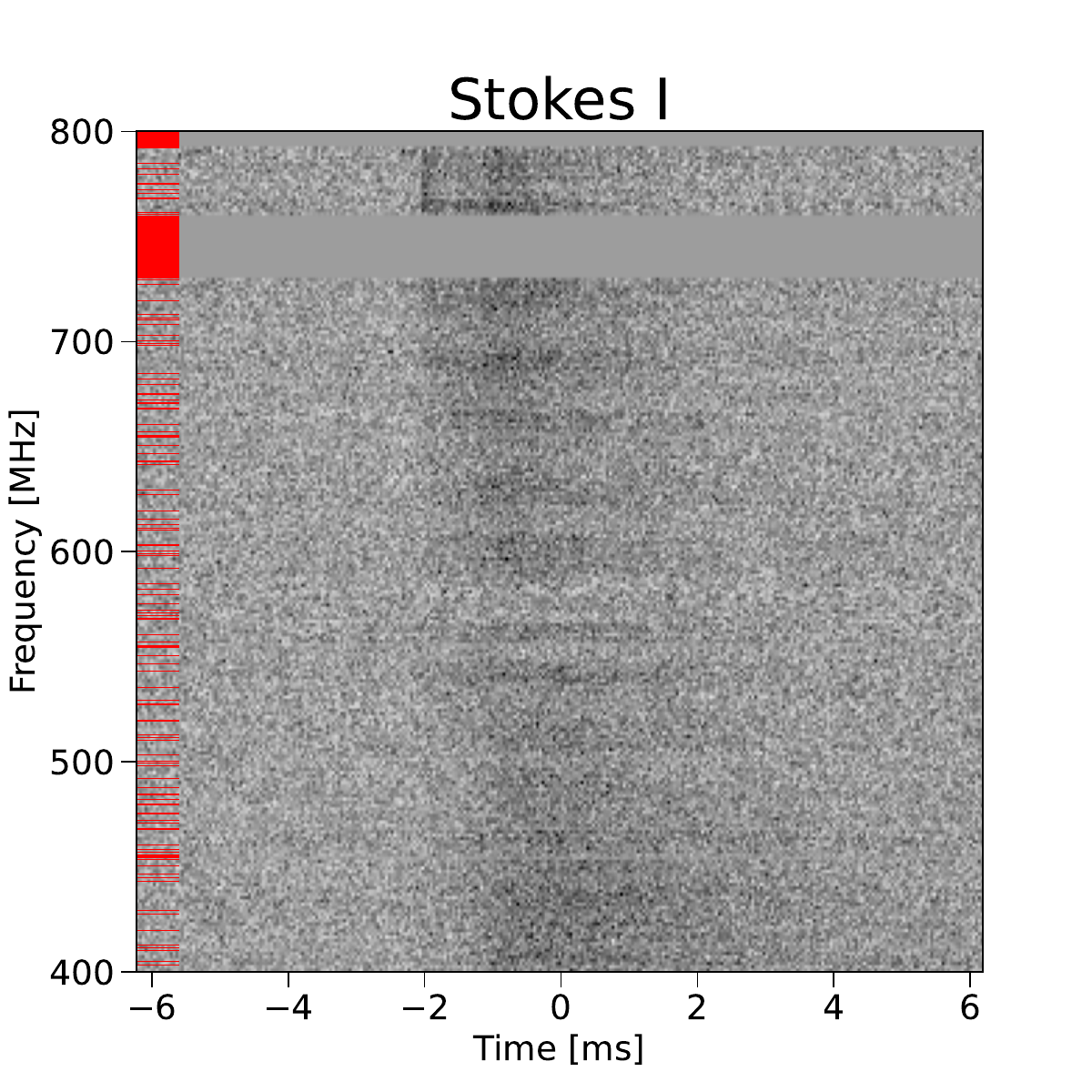}
    \includegraphics[width=0.325\textwidth]{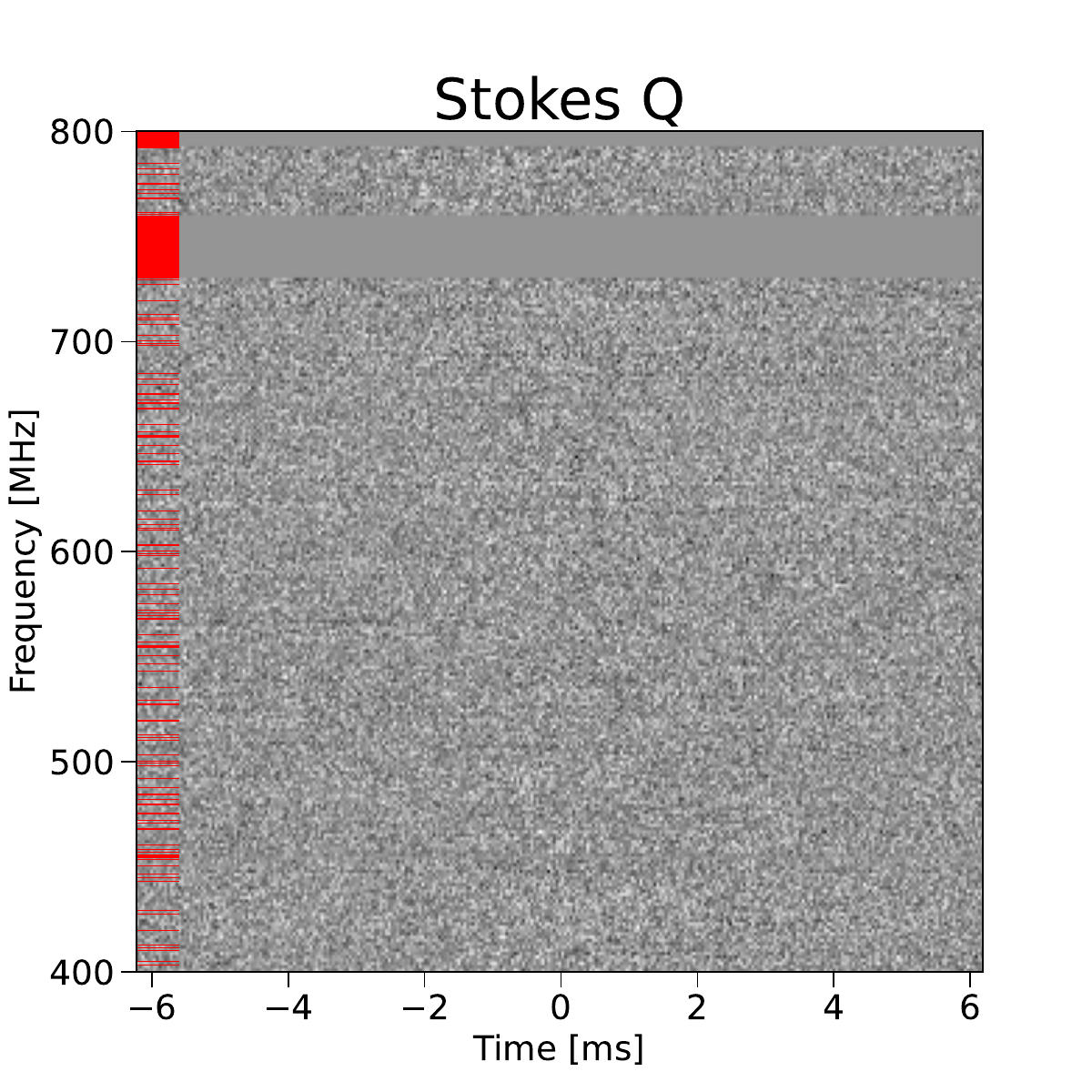}
    \includegraphics[width=0.325\textwidth]{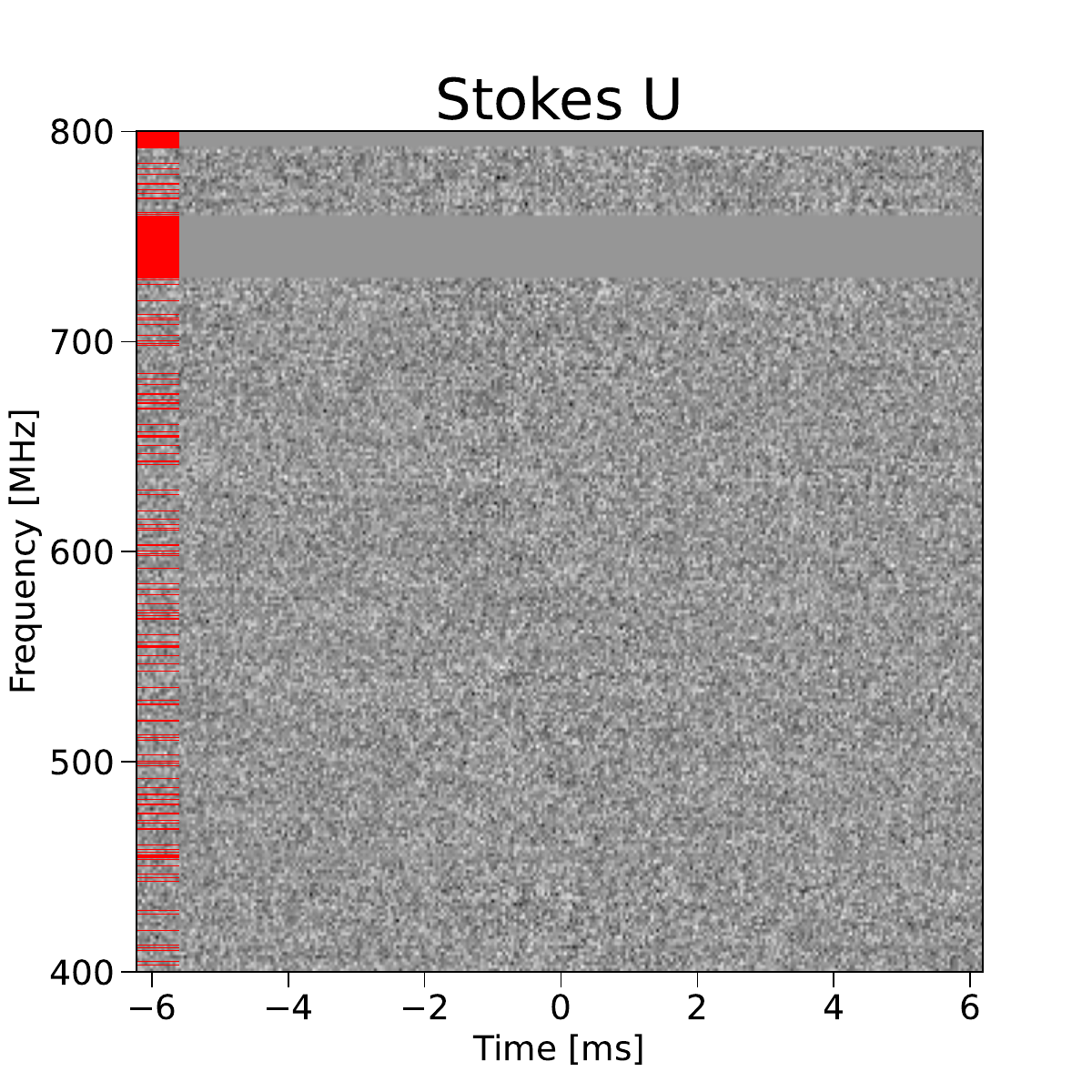}
    \caption{Stokes $I$ (left), $Q$ (middle), and $U$ (right) waterfalls for FRB 20190502A, which is unpolarized with an upper limit of $L/I \leq 0.12$. As in Figure \ref{fig:waterfalls}, the red streaks to the left of each waterfall plot highlight frequency channels that are masked out due to radio frequency interference. Here, the data have been downsampled by a factor of $n_\mathrm{down} = 16$ and thus the resolution of the flagged radio frequency interference channels and Stokes images do not match exactly.}
    \label{fig:unpol_example}
\end{center}
\end{figure*}

One example of the Stokes $I$, $Q$, and $U$ waterfall plots for an unpolarized FRB (FRB 20190502A) is shown in Figure \ref{fig:unpol_example}. While the FRB is clearly visible in Stokes $I$, there does not appear to be much corresponding signal in the Stokes $Q$ or $U$ waterfalls. The upper limit on the time and frequency averaged linear polarization fraction of this FRB is $L/I \leq 0.12$.

\setlength{\tabcolsep}{4.15pt}
\setlength{\LTcapwidth}{1.0\textwidth}
\begin{center}
\begin{longtable*}{cccccccccc}
\caption{A summary of polarization properties analyzed in this work for the 89 well-fit FRBs and 29 unpolarized FRBs. The Transient Name Server (TNS; \url{https://www.wis-tns.org/}) names for each FRB are listed in column 1. Column 2 provides the downsampling factor $n_\mathrm{down}$ applied to the de-dispersed data; this factor determines the time resolution of the waterfall plots in Figure \ref{fig:waterfalls} such that the time resolution is $n_\mathrm{down} \times 2.56~\mu\mathrm{s}$. The average linear polarization fraction across the burst envelope and emitting band, $L/I$, is presented in column 3. For the unpolarized bursts, we provide an upper limit on $L/I$ derived as the fraction required to produce a $6 \sigma$ $\mathrm{RM}_\mathrm{obs}$ detection in the FDF. Column 4 provides the structure-optimized dispersion measure, $\mathrm{DM}_\mathrm{obs,struct}$, used in this work \citep[based on work by][]{2023ApJ...950...12M} and the foreground Galactic DM estimate, $\mathrm{DM}_\mathrm{MW} = \mathrm{DM}_\mathrm{disk} + \mathrm{DM}_\mathrm{halo}$, is given in columns 5 \citep[with the MW disk component from YMW16 and assuming a constant MW halo contribution of $30$~pc~cm$^{-3}$;][]{2015MNRAS.451.4277D, 2020ApJ...888..105Y, 2023ApJ...946...58C}. The RM derived with RM-synthesis $\mathrm{RM}_\mathrm{obs,FDF}$ and QU-fitting $\mathrm{RM}_\mathrm{obs,QU}$ are presented in columns 6 and 7, respectively. The foreground MW contributions to the $\mathrm{RM}_\mathrm{obs}$, $\mathrm{RM}_\mathrm{MW} = \mathrm{RM}_\mathrm{disk} + \mathrm{RM}_\mathrm{halo}$ towards each FRB are estimated in column 8 \citep[following][]{2022A&A...657A..43H}. The reduced-chi squared statistic from fitting to a constant PA, $\chi_\nu^2$, (see Section \ref{sec:pa_var}) is given in column 9. In Section \ref{sec:results_depol}, we characterize the depolarization by taking the fraction between the $L/I$ in the bottom half and top half of the CHIME/FRB frequency band, $(L/I)_{500}$ and $(L/I)_{700}$, respectively. The ``depolarization ratio'' $f_\mathrm{depol} = (L/I)_{700} / (L/I)_{500}$ (see Section \ref{sec:results_depol_frac}) is presented in column 10. Where available, the uncertainties on the parameters are provided in parentheses indicating the error margin on the last significant figure listed (e.g., $L/I = 0.90(1)$ is equivalent to $L/I = 0.90 \pm 0.01$). The 10 FRBs that are contaminated by instrumental polarization are not tabulated as we cannot be fully confident in their polarization outputs. These polarization results are also tabulated in the live online table for the CHIME/FRB baseband catalog (\url{https://www.chime-frb.ca/baseband-catalog-1}). In the online version of the manuscript, we provide a machine readable version of this table.} \label{tb:pol_results}\\
\hline
\hline
TNS Name & $n_\mathrm{down}$ & $L/I$ & $\mathrm{DM}_\mathrm{obs,struct}$ & $\mathrm{DM}_\mathrm{MW}$ & $\mathrm{RM}_\mathrm{obs,FDF}$ & $\mathrm{RM}_\mathrm{obs,QU}$ & $\mathrm{RM}_\mathrm{MW}$ & PA $\chi_\nu^2$ & $f_\mathrm{depol}$\\
 &  &  & (pc~cm$^{-3}$) & (pc~cm$^{-3}$) & (rad~m$^{-2}$) & (rad~m$^{-2}$) & (rad~m$^{-2}$) &  & \\
\hline
\endfirsthead

\multicolumn{10}{c}
{{\bfseries \tablename\ \thetable{} -- continued from previous page}} \\
\hline
TNS Name & $n_\mathrm{down}$ & $L/I$ & $\mathrm{DM}_\mathrm{obs,struct}$ & $\mathrm{DM}_\mathrm{MW}$ & $\mathrm{RM}_\mathrm{obs,FDF}$ & $\mathrm{RM}_\mathrm{obs,QU}$ & $\mathrm{RM}_\mathrm{MW}$ & PA $\chi_\nu^2$ & $f_\mathrm{depol}$\\
 &  &  & (pc~cm$^{-3}$) & (pc~cm$^{-3}$) & (rad~m$^{-2}$) & (rad~m$^{-2}$) & (rad~m$^{-2}$) &  & \\
\hline
\endhead

\hline \multicolumn{2}{c}{{Continued on next page}} \\ \hline
\endfoot
\hline
\endlastfoot

\hline \multicolumn{10}{c}{{\textbf{Polarized FRBs}}} \\ \hline
FRB 20181209A &                  1 & 0.90(1) &       328.59(1) &     47 &  $-$110.28(5) &   106.81(1) &    $-$21(8) &            1.79 &    -- \\
FRB 20181213A &                  4 & 1.09(3) &       678.69(1) &     44 &    10.20(9) &    10.2(1) &    $-$12(6) &            1.92 &        0.88(6) \\
FRB 20181214C &                 32 & 0.60(3) &       632.832(3) &     41 &    23.6(2) &    23.8(2) &      6(2) &            2.07 &       1.8(2)\\
FRB 20181215B &                  1 & 1.017(9) &      494.044(6) &     43 &     8.59(2) &     8.18(3) &     14(4) &            4.43 &       1.06(2)\\
FRB 20181220A$^{\dagger}$ &                  8 & 0.43(2) &       209.525(8) &     46 &    97.5(4) &    $-$0.64(7) &    $-$23(11) &            4.09 &       -- \\
FRB 20181221A &                 32 & 0.42(2) &       316.25(5) &     42 &    39.0(4) &   $-$44.3(4) &      7(3) &           10.56 &       -- \\
FRB 20181222E &                 32 & 0.70(3) &       327.989(4) &     45 &   $-$91.8(1) &   $-$92.9(1) &    $-$11(8) &            1.50 &       -- \\
FRB 20181224E &                  8 & 0.62(2) &       581.84(1) &     44 &     1.65(8) &     0.67(9) &      8(6) &            4.70 &       -- \\
FRB 20181226D &                  1 & 1.00(1) &       385.338(5) &     66 &    64.63(5) &    64.73(5) &     25(8) &            1.49 &      1.09(4)\\
FRB 20181226E &                 16 & 0.57(2) &       308.78(1) &     45 &    $-$1.0(2) &    $-$0.2(2) &      0(15) &            1.42 &       -- \\
FRB 20181228B &                256 & 0.58(5) &       568.538(6) &     43 &    $-$0.1(5) &   $-$14.1(9) &     $-$1(7) &            1.49 &       -- \\
FRB 20181231B &                  2 & 0.78(1) &       197.366(9) &     45 &    $-$9.66(5) &    10.51(1) &    $-$16(3) &            1.32 &       -- \\
FRB 20190102A &                 16 & 0.74(1) &       699.1(4) &     44 &   198.8(2) &  $-$198.40(4) &    $-$52(8) &            1.36 &       -- \\
FRB 20190102B &                 32 & 0.86(3) &       367.07(4) &     43 &   $-$13.4(1) &   $-$13.31(3) &    $-$41(9) &            0.76 &       1.10(8)\\
FRB 20190106B &                  1 & 0.910(9) &      316.536(2) &     46 &   $-$73.18(3) &    72.99(3) &   $-$133(36) &           11.54 &       -- \\
FRB 20190110A &                  1 & 0.92(1) &       472.788(3) &     54 &   $-$31.14(5) &   $-$30.74(6) &    $-$36(23) &            3.54 &       -- \\
FRB 20190110C &                 32 & 0.96(6) &       222.01(1) &     42 &   118.5(3) &   118.4(3) &      8(2) &            2.15 &       -- \\
FRB 20190111B &                  2 & 0.63(1) &      1336.87(1) &     46 &   332.0(1) &   331.9(1) &     37(15) &            2.45 &       -- \\
FRB 20190118A$^{\dagger}$ &                  1 & 0.453(4) &      225.108(5) &     46 &    85.99(3) &    $-$0.670(1) &     41(13) &            2.48 &       -- \\
FRB 20190121A &                 32 & 0.97(1) &       425.28(3) &     46 &    95.47(5) &    95.41(6) &     10(15) &            2.48 &       1.01(3)\\
FRB 20190122C &                  4 & 1.07(1) &       690.032(8) &     41 &    50.64(2) &    50.71(2) &      2(3) &            3.73 &       1.07(2)\\
FRB 20190124F &                  2 & 0.72(1) &       254.799(4) &     42 &     5.00(8) &     4.62(9) &     $-$4(11) &            3.73 &       -- \\
FRB 20190130B &                  4 & 1.01(2) &       988.75(1) &     41 &    75.54(6) &    77.54(3) &      9(7) &            1.72 &       -- \\
FRB 20190131E$^{\dagger}$ &                  2 & 0.38(1) &       279.798(6) &     43 &   174.3(1) &    $-$0.56(1) &    $-$29(12) &            1.25 &       -- \\
FRB 20190201B &                 32 & 0.77(3) &       749.07(2) &     48 &  $-$157.2(2) &  $-$156.62(9) &     $-$3(5) &            3.99 &       -- \\
FRB 20190202B &                  2 & 0.31(1) &       464.839(4) &     66 &  $-$572.0(3) &   571.3(2) &      2(4) &           20.42 &       -- \\
FRB 20190203A &                 16 & 0.64(2) &       420.586(6) &     44 &  $-$341.6(1) &  $-$337.8(1) &    $-$16(4) &            2.39 &       -- \\
FRB 20190204B &                 64 & 0.85(5) &      1464.842(6) &     44 &   391.5(3) &   391.7(3) &    $-$23(16) &            2.67 &       -- \\
FRB 20190206A$^{\dagger}$ &                 64 & 0.57(2) &       188.353(3) &     45 &   $-$11.8(1) &    $-$0.30(6) &     13(7) &            3.05 &       -- \\
FRB 20190208C &                  4 & 0.73(3) &       238.323(5) &     44 &   $-$77.6(1) &   $-$80.7(1) &    $-$15(7) &            0.88 &       -- \\
FRB 20190210B &                  4 & 0.81(1) &       624.24(1) &     76 &  $-$359.3(1) &  $-$359.42(9) &     32(17) &           25.44 &       -- \\
FRB 20190212B &                  4 & 0.73(2) &       600.185(3) &     43 &   175.8(2) &  $-$175.7(2) &    $-$10(3) &            2.17 &       -- \\
FRB 20190213D &                256 & 0.89(4) &      1346.7(4) &     47 &  $-$319.4(4) &   318.3(5) &   $-$126(33) &            3.33 &       -- \\
FRB 20190214C &                128 & 0.84(3) &       532.96(1) &     42 &  1169.0(1) & $-$1169.6(1) &      8(5) &            6.10 &       -- \\
FRB 20190217A &                256 & 0.94(6) &       798.14(4) &     62 &   594.0(2) &   595.1(3) &      6(11) &            2.83 &       0.8(1)\\
FRB 20190224C &                128 & 0.54(2) &       497.12(2) &     63 &     0.6(3) &     0.8(3) &     23(9) &            6.42 &       -- \\
FRB 20190224D &                  2 & 0.81(2) &       752.892(6) &     44 &   $-$42.49(9) &   $-$41.86(9) &    $-$16(8) &           24.68 &       1.04(8)\\
FRB 20190226A &                  4 & 0.65(2) &       601.546(7) &     55 &   234.4(2) &  $-$234.1(1) &     23(21) &            2.48 &       0.96(7)\\
FRB 20190303B &                  1 & 0.624(3) &      193.429(5) &     44 &    31.08(1) &    31.074(3) &    $-$23(3) &            3.55 &       -- \\
FRB 20190304A &                  8 & 0.57(2) &       483.521(8) &     44 &   $-$78.0(1) &    78.0(1) &    $-$12(5) &            1.83 &       -- \\
FRB 20190304B &                 64 & 1.03(5) &       469.90(2) &     41 &    39.5(2) &   $-$39.0(1) &      5(1) &            1.12 &       -- \\
FRB 20190320B &                  1 & 0.98(1) &       489.501(8) &     42 &    53.34(6) &    53.36(6) &     21(3) &            1.91 &       0.99(3)\\
FRB 20190320E &                 32 & 0.95(3) &       299.09(2) &     44 &   $-$75.1(1) &    75.74(1) &    $-$17(7) &            1.63 &     0.85(6)\\
FRB 20190323B &                  1 & 0.911(5) &      789.527(7) &     43 &   229.46(2) &   229.058(3) &    $-$15(8) &            2.57 &     0.94(1)\\
FRB 20190327A &                  8 & 0.48(1) &       346.579(7) &     44 &    11.2(1) &    11.2(1) &     61(19) &            3.03 &       1.18(7)\\
FRB 20190405B &                 64 & 0.46(3) &      1113.72(7) &     44 &     1.8(3) &     2.0(3) &    $-$22(8) &            5.90 &       -- \\
FRB 20190411C &                  1 & 0.804(4) &      233.714(8) &     43 &   $-$21.44(2) &   $-$20.42(2) &    $-$23(2) &          149.03 &       -- \\
FRB 20190412A &                 16 & 0.74(2) &       364.55(1) &     43 &  $-$168.7(2) &   167.58(7) &    $-$13(7) &           10.82 &       1.0(1)\\
FRB 20190417C &                  1 & 0.916(1) &      320.266(4) &     47 &   475.373(4) &  $-$475.4005(2) &      5(11) &           16.88 &       -- \\
FRB 20190419B$^{**}$ &                  8 & 0.23(1) &       165.13(1) &     44 &     4.1(3) &   $-$10.63(8) &    $-$31(10) &             --&       -- \\
FRB 20190423A &                  1 & 0.949(2) &      242.600(8) &     41 &    23.097(6) &   $-$22.933(7) &      9(2) &            8.37 &       -- \\
FRB 20190425A &                  1 & 0.949(3) &      128.14(1) &     44 &    57.278(4) &    57.043(2) &     49(15) &           10.78 &    1.026(5) \\
FRB 20190427A &                 32 & 0.61(3) &       455.78(1) &     72 &  $-$523.8(2) &  $-$524.1(3) &     79(31) &            1.61 &       1.0(1)\\
FRB 20190430C$^{\ddagger}$ &                 16 & 1.04(2) &       400.3(3)  &     45 &  $-$76.1(9)  &  $-$70.4(5)  &     94(27) &            1.17 &       -- \\
FRB 20190501B &                 16 & 0.88(2) &       783.967(4) &     43 &  $-$121.09(8) &   121.35(9) &     12(6) &            3.80 &       -- \\
FRB 20190502B &                256 & 0.50(3) &       918.6(2) &     42 &   126.7(7) &   126.3(7) &     50(8) &            2.75 &       -- \\
FRB 20190502C &                  2 & 0.60(2) &       396.878(9) &     44 &   $-$35.5(2) &    36.5(1) &    $-$13(5) &            4.52 &       -- \\
FRB 20190519E &                  8 & 1.00(6) &       693.622(7) &     37 &   $-$17.0(3) &   $-$17.0(4) &     20(5) &            2.86 &       -- \\
FRB 20190519H &                  1 & 0.954(5) &     1170.878(6) &     45 &   $-$24.84(3) &   $-$25.03(3) &    $-$17(10) &            3.32 &       -- \\
FRB 20190604G &                 16 & 0.33(1) &       232.998(7) &     47 &   364.4(3) &   364.6(4) &     $-$8(4) &            9.83 &       1.2(2)\\
FRB 20190605C$^{\dagger}$ &                  1 & 0.319(7) &      187.713(5) &     43 &   $-$64.68(9) &    $-$0.10(4) &     $-$2(7) &            4.28 &       -- \\
FRB 20190606B &                128 & 0.86(4) &       277.67(3) &     44 &    16.5(1) &    16.7(2) &    $-$28(10) &            1.04 &       0.95(9)\\
FRB 20190609A &                  8 & 0.86(3) &       316.684(3) &     45 &    42.4(1) &   $-$44.2(2) &    $-$16(11) &            1.23 &       -- \\
FRB 20190609B &                  2 & 0.300(8) &      292.174(7) &     44 &    28.1(1) &    29.7(1) &    $-$26(10) &            4.18 &       -- \\
FRB 20190609C &                  2 & 1.04(6) &       479.852(5) &     66 &    $-$5.4(3) &     5.1(3) &    $-$33(13) &            4.14 &       -- \\
FRB 20190609D &                 64 & 0.91(5) &       511.56(2) &     51 &   $-$50.4(2) &   $-$48.6(2) &      3(6) &            4.36 &       -- \\
FRB 20190612A$^{*}$ &                256 & 0.78(4) &       433.14 &     43 &   $-$16.9(2) &   $-$18.7(5) &    $-$19(6) &            1.09 &       -- \\
FRB 20190612B &                  1 & 0.92(2) &       187.524(7) &     43 &     2.05(6) &    $-$0.88(8) &      5(6) &            1.86 &       -- \\
FRB 20190613B &                  1 & 0.910(9) &      285.088(5) &     56 &   $-$22.18(3) &   $-$22.36(3) &    $-$11(18) &            2.99 &       -- \\
FRB 20190614A &                 32 & 0.40(2) &      1063.917(6) &     44 &    $-$1.0(2) &    $-$1.7(2) &    $-$22(9) &            1.37 &       -- \\
FRB 20190617A &                  1 & 0.869(2) &      195.749(6) &     44 &   $-$93.089(7) &    94.068(1) &    $-$15(7) &           48.31 &       -- \\
FRB 20190617B &                 32 & 0.45(1) &       272.73(7) &     58 &     1.7(2) &     2.9(1) &     45(14) &            7.33 &       -- \\
FRB 20190618A &                  1 & 0.550(6) &      228.920(6) &     44 &   $-$61.09(5) &    62.958(1) &    $-$75(12) &            2.66 &       -- \\
FRB 20190619B &                 64 & 0.48(3) &       270.549(3) &     44 &   $-$43.2(3) &   $-$43.1(1) &    $-$22(11) &            1.33 &       -- \\
FRB 20190619C &                  8 & 0.97(3) &       488.072(3) &     47 &  $-$226.9(2) &   226.9(3) &    $-$39(7) &            0.97 &       1.0(1)\\
FRB 20190621C &                  1 & 0.61(1) &       570.342(7) &     42 &    $-$0.18(6) &    $-$2.06(7) &      1(5) &            7.12 &       -- \\
FRB 20190621D &                 32 & 0.48(2) &       647.32(4) &     44 &   760.0(2) &  $-$759.2(2) &    $-$46(13) &            3.71 &       -- \\
FRB 20190623A &                 16 & 0.86(4) &      1082.16(1) &     45 &   162.5(2) &  $-$161.54(7) &    108(19) &            1.41 &       -- \\
FRB 20190624B &                  1 & 0.802(1) &      213.947(8) &     45 &   $-$16.527(2) &     4.00503(6) &      9(14) &          233.93 &       -- \\
FRB 20190627A &                 64 & 1.02(8) &       404.3(1) &     42 &   $-$48.3(6) &    49.4(7) &    $-$10(6) &            0.69 &       -- \\
FRB 20190627C &                  8 & 0.93(2) &       968.50(1) &     44 &    69.09(6) &    68.47(6) &     12(20) &            1.40 &       1.09(4)\\
FRB 20190628A &                 64 & 0.93(6) &       745.790(8) &     41 &    23.6(3) &    23.7(3) &      7(1) &            1.23 &       0.9(1)\\
FRB 20190628B &                128 & 0.80(5) &       407.99(2) &     44 &    19.2(3) &    19.7(3) &    $-$23(13) &            3.91 &       -- \\
FRB 20190630B &                 64 & 0.98(1) &       651.7(3) &     46 &     6.65(4) &     7.00(5) &   $-$205(102) &            3.90 &       -- \\
FRB 20190630C &                 32 & 0.47(2) &      1660.21(1) &     45 &   641.0(4) &   641.7(2) &     $-$8(8) &            2.04 &       -- \\
FRB 20190630D &                 16 & 0.59(3) &       323.540(3) &     48 &     5.1(2) &     4.8(2) &     12(8) &           12.94 &       -- \\
FRB 20190701A &                 16 & 0.92(5) &       637.091(9) &     44 &  $-$154.5(2) &   154.3(2) &     16(11) &            2.28 &       1.1(1)\\
FRB 20190701B &                  8 & 0.67(3) &       749.093(8) &     45 &  $-$534.3(2) &  $-$533.5(5) &      4(10) &            1.66 &       -- \\
FRB 20190701D &                 64 & 0.75(2) &       933.32(3) &     46 &  $-$138.67(8) &   137.01(4) &    $-$20(10) &            1.51 &       -- \\
\hline \multicolumn{10}{c}{{\textbf{Unpolarized FRBs}}} \\ \hline
FRB 20181219C &                128 & $<$~0.21(3) &       647.68(4) &    43 &     -- &    -- &    $-$9(2) &             -- &       --  \\
FRB 20181223C &                 32 & $<$~0.14(3) &       112.45(1) &    40 &     -- &    -- &    7(6) &             -- &       --  \\
FRB 20181229A &                 64 & $<$~0.11(3) &       955.45(2) &    44 &     -- &    -- &    5(1) &             -- &       --  \\
FRB 20181231A &                256 & $<$~0.7(1) &      1376.9(3) &    44 &     -- &    -- &    $-$33(6) &             -- &       --  \\
FRB 20181231C &                128 & $<$~0.23(5) &       556.03(2) &    42 &     -- &    -- &    11(5) &             -- &       --  \\
FRB 20190103C &                128 & $<$~0.14(2) &      1349.3(1) &    75 &     -- &    -- &    33(22) &             -- &       --  \\
FRB 20190115B &                 64 & $<$~0.10(4) &       748.18(3) &    45 &     -- &    -- &    $-$13(8) &             -- &       --  \\
FRB 20190227A &                  8 & $<$~0.136(8) &       394.031(8) &    50 &     -- &    -- &    24(11) &             -- &       --  \\
FRB 20190320A &                256 & $<$~0.16(4) &       614.2(1) &    49 &     -- &    -- &    $-$29(19) &             -- &       --  \\
FRB 20190411B &                256 & $<$~0.23(4) &      1229.417(7) &    40 &     -- &    -- &    34(8) &             -- &       --  \\
FRB 20190418A &                 64 & $<$~0.28(5) &       184.473(3) &    63 &     -- &    -- &    $-$40(24) &             -- &       --  \\
FRB 20190423D &                256 & $<$~0.20(4) &       496(1) &    45 &     -- &    -- &    $-$10(8) &             -- &       --  \\
FRB 20190425B &                  4 & $<$~0.26(1) &      1031.63(1) &    44 &     -- &    -- &    $-$25(9) &             -- &       --  \\
FRB 20190502A &                 16 & $<$~0.12(1) &       625.74(1) &    41 &     -- &    -- &    4(2) &             -- &       --  \\
FRB 20190518C &                  8 & $<$~0.14(2) &       443.964(6) &    45 &     -- &    -- &    11(5) &             -- &       --  \\
FRB 20190607B &                128 & $<$~0.37(7) &       289.331(2) &    49 &     -- &    -- &    $-$1(12) &             -- &       --  \\
FRB 20190608A &                 32 & $<$~0.21(4) &       722.14(1) &    43 &     -- &    -- &    $-$32(6) &             -- &       --  \\
FRB 20190613A &                128 & $<$~0.29(5) &       714.71(3) &    45 &     -- &    -- &    49(14) &             -- &       --  \\
FRB 20190614C &                256 & $<$~0.26(7) &       589.1(1) &    45 &     -- &    -- &    $-$75(10) &             -- &       --  \\
FRB 20190616A &                  8 & $<$~0.15(2) &       212.511(5) &    42 &     -- &    -- &    6(2) &             -- &       --  \\
FRB 20190617C &                256 & $<$~0.6(7) &       638.90(2) &    47 &     -- &    -- &    15(4) &             -- &       --  \\
FRB 20190619A &                 32 & $<$~0.10(3) &       899.82(1) &    42 &     -- &    -- &    $-$12(3) &             -- &       --  \\
FRB 20190621B &                256 & $<$~0.22(7) &      1061.14(2) &    41 &     -- &    -- &    16(4) &             -- &       --  \\
FRB 20190622A &                 32 & $<$~0.28(6) &      1122.807(9) &    44 &     -- &    -- &    $-$20(9) &             -- &       --  \\
FRB 20190623C &                256 & $<$~0.25(8) &      1049.94(1) &    44 &     -- &    -- &    $-$17(8) &             -- &       --  \\
FRB 20190624A &                128 & $<$~0.23(4) &       973.9(1) &    42 &     -- &    -- &    $-$13(5) &             -- &       --  \\
FRB 20190627D &                256 & $<$~0.6(1) &      2000.31(3) &    45 &     -- &    -- &    $-$59(40) &             -- &       --  \\
FRB 20190628C &                256 & $<$~0.6(1) &      1746.8(3) &    46 &     -- &    -- &    5(13) &             -- &       --  \\
FRB 20190701C &                 64 & $<$~0.43(5) &       973.79(1) &    45 &     -- &    -- &    $-$21(10) &             -- &       --  \\

\hline
\hline
\multicolumn{10}{l}{$^{*}$ DM was obtained by maximizing the S/N of the burst as a large fraction of the signal falls outside the time range of the} \\
\multicolumn{10}{l}{saved baseband data.} \\
\multicolumn{10}{l}{$^{**}$ Not enough points on the PA curve exceed the $\mathrm{S/N}(L)_\mathrm{thresh} = 5$ requirement to derive a $\chi_\nu^2$ fit.} \\
\multicolumn{10}{l}{$^{\dagger}$ Corrected for instrumental polarization by re-fitting the $\mathrm{RM}_\mathrm{obs,FDF}$ to a secondary peak.} \\
\multicolumn{10}{l}{$^{\ddagger}$ A manual frequency mask was applied due to a few frequency channels containing severe radio frequency interference} \\
\multicolumn{10}{l}{that were not automatically flagged during raw data processing.}
\end{longtable*}
\end{center}

\subsection{Linear polarization} \label{sec:results_li}
We plot a histogram of the $L/I$ distribution for the 84 FRBs with well fit polarization properties (removing the five FRBs that are corrected for instrumental polarization) in the top left panel of Figure \ref{fig:all_hist_cdf} in red and overlay a Gaussian kernel density estimate (KDE). The $L/I$ distribution is skewed towards $L/I = 1.0$, with a mean $L/I$ of $0.764$ (plotted as a dashed line in Figure \ref{fig:all_hist_cdf}) and a median of $0.805$ (plotted as a dash-dotted line). For the 29 FRBs with no significant polarized detection, we place upper limits on their $L/I$ given our $6\sigma$ detection threshold and plot them as a grey histogram. The mean and median $L/I$, after accounting for the unpolarized FRB upper limits, are $L/I = 0.633$ and $L/I = 0.647$, respectively.

The normalized $L/I$ distribution CDF is computed using the Kaplan-Meier method and is presented in the top right panel of Figure \ref{fig:all_hist_cdf} as a red curve. This CDF includes both the polarized FRBs (with a measured $L/I$) and the unpolarized FRBs (with a upper limit on $L/I$) and the $95$\% confidence interval (CI) is represented by the red shaded region. For each repeating FRB in our sample, we take the median $L/I$ across all their polarized bursts and overlay a normalized CDF of the overall repeating FRB $L/I$ distribution as a blue curve. The full range of $L/I$ values for each repeater is depicted by the blue shaded region. The mean and median of the repeater $L/I$ distribution are $0.681$ and $0.728$, respectively (see Table \ref{tb:summary_stats}). Note that we have not accounted for individual bursts from repeating sources that are unpolarized, as they have not always been reported, and the mean/median repeater $L/I$ may be slightly lower if we were able to fully account for these bursts.

We apply the AD and KS tests to the polarized repeating and non-repeating $L/I$ distributions to discern whether they arise from the same underlying population. The summary statistics and associated $p$-values from these tests are listed in Table \ref{tb:ad_ks_tests}. We find no conclusive evidence that the two polarized $L/I$ distributions are statistically different using the AD and KS tests. Taking into account the unpolarized non-repeaters, the Peto \& Peto and Log-Rank tests return $p_\mathrm{PP} = 0.962$ and $p_\mathrm{LR} = 0.769$, which suggests that there is no evidence for a dichotomy in repeater and non-repeater $L/I$ distributions.

\subsection{Rotation measure and dispersion measure} \label{sec:results_rm}
In the second and third row of the left columns of Figure \ref{fig:all_hist_cdf}, we plot the foreground subtracted $|\mathrm{RM}_\mathrm{EG}|$ and $\mathrm{DM}_\mathrm{EG}$ distributions of our 89 non-repeating FRBs as a histogram and overlay a Gaussian KDE. Note that we opt to analyze the $|\mathrm{RM}_\mathrm{EG}|$ here, as opposed to $\mathrm{RM}_\mathrm{EG}$, because the sign of the RM is determined by the projected LoS direction of $B_\parallel$, which we do not expect to have any preferred orientation with the observer and, therefore, does not provide any further insight regarding the host environment for non-repeating FRBs. Separately, we provide a histogram of the $\mathrm{RM}_\mathrm{EG}$ distribution for our 89 non-repeating FRBs in Figure \ref{fig:rm_nonabs}. As expected, the $\mathrm{RM}_\mathrm{EG}$ distribution is approximately symmetric about $0~\mathrm{rad}~\mathrm{m}^{-2}$. Both distributions appear to be approximately log-normally distributed. The mean, median, and $16^\mathrm{th} - 84^\mathrm{th}$ percentile range (PR) of both the $|\mathrm{RM}_\mathrm{EG}|$ and $\mathrm{DM}_\mathrm{EG}$ distributions are tabulated in Table \ref{tb:summary_stats}.

The normalized CDFs of the non-repeater $|\mathrm{RM}_\mathrm{EG}|$ and $\mathrm{DM}_\mathrm{EG}$ distributions are shown as red curves in the second and third row of the right panels of Figure \ref{fig:all_hist_cdf}. The bursts with no significant $L$ detections are not plotted. Analogous to Section \ref{sec:results_rm}, we take the median $|\mathrm{RM}_\mathrm{EG}|$ and $\mathrm{DM}_\mathrm{EG}$ across all available bursts of each repeater as their representative value for the normalized CDFs (blue curves). The mean and median values of $|\mathrm{RM}_\mathrm{EG}|$ and $\mathrm{DM}_\mathrm{EG}$ for both repeaters and non-repeaters are tabulated in Table \ref{tb:summary_stats}. The AD and KS test statistics and $p$-values for repeating versus non-repeating $|\mathrm{RM}_\mathrm{EG}|$ and $\mathrm{DM}_\mathrm{EG}$ distributions are listed in Table \ref{tb:ad_ks_tests}.

For $|\mathrm{RM}_\mathrm{EG}|$, we find $p$-values of $p_\mathrm{AD} = 0.172$ and $p_\mathrm{KS} = 0.343$ with the AD and KS tests, respectively. There is no statistical evidence for a dichotomy between the two populations with respect to $|\mathrm{RM}_\mathrm{EG}|$. In regards to the $\mathrm{DM}_\mathrm{EG}$ distributions, we find $p_\mathrm{AD} = 0.064$ and $p_\mathrm{KS} = 0.290$, which does not suggest a difference between the two $\mathrm{DM}_\mathrm{EG}$ distributions. These results, however, are in contrast with \cite{2023ApJ...947...83C}, who find a difference in the extragalactic DM of the repeating and non-repeating FRB populations at the $\sim 99$\% level with the AD test and a $\sim 96$\% level with the KS test. The cause for such a disparity in $\mathrm{DM}_\mathrm{EG}$ results between this work and those by \cite{2023ApJ...947...83C} is two-fold. First, the memory buffer on CHIME/FRB baseband data is $\sim 20$~seconds, which results in loss of frequency channels for FRBs with $\mathrm{DM}_\mathrm{obs} \gtrsim 1000~\mathrm{pc}~\mathrm{cm}^{-3}$, and for the most dispersed events we are not able to save any data. This leads to a bias in sensitivity against high DM events in baseband data (reported on here) that does not exist in intensity-only data (in which the dichotomy was seen). Having fewer frequency channels available leads to the potential loss of polarized signal, which may make it more difficult to detect the polarization properties for a high DM FRB. Thus, a higher fraction of unpolarized FRBs have $\mathrm{DM}_\mathrm{obs} \gtrsim 1000~\mathrm{pc}~\mathrm{cm}^{-3}$ than polarized FRBs. However, the polarization properties of the polarized FRB sample are largely unaffected except in extreme cases where we lose most of the CHIME/FRB band (and thus most of the polarized signal). In these cases, we have larger uncertainties for the $L/I$, $\mathrm{RM}_\mathrm{obs}$, and individual PA measurements across the burst (which is then propagated into the PA $\chi_\nu^2$ metric). Secondly, the sensitivities of the AD and KS tests scale with the number of data points in the input distributions. The intensity data used by \cite{2023ApJ...947...83C} is a substantially larger sample (305 non-repeaters and 40 repeaters) than that used in this work. Both these biases lead to repeaters and non-repeaters having more similar $\mathrm{DM}_\mathrm{EG}$ distributions in this work than those seen by \cite{2023ApJ...947...83C}.

Figure \ref{fig:dm_rm} shows $\mathrm{DM}_\mathrm{EG}$ plotted against $|\mathrm{RM}_\mathrm{EG}|$ for the repeating and non-repeating FRBs. We apply Spearman's rank correlation coefficient to the $\mathrm{log}_{10}\left(|\mathrm{RM}_\mathrm{EG}|\right)$ -- $\mathrm{log}_{10}\left(\mathrm{DM}_\mathrm{EG}\right)$ relation in three distinct sets of data: (i) only the non-repeaters, (ii) only the repeaters, and (iii) the combined repeater and non-repeater set. We find no statistically significant linear correlation in the repeater only data, but in the non-repeater and combined data sets we find marginal evidence for a positive monotonic correlations with $p_\mathrm{SR} = 0.006$ and $p_\mathrm{SR} = 0.010$, respectively. The lack of correlation in the repeater only sample may be partially caused by the smaller sample size when compared to the non-repeater data. The test results are summarized in Table \ref{tb:spearmanr_tests}.

\subsection{LoS magnetic field lower limits} \label{sec:results_bpar}
\begin{figure*}[ht]
    \centering
    \includegraphics[width=0.775\textwidth]{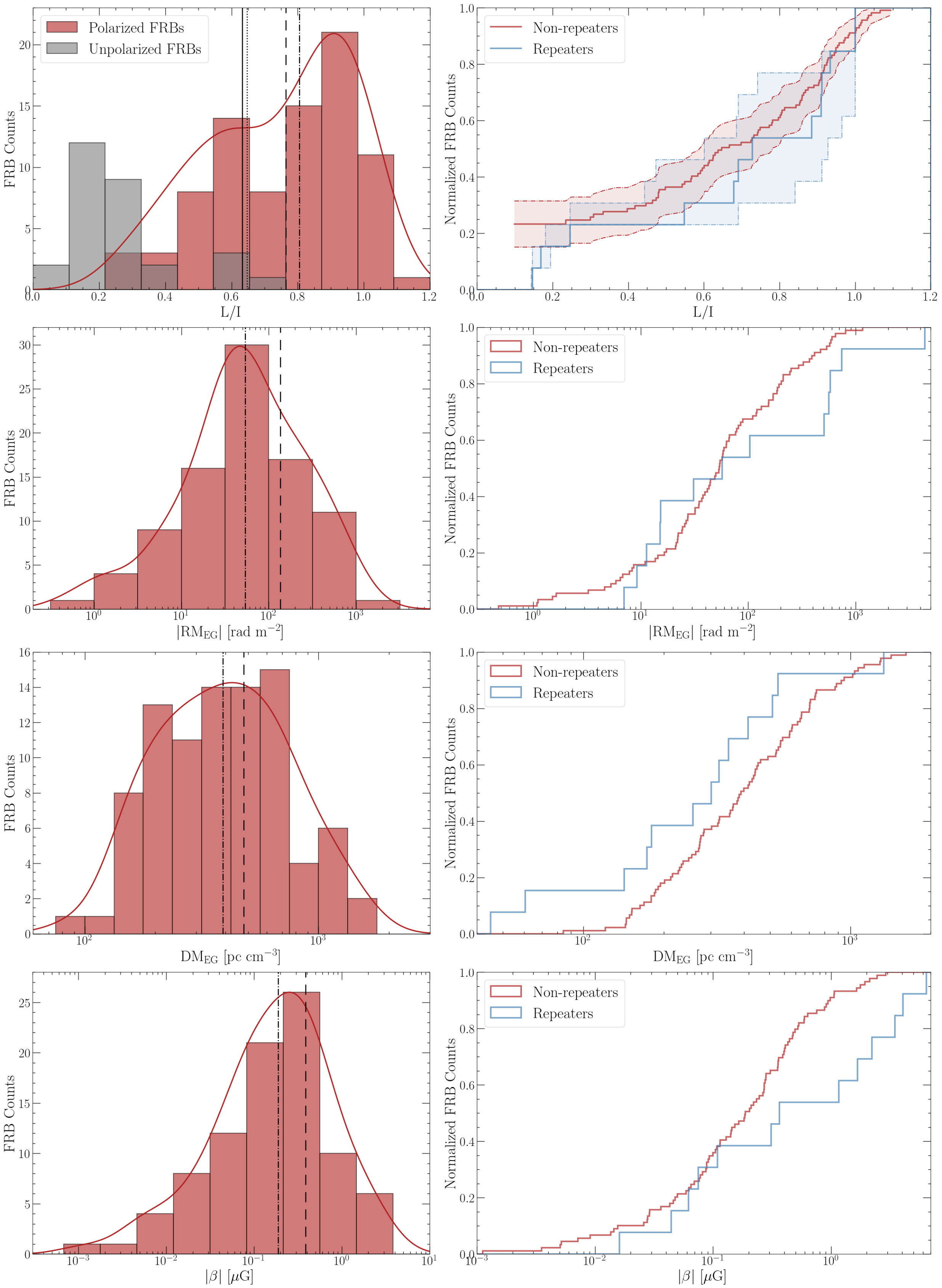}
    \caption{(Left) Distribution of $L/I$, $|\mathrm{RM}_\mathrm{EG}|$, DM$_\mathrm{EG}$, and $\left|\beta\right|$ for the non-repeating FRBs with a significant linearly polarized detection (89 in total but 84 for $L/I$ after removing the instrumental polarization corrected FRBs). In the top row, we also include the $L/I$ upper limit distribution of the $29$ unpolarized FRBs in grey. The dashed and dash-dotted black lines in each panel correspond to the respective mean and median values and a KDE of each distribution is overplotted. In the top left panel, the solid and dotted lines represent the mean and median, respectively, when accounting for the $29$ unpolarized non-repeating FRB $L/I$ upper limits. (Right) The normalized $L/I$, $|\mathrm{RM}_\mathrm{EG}|$, DM$_\mathrm{EG}$, and $\left|\beta\right|$ CDFs for the repeating (solid blue line) and non-repeating (solid red line) FRBs. For the non-repeating sources in the top right panel, we compute the CDF using the Kaplan-Meier method and the $95$\% CI is shown as the shaded red region. For the repeating FRB sources in the top right panel, the median value across all bursts is used and the full range of $L/I$ is encompassed in the shaded blue region.}
    \label{fig:all_hist_cdf}
\end{figure*}

\begin{figure}[ht]
    \includegraphics[width=0.47\textwidth]{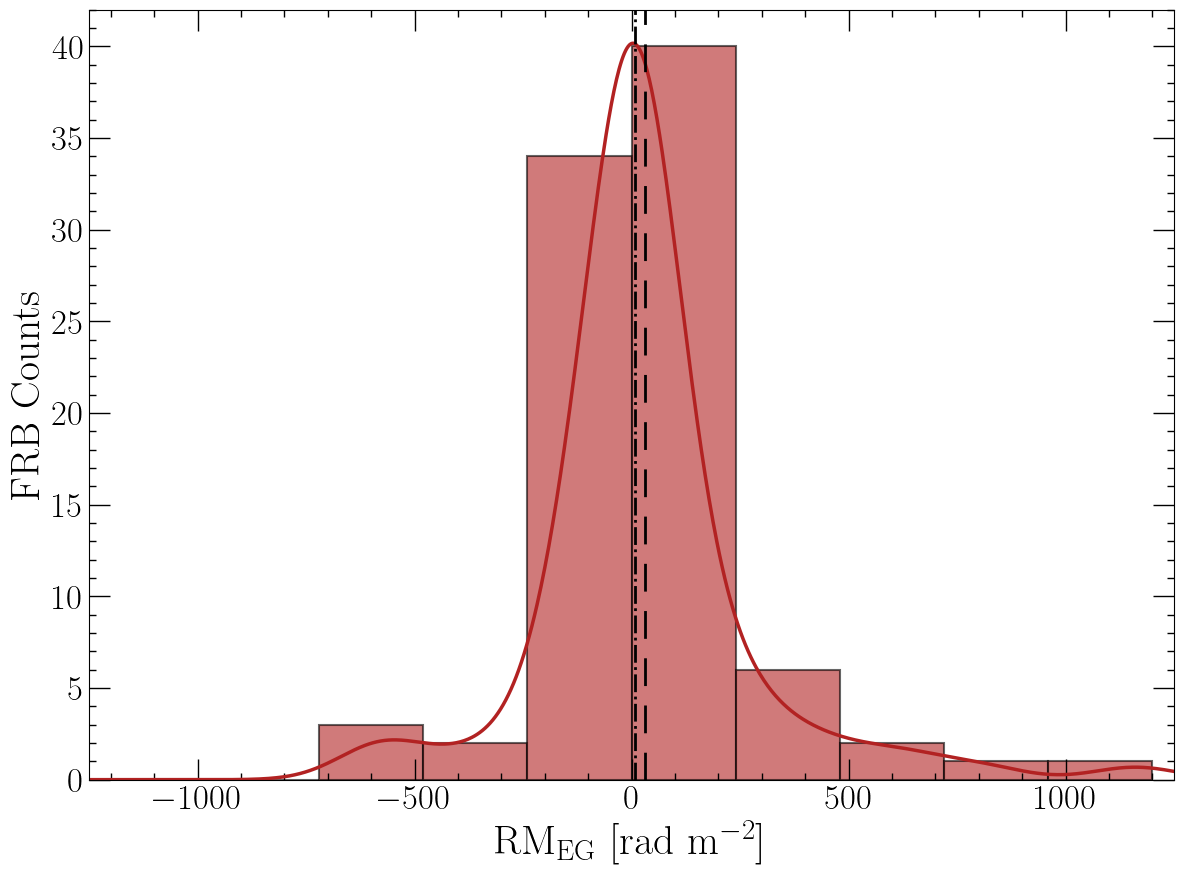}
    \caption{The distribution of $\mathrm{RM}_\mathrm{EG}$ for the 89 polarized non-repeating FRBs in our sample. The mean and median values are presented as vertical dashed and dash-dotted lines, respectively, and a Gaussian kernel density estimate is overplotted as a red curve. The distribution appears to be symmetric about $0~\mathrm{rad}~\mathrm{m}^{-2}$.}
    \label{fig:rm_nonabs}
\end{figure}

\begin{figure}[ht]
    \includegraphics[width=0.47\textwidth]{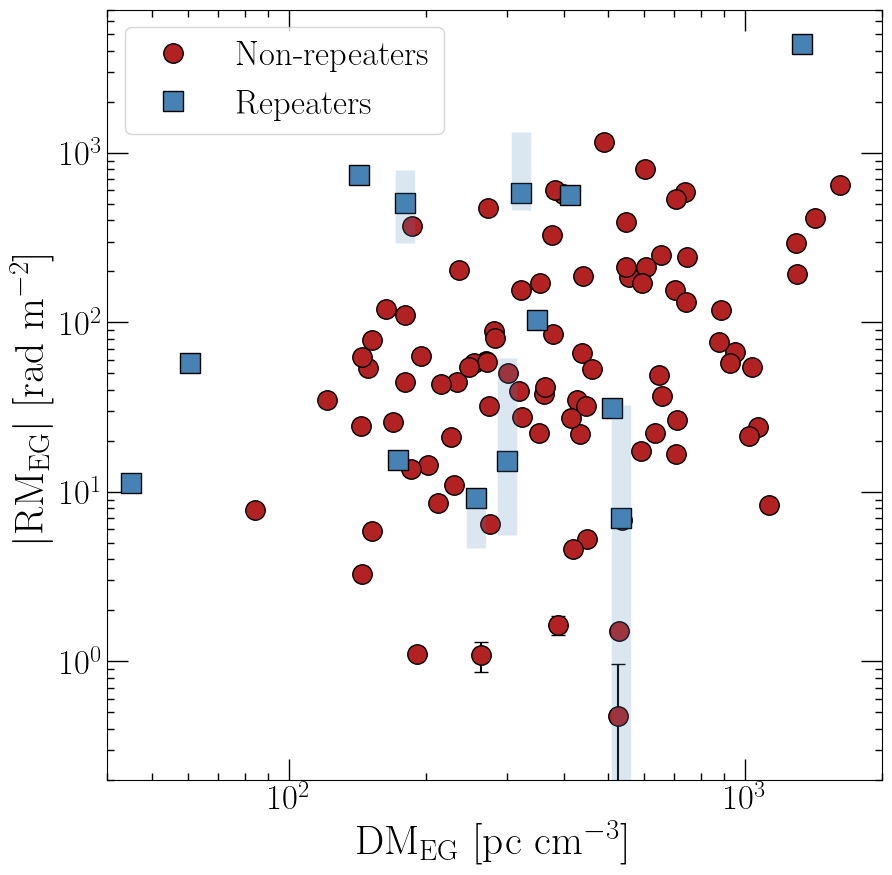}
    \caption{The foreground subtracted $|\mathrm{RM}_\mathrm{EG}|$ plotted as a function of the foreground subtracted DM, $\mathrm{DM}_\mathrm{EG}$ for our sample of non-repeating FRBs (red circles) and repeating FRBs (blue squares). For the non-repeating FRBs, the measurement uncertainties are plotted as black error bars and are typically smaller than the markers. The shaded blue regions represent the intrinsic spread in the observed parameters over time, which dominate the measurement uncertainties for repeating sources.}
    \label{fig:dm_rm}
\end{figure}

Following the steps described in Section \ref{sec:bpar_lim}, we derive observer frame lower limits on the mean magnetic field strength of the host galaxy environment parallel to the LoS, $\left|\beta\right|$. Conducting the same analysis that we have applied to $L/I$, $|\mathrm{RM}_\mathrm{EG}|$, and $\mathrm{DM}_\mathrm{EG}$, we plot the histogram and KDE of the $\left|\beta\right|$ distribution in the bottom left panel of Figure \ref{fig:all_hist_cdf} and the respective CDFs of the repeating and non-repeating populations in the bottom right panel. Similar to $|\mathrm{RM}_\mathrm{EG}|$, the $\left|\beta\right|$ distribution seems to be log-normally distributed, but with a more prominent tail extending towards lower values. We present the mean and median $\left|\beta\right|$ for repeaters and non-repeaters in Table \ref{tb:summary_stats}. Further, we see marginal evidence for a dichotomy in the $\left|\beta\right|$ distributions with $p_\mathrm{AD} = 0.017$ and $p_\mathrm{KS} = 0.042$, respectively, with repeating FRBs having higher $\left|\beta\right|$ on average. While both of these tests have relatively small $p$-values, we caution against the overinterpretation of these results for a few reasons which are discussed in Section \ref{sec:discussion_rep_nonrep}.

In Section \ref{sec:results_rm}, we detailed why our baseband sample may be biased against high $\mathrm{DM}_\mathrm{obs}$ events which, based on the results by \cite{2023ApJ...947...83C}, would preferentially be composed of non-repeating events. If indeed an unbiased baseband sample resulted in non-repeaters having, on average, higher $\mathrm{DM}_\mathrm{obs}$ but the same $|\mathrm{RM}_\mathrm{obs}|$, then their $\left|\beta\right|$ would be lower and the dichotomy in the $\left|\beta\right|$ distributions would be even more significant. 

\begin{table}[ht]
\setlength{\tabcolsep}{3.5pt}
\centering
\caption{Mean, median, and the $16^\mathrm{th} - 84^\mathrm{th}$ percentile range of the $L/I$ (only polarized FRBs), $L/I$ (including unpolarized upper limits), $|\mathrm{RM}_\mathrm{EG}|$, DM$_\mathrm{EG}$, and $\left|\beta\right|$ distributions for both our non-repeating and repeating FRB samples.} \label{tb:summary_stats}
\begin{tabular}{cccc} 
\hline
\hline
Parameter & Mean & Median & $16^\mathrm{th} - 84^\mathrm{th}$ PR\\
\hline
\multicolumn{4}{c}{\textbf{Non-repeating FRBs}}\\
\hline
$L/I$ & $0.764$ & $0.805$ & $\left[0.545,0.967\right]$\\
$L/I$ (with upper limits) & 0.633 & 0.647 & $\left[<0.234,0.949\right]$ \\
$|\mathrm{RM}_\mathrm{EG}|$ (rad~m$^{-2}$) & $135.4$ & $53.8$ & $\left[11.2,241.9\right]$\\
$\mathrm{DM}_\mathrm{EG}$ (pc~cm$^{-3}$) & $479.1$ & $389.3$ & $\left[191.0,734.3\right]$\\
$\left|\beta\right|$ ($\mu$G)& $0.386$ & $0.188$ & $\left[0.037,0.594\right]$\\
\hline
\multicolumn{4}{c}{\textbf{Repeating FRBs}}\\
\hline
$L/I$ & $0.681$ & $0.728$ & $\left[0.240,0.939\right]$\\
$|\mathrm{RM}_\mathrm{EG}|$ (rad~m$^{-2}$) & $541.7$ & $57.2$ & $\left[11.2,592.8\right]$\\
$\mathrm{DM}_\mathrm{EG}$ (pc~cm$^{-3}$) & $355.8$ & $301.3$ & $\left[135.8,512.7\right]$\\
$\left|\beta\right|$ ($\mu$G) & $1.539$ & $0.365$ & $\left[0.061,3.523\right]$\\
\hline
\hline
\end{tabular}
\end{table}

\begin{table}[ht]
\centering
\caption{Summary of AD, KS, Peto \& Peto, and Log-Rank tests on whether the $L/I$, $|\mathrm{RM}_\mathrm{EG}|$, $\mathrm{DM}_\mathrm{EG}$, and $\left|\beta\right|$ of non-repeating and repeating FRBs are drawn from the same underlying distribution. Recall that, without host galaxy redshifts, the values of $\left|\beta\right|$ have an uncorrected dependence on the distance of FRB sources. Both the test statistic and associated significance level are reported for each test. The AD and KS tests are applied to only the $84$ polarized FRBs in our sample while the Peto \& Peto and Log-Rank tests additionally factor in the $L/I$ upper limits for the $29$ unpolarized non-repeaters. The $p$-values that are $< 0.05$ are presented in bold face.} \label{tb:ad_ks_tests}
\begin{tabular}{ccccc} 
\hline
\hline
 & \multicolumn{2}{c}{\textbf{AD Test}} & \multicolumn{2}{c}{\textbf{KS Test}}\\
\hline
Parameter & $S_\mathrm{AD}$ & $p_\mathrm{AD}$ & $S_\mathrm{KS}$ & $p_\mathrm{KS}$\\
\hline
$L/I$ & 0.179 & $\geq 0.250$ & 0.220 & 0.560\\
$|\mathrm{RM}_\mathrm{EG}|$ (rad~m$^{-2}$) & 0.687 & 0.172 & 0.263 & 0.343\\
$\mathrm{DM}_\mathrm{EG}$ (pc~cm$^{-3}$) & 1.71 & 0.064 & 0.277 & 0.290\\
$\left|\beta\right|$ ($\mu$G) & 3.18 & \textbf{0.017} & 0.394 & \textbf{0.042}\\
\hline
 & \multicolumn{2}{c}{\textbf{Peto \& Peto}} & \multicolumn{2}{c}{\textbf{Log-Rank}}\\
\hline
Parameter & \multicolumn{2}{c}{$p_\mathrm{PP}$} &  \multicolumn{2}{c}{$p_\mathrm{LR}$} \\
\hline
$L/I$ (with upper limits) & \multicolumn{2}{c}{0.962} & \multicolumn{2}{c}{0.768} \\
\hline
\hline
\end{tabular}
\end{table}

\subsection{FRB host rotation measure and LoS magnetic field strength estimates} \label{sec:results_z_corr}
While we do not have redshift information for individual FRBs, we attempt to apply a statistical correction and derive $\left|\mathrm{RM}_\mathrm{host}\right|$ and $\left|\left<B_{\parallel,\mathrm{host}}\right>\right|$ distributions for our non-repeating sample. To do this, we draw a $\mathrm{DM}_\mathrm{host}$ distribution matching the size of our polarized non-repeating sample from a log-normal distribution with a mean ($1.93/\mathrm{log}_{10}(e)$) and standard deviation ($0.41/\mathrm{log}_{10}(e)$) following \cite{2023ApJ...944..105S}. Subtracting this $\mathrm{DM}_\mathrm{host}$ from our $\mathrm{DM}_\mathrm{EG}$, we obtain a $\mathrm{DM}_\mathrm{IGM}(z)$ distribution that is then used to compute a $z$ distribution by assuming $z \sim \mathrm{DM}_\mathrm{IGM}(z)/1000$ \citep{2020Natur.581..391M}. With a $z$ distribution in hand, we obtain $\left|\mathrm{RM}_\mathrm{host}\right| = \left|\mathrm{RM}_\mathrm{EG}\right|(1+z)^2$ and then $\left|\left<B_{\parallel,\mathrm{host}}\right>\right| = 1.232 \left|\mathrm{RM}_\mathrm{host}\right|/\mathrm{DM}_\mathrm{host}$. We repeat this process over $1,000$ trials and derive the mean $\left|\mathrm{RM}_\mathrm{host}\right|$ and $\left|\left<B_{\parallel,\mathrm{host}}\right>\right|$ distributions. We reinforce that we do not apply this type of $z$ correction to individual FRBs and only use it as a statistical correction to make conclusions on the polarized, non-repeating population as a whole.

\begin{figure*}[ht!]
\begin{center}
    \includegraphics[width=0.99\textwidth]{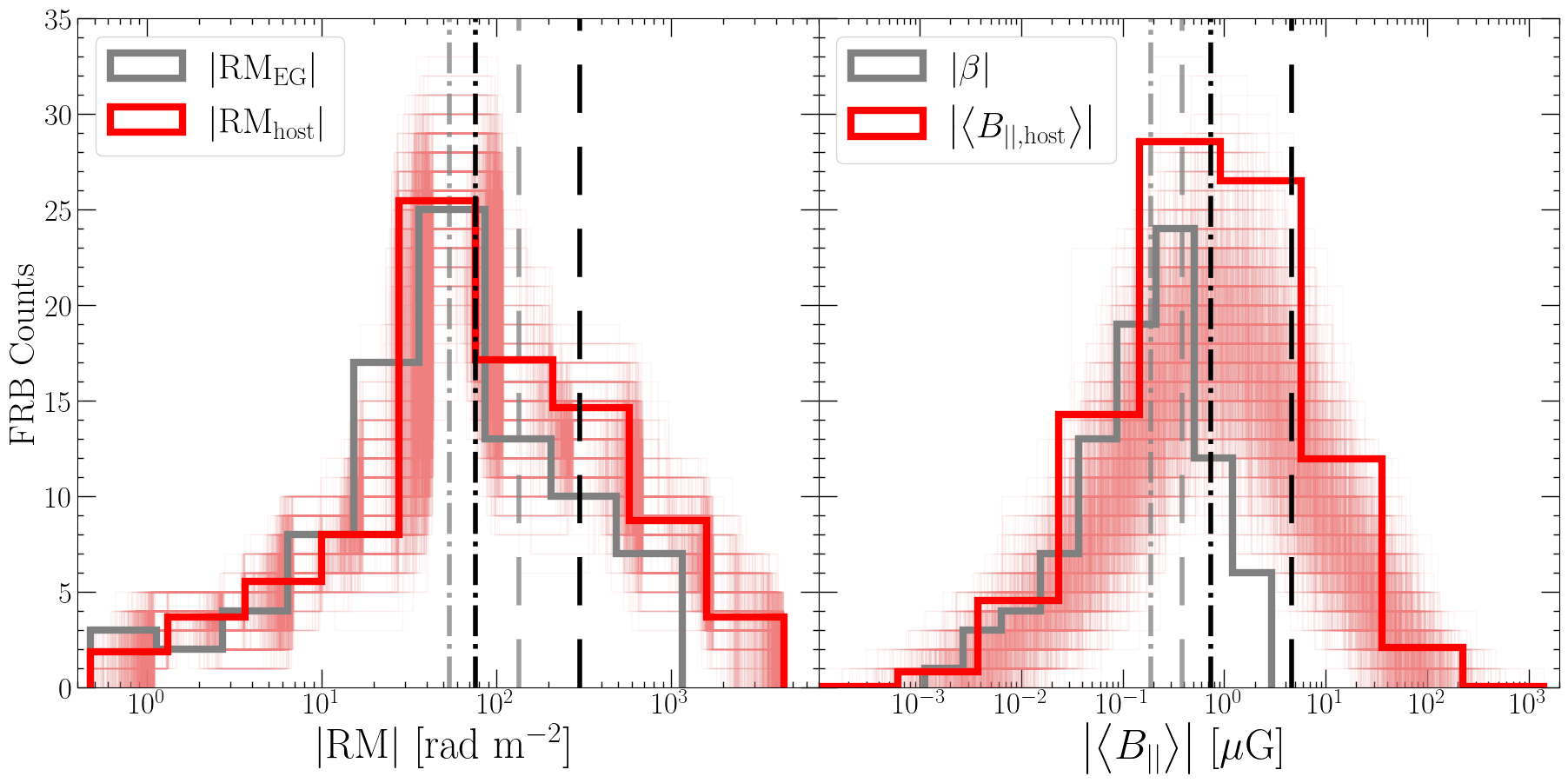}
    \caption{(Left) Mean $\left| \mathrm{RM}_\mathrm{host} \right|$ distribution (solid red line) of our non-repeating FRB sample after applying a redshift correction based on the $\mathrm{DM}_\mathrm{host}$ distribution derived by \cite{2023ApJ...944..105S}. The mean and median of the distribution are plotted as black dashed and dash-dotted lines, respectively. All 1000 individual trials, each with an independently drawn $\mathrm{DM}_\mathrm{host}$ distribution, are shown as faint red histograms in the background. The observer frame $\left| \mathrm{RM}_\mathrm{EG} \right|$ distribution with no redshift correction is overplotted (solid grey line) with its respective mean and median shown as grey dashed and dash-dotted lines, respectively. (Right) An analogous setup to the left panel but for comparing the rest frame $\left|\left<B_{\parallel,\mathrm{host}}\right>\right|$ and observer frame lower limits, $\left| \beta \right|$.}
    \label{fig:rm_b_z_corr}
\end{center}
\end{figure*}

In Figure \ref{fig:rm_b_z_corr}, we plot the mean $\left|\mathrm{RM}_\mathrm{host}\right|$ and $\left|\left<B_{\parallel,\mathrm{host}}\right>\right|$ distributions derived from this approach and contrast them to the observer frame $\left|\mathrm{RM}_\mathrm{EG}\right|$ and $\left| \beta \right|$ distributions. As expected, both the $\left|\mathrm{RM}_\mathrm{host}\right|$ and $\left|\left<B_{\parallel,\mathrm{host}}\right>\right|$ distributions are shifted to higher values compared to the observer frame $\left|\mathrm{RM}_\mathrm{EG}\right|$ and $\left| \beta \right|$ distributions. The mean and median of the $\left|\mathrm{RM}_\mathrm{host}\right|$ distribution is $302.7~\mathrm{rad}~\mathrm{m}^{-2}$ and $75.9~\mathrm{rad}~\mathrm{m}^{-2}$, respectively, with a $16^\mathrm{th} - 84^\mathrm{th}$ PR of $[15.0, 494.3]~\mathrm{rad}~\mathrm{m}^{-2}$. The mean and median of the $\left|\left<B_{\parallel,\mathrm{host}}\right>\right|$ distribution is $4.62~\mu\mathrm{G}$ and $0.740~\mu\mathrm{G}$, respectively, with a $16^\mathrm{th} - 84^\mathrm{th}$ PR of $[0.088, 5.73]~\mu\mathrm{G}$. 

\subsection{PA variability} \label{sec:results_pa}
\begin{figure}[ht]
    \includegraphics[width=0.47\textwidth]{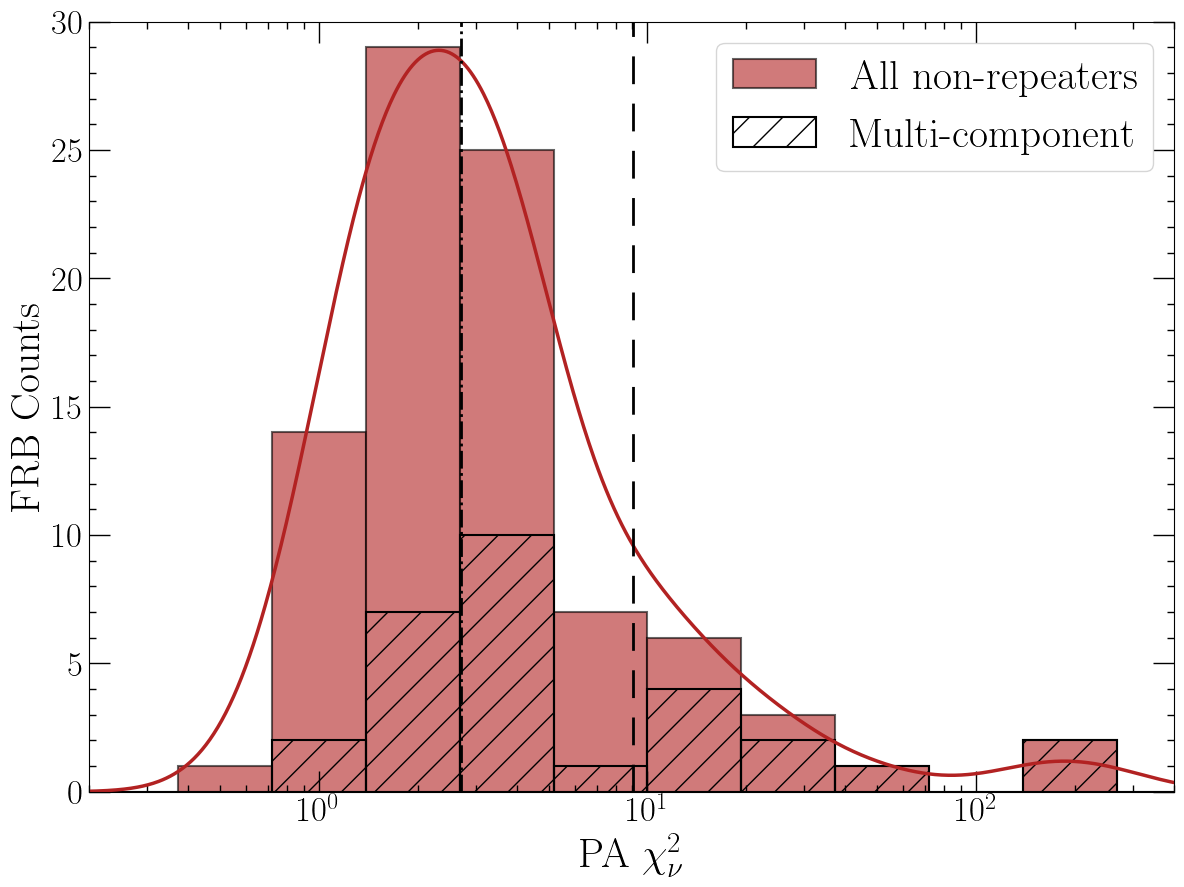}
    \caption{The distribution of $\chi_\nu^2$ values from comparing the temporal PA variations of all 88 non-repeating FRBs against a flat PA profile. The mean and median values are presented as vertical dashed and dash-dotted lines, respectively, and a Gaussian kernel density estimate is overplotted as a red curve. FRBs with multiple components are highlighted as a hatched histogram.}
    \label{fig:pa_chi2}
\end{figure}

\begin{figure*}[ht]
    \centering
    \includegraphics[width=0.87\textwidth]{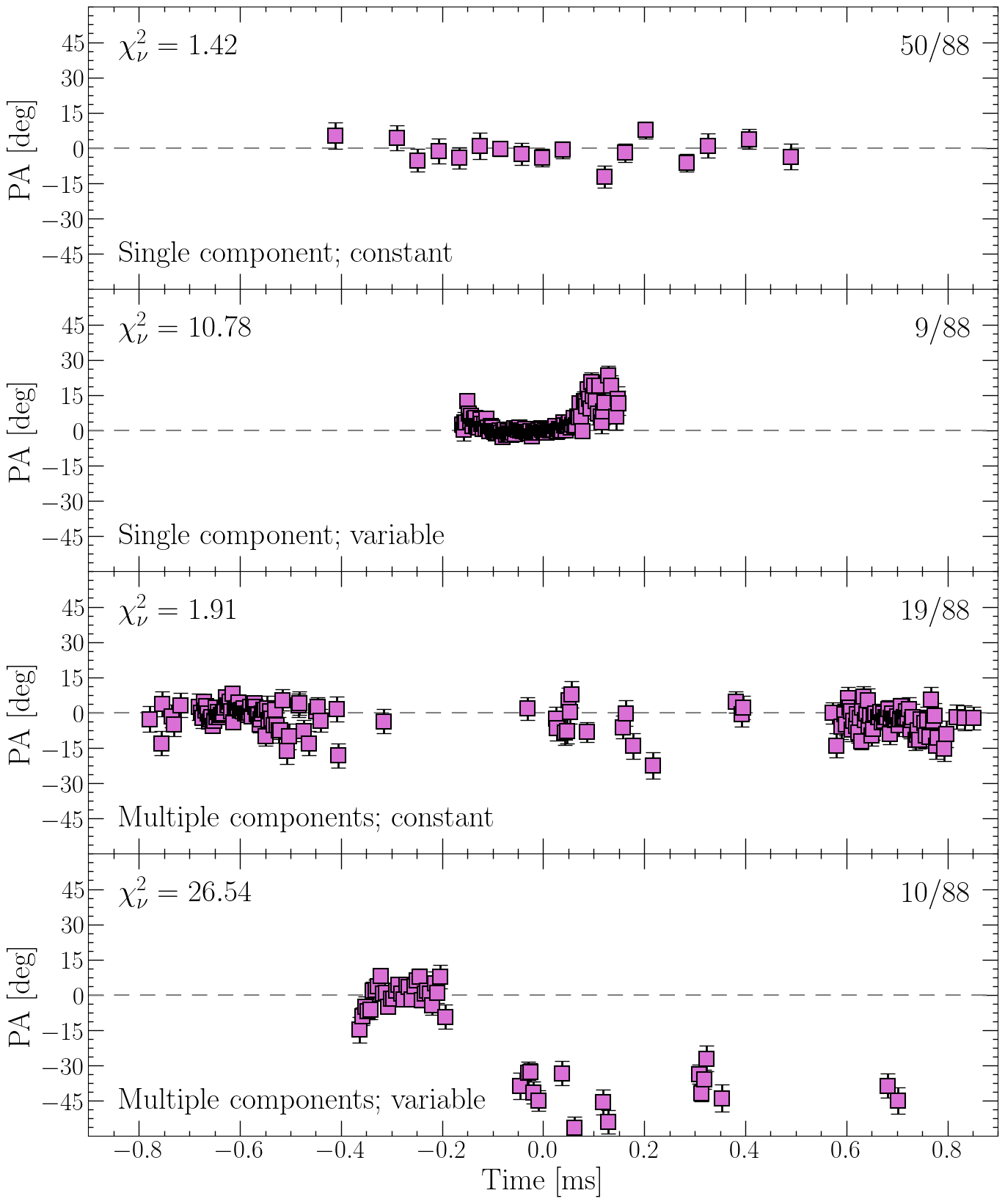}
    \caption{PA profiles of four FRBs in our sample (from top to bottom: FRB 20181226E, FRB 20190425A, FRB 20190320B, and FRB 20190224D) with distinct temporal behaviors exemplifying the archetypes presented in Section \ref{sec:results_pa}: (i) single component with a constant PA, (ii) single component with a variable PA, (iii) multiple components with a constant PA, and (iv) multiple components with a variable PA. Measurement uncertainties are presented as black lines but are, most of the time, smaller than the marker size. The $\chi_\nu^2$ value from fitting the PA profile to a flat PA at $0~\mathrm{deg}$ is listed at the top left of each panel. The text in the bottom left of each panel describes the classification of the PA variability into one of these four archetypes. The number of FRBs belonging to each archetype is presented in the top right corner as a fraction of the total polarized sample. Note that there is one FRB (FRB 20190419B) for which we cannot derive a $\chi_\nu^2$ value as it does not have a sufficient number of data points in its PA profile that exceed the $\mathrm{S/N}(L)_\mathrm{thresh} = 5$ S/N threshold, so it is not included in any of the four archetypes described here.}
    \label{fig:pa_examples}
\end{figure*}

In Section \ref{sec:pa_var}, we describe a methodology to quantify the magnitude of PA variability for an FRB as a function of time using a reduced $\chi_\nu^2$ test. We apply this technique to our sample of 89 non-repeating FRBs. A histogram and KDE of the resulting reduced $\chi_\nu^2$ values are presented in Figure \ref{fig:pa_chi2} and are included in Table \ref{tb:pol_results}. Note that one FRB in our sample (FRB 20190419B) did not have a sufficient number of points on its PA curve to produce a reasonable $\chi_\nu^2$ fit due to a combination of low S/N and low linear polarization. For this FRB, we do not report a $\chi_\nu^2$ value and it is excluded from the categorization below. The mean and median $\chi_\nu^2$ are $9.01$ and $2.71$, respectively, for the 88 non-repeating FRBs with PA $\chi_\nu^2$ fits. We define a conservative PA variability threshold of $\chi_\nu^2 \geq 5$, above which we consider the PA profile to be variable. We present four illustrative PA profiles in Figure \ref{fig:pa_examples} corresponding to four qualitative archetypes for PA behavior that we define below \citep[in consonance with the classification system deployed by][]{2023arXiv230806813S}. In these archetypes, we also classify bursts based on whether they are single component or multi-component (determined through visual inspection after de-dispersing to the respective $\mathrm{DM}_\mathrm{obs,struct}$).
\begin{enumerate}
    \item \textit{Single component, constant PA.} This category encompasses all single-component FRBs in our non-repeating sample that have $\chi_\nu^2 < 5$ across their PA profiles. An example of this type of PA behavior is presented in the first panel of Figure \ref{fig:pa_examples} (FRB 20181226E), where the PA remains constant across the $\sim 1$~ms burst duration. A single-component, constant PA behavior is the most common out of the four qualitative archetypes we define, with 50/88 FRBs falling under this categorization.
    \item \textit{Single component, variable PA.} This second category includes all single-component bursts with $\chi_\nu^2 \geq 5$. These FRBs display PA variations that are continuous as a function of time across a single burst. The second panel of Figure \ref{fig:pa_examples} shows one such example (FRB 20190425A) wherein the PA rises $\sim 25$~deg in $0.35$~ms. Most FRBs under this umbrella are similar to FRB 20190425A in that they have modest PA variations and do not show, for example, the large S-shaped swings typically seen in pulsar emission \citep{2012hpa..book.....L} and seen infrequently in FRBs \citep[e.g., see][]{2024arXiv240209304M}. This second category constitutes only 9/88 FRBs in our sample.  
    \item \textit{Multiple component, constant PA.} Of the 29 multi-component bursts, 19 display a constant PA, with $\chi_\nu^2 < 5$, across their entire burst envelope. One example of this behavior is shown in Figure \ref{fig:pa_examples}; in the third panel we see a constant PA across $3+$ components spanning $\sim 1.7$~ms (FRB 20190320B).
    \item \textit{Multiple components, variable PA.} The final category describes all multi-component bursts whose PAs vary component-to-component, leading to $\chi_\nu^2 \geq 5$. This classification is distinct from the single component, variable PA category as the PA profile, in this case, may sometimes remain constant within each component but the PA between one or more components is variable and we do not see a smooth, continuous change in the PA. For the FRBs in this archetype it is possible that either: (i) the PA variations are discontinuous between components or (ii) the PA variation is continuous but the emission bridging components is too faint to accurately measure PAs in that time range. The fourth panel of Figure \ref{fig:pa_examples} shows the PA of FRB 20190224D which appears to have $3-4$ components, each with widths of $0.05 - 0.25$~ms. Here, the PA of the first component differs from the rest by $\sim 45$~deg, while the intra-component PA for each of them remains somewhat constant. In total, 10/88 FRBs fall into under this classification.
\end{enumerate}

We compare the $L/I$, $|\mathrm{RM}_\mathrm{EG}|$, and $\left|\beta\right|$ of the FRBs constituting the four archetypes described above but find no evidence for any differentiation between the subpopulations. Further, we find no correlation or anti-correlation between the PA $\chi_\nu^2$ values and $L/I$ and $|\mathrm{RM}_\mathrm{EG}|$. 

\subsubsection{Rotating vector model} \label{sec:results_rvm}
Like FRBs, pulsars display a rich phenomenology of behavior in their PA evolution. However, the PA behavior of pulsars is markedly different to that of FRBs; FRB PA curves tend to be flatter than those encountered in time-averaged profiles of pulsars and excursions in the PA, when present, tend to be more erratic in FRBs. For pulsars, a geometric model known as the rotating vector model \citep[RVM;][]{1969ApL.....3..225R} is often invoked to explain the smooth swing in the PA over pulse phase as a projection effect of the neutron star's rotating dipolar field. Over the years, polarimetric studies of pulsars have demonstrated the RVM to be a powerful tool with approximately $\sim$60\% of the population displaying PA behavior that is well described by the RVM \citep[e.g.,][]{Johnston2023}. The RVM provides information on important geometrical parameters, such as the inclination angle between the neutron star's magnetic axis and rotation axis ($\alpha$) and the impact angle between the LoS and the magnetic axis ($\gamma$; often referred to as $\beta$ elsewhere).

\begin{figure*}[ht]
    \includegraphics[width=0.99\textwidth]{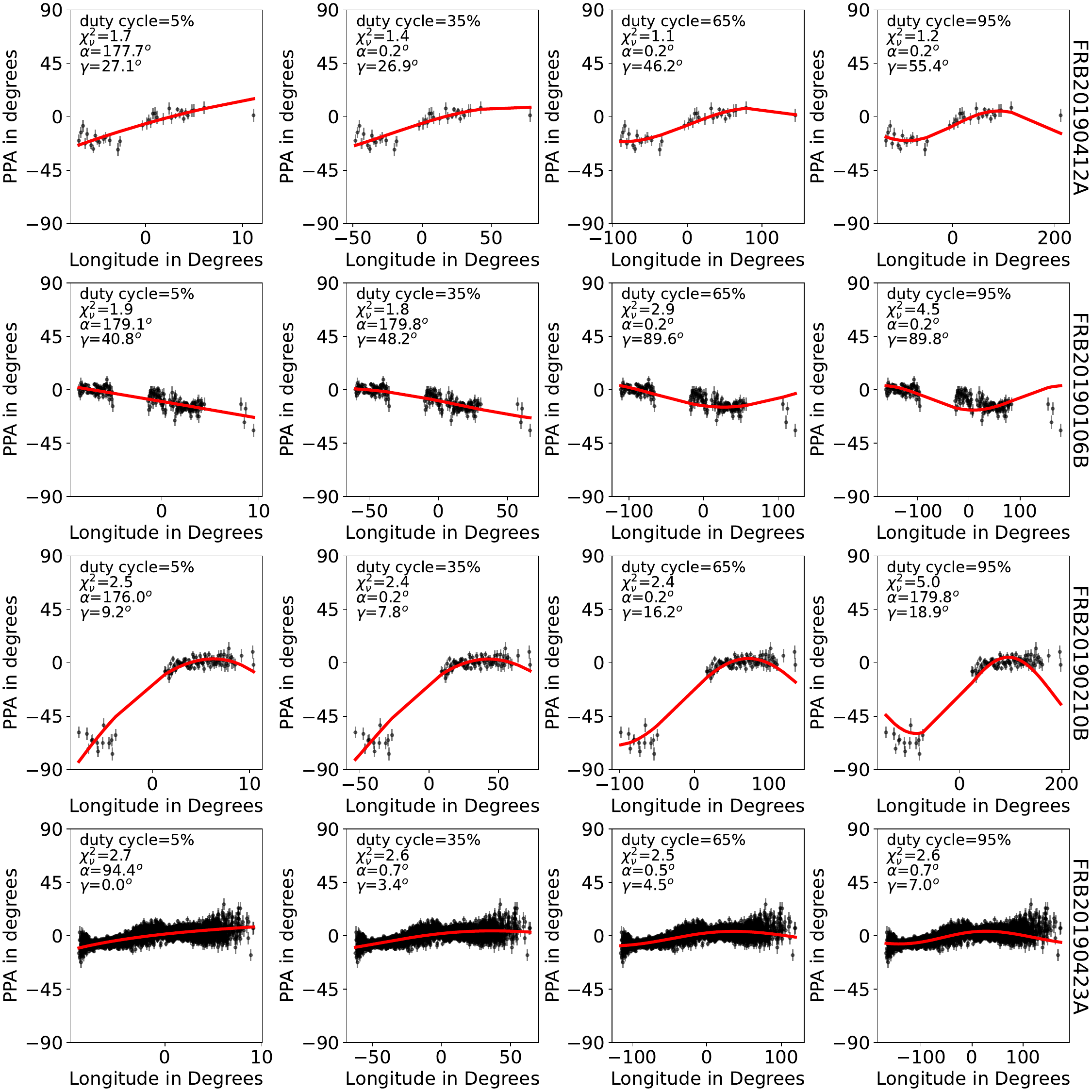}
    \caption{PA curves (black points) and associated best-fit RVM models (red) for a series of different assumed duty cycles (5,35,65,95$\%$) for a sample of non-repeating FRBs (FRBs 20190412A, 20190106B, 20190210B, and 20190423A) that display significant PA variation. Quality of fits are indicated by reduced chi-squared values ($\chi^2_\nu$; top left corner) along with best-fit $\alpha,\gamma$ parameters. These four FRBs were chosen as they have the smallest RVM $\chi_\nu^2$ values from the analyzed sample. All four FRBs are associated with the multiple component, variable PA archetype.}
    \label{fig:pa_rvm_tile}
\end{figure*}

While growing evidence does exist for magnetospheric origins of some FRBs \cite[e.g.,][]{2020Natur.586..693L,2022NatAs...6..393N}, the suitability of a geometric interpretation of the RVM remains an open question. If the non-repeating FRB sample reported here does arise within the magnetosphere of rotating neutrons stars then the substantial differences in PA behavior of FRBs and pulsars suggests that, at the very least, FRB emission occurs at very different regions of the magnetosphere than that of the pulsar sample. In an attempt to study this more systematically, we fit the RVM to a subsample that displays significant PA variations ($n=19$, i.e., the single component, variable PA \& multiple component variable PA archetypes from Section~\ref{sec:results_pa}) and find best-fit reduced $\chi_{\nu}^2$ values that are regularly between $2-3$. PA curves and their associated best-fit RVMs are displayed in Figure~\ref{fig:pa_rvm_tile} for a selection of this subsample with the lowest RVM $\chi_{\nu}^2$ values. Importantly, without information on the period of the source (in the rest frame of the source), many geometric configurations can give rise to the observed PA behavior. This ambiguity is captured by the four columns of Figure~\ref{fig:pa_rvm_tile}, which display the best-fitting RVMs at four different assumed spin periods corresponding to duty cycle trials of 5, 35, 65 \& 95$\%$. While the fit quality obtained for this sample is not inconsistent with that attained from pulsars that are considered well described by the RVM \citep[e.g.,][]{2023RAA....23j4002W}, we caution against concluding this as proof of an RVM-like scenario operating in FRBs. Indeed, the RVM can replicate a wide variety of PA behavior and is thus susceptible to spurious ``fits" that may potentially mislead the interpretation of the physical mechanism producing the observed PA evolution. Good quality (but likely spurious) RVM fits can be easily obtained in cases where the PA evolution is modest/quasi-linear and it is thus unsurprising that many of the FRBs in our sample with the lowest $\chi_{\nu}^2$ are those exhibiting these properties.

In the absence of further information, namely the spin period of the source and detection of repetition, we are unable to make any strong claims of whether PA evolution of this sample can be described by an RVM-like scenario and which set of geometric $\alpha,\gamma$ parameters are preferred. A continuous ``S"-shaped PA swing, common to pulsars and recently observed in at least one nearby CHIME detected FRB \citep{2024arXiv240209304M}, has not been observed in this sample. However, if we assume an RVM-like scenario for our sample, then the overwhelming preponderance of flat PA curves of our non-repeating sample does imply a preference for geometries where $\alpha$ clusters near $\alpha=0^{\circ}$ or $\alpha=180^{\circ}$. This indicates a much higher degree of alignment/anti-alignment of the neutron star's magnetic and rotation axis than what is commonly inferred for the pulsar population. This scenario remains quite speculative with the data in hand, but is discussed further in Section~\ref{sec:discussion_pulsars_magnetars}. 

\subsection{Depolarization} \label{sec:results_depol}
Possible frequency-dependent depolarization has been seen in some prolific repeating FRBs and it has been suggested to arise from scattering in the magnetoionic medium surrounding the FRB source \citep{2022Sci...375.1266F}. In this Section, we investigate whether frequency-dependent depolarization is present in our non-repeating FRB sample to better inform whether or not this spectral depolarization is a ubiquitous feature across the entire FRB population.

\subsubsection{Depolarization within the CHIME/FRB band} \label{sec:results_depol_frac}
First, we attempt to constrain any changes in the linear polarization fraction within the CHIME/FRB band itself, in the observer frame. For this purpose, we select a subset of FRBs that emit over the full $400$~MHz bandwidth of CHIME/FRB; in total this condition is met for 23 FRBs. Note that this subset does not necessarily cover all FRBs from the first CHIME/FRB catalog \citep{2021ApJS..257...59C} with $400$~MHz bandwidth for which baseband data exist. This is because of the limited baseband buffer ($\sim 20$~seconds), which can lead to missing frequency channels for some heavily dispersed events. In addition, during the pre-commissioning stage (up to 2018 August 27), the uncertainties in DM and FRB time of arrival were not correctly accounted for in the baseband system, leading to partial or full loss of baseband data for some events.

For the subset of 23 FRBs with $400$~MHz bandwidth in the baseband data, we derive a depolarization ratio $f_\mathrm{depol}$ between the band-averaged linear polarization fraction at $400-600$~MHz, $(L/I)_{500}$, and at $600-800$~MHz, $(L/I)_{700}$,
\begin{equation}
f_\mathrm{depol} = \frac{(L/I)_{500}}{(L/I)_{700}}\,. \label{eq:f_depol}
\end{equation}
Note that, while we refer to $f_\mathrm{depol}$ as a ``depolarization ratio'', it is possible to observe $f_\mathrm{depol} > 1$ (i.e., in that case the polarization fraction increases towards lower frequencies). Characterizing the depolarization in this manner affords a model-independent determination of any decrease in $L/I$ across the CHIME frequency range. In Table \ref{tb:pol_results}, we present $f_\mathrm{depol}$ for all 23 of the FRBs occupying the full $400-800$~MHz band.

\begin{figure}[ht]
    \includegraphics[width=0.46\textwidth]{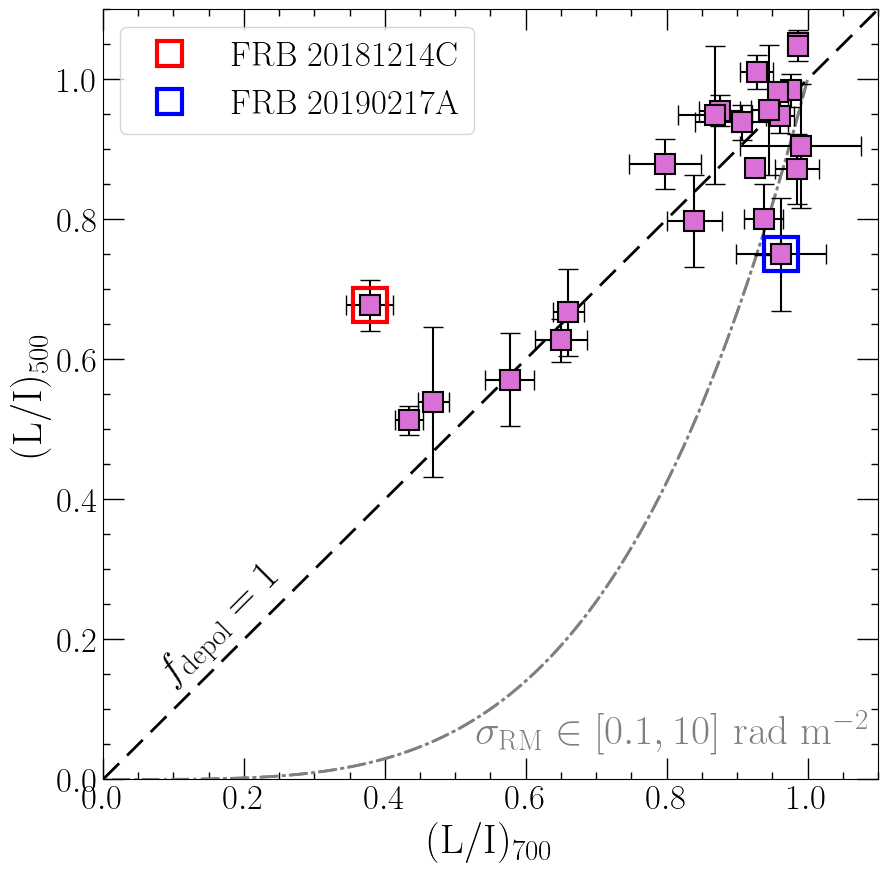}
    \includegraphics[width=0.46\textwidth]{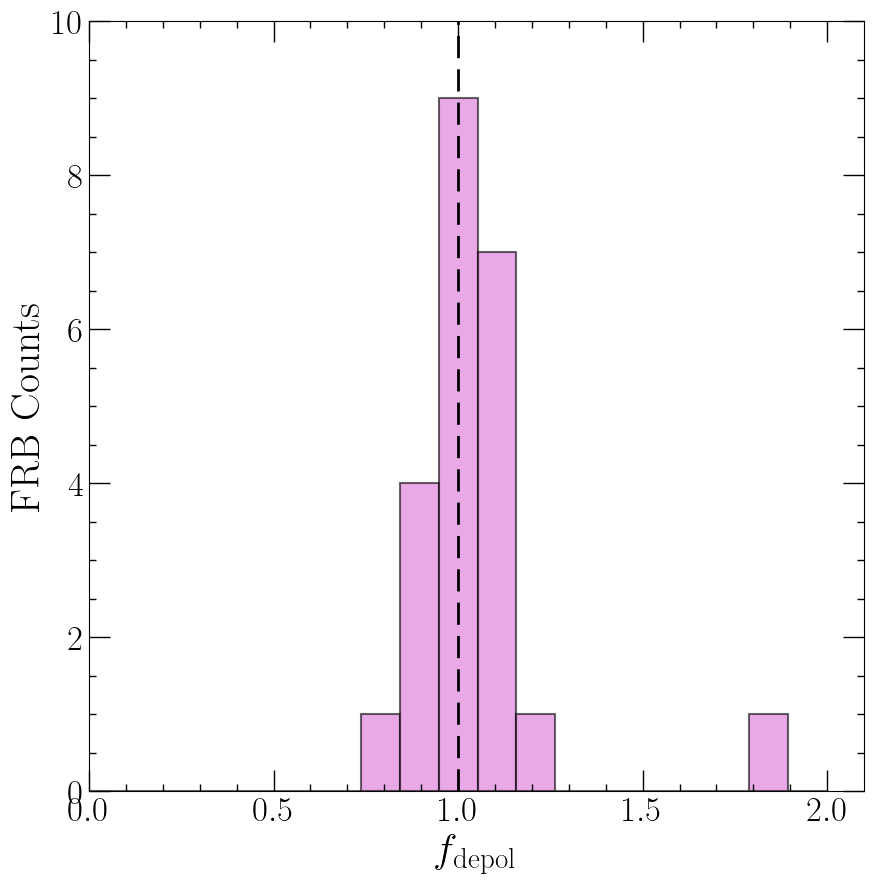}
    \caption{(Top) The linear polarization fraction averaged over $400-600$~MHz plotted against the average fraction over $600-800$~MHz for 23 FRBs that emit over the full $400$~MHz CHIME/FRB band. The black dashed line indicated a depolarization ratio of 1 (i.e., no change in the linear polarization fraction over the observed band). Two outliers are highlighted: FRB 20181214C with $f_\mathrm{depol} = 1.8 \pm 0.2$ in red and FRB 20190217A which has $f_\mathrm{depol} = 0.8 \pm 0.1$ in blue. The grey dot-dashed line is the expected distribution of $(L/I)_{500}$ versus $(L/I)_{700}$ due to RM scattering from multi-path propagation for a range of $\sigma_\mathrm{RM}$ between $0.1 - 10~\mathrm{rad}~\mathrm{m}^{-2}$ (with $\sigma_\mathrm{RM} = 0.1~\mathrm{rad}~\mathrm{m}^{-2}$ being at the top of the curve and $\sigma_\mathrm{RM} = 10~\mathrm{rad}~\mathrm{m}^{-2}$ towards the origin). (Bottom) A histogram of the depolarization ratio $f_\mathrm{depol}$ with a dashed line indicating $f_\mathrm{depol} = 1$.}
    \label{fig:depol}
\end{figure}

The top panel of Figure \ref{fig:depol} shows $(L/I)_{500}$ plotted as a function of $(L/I)_{700}$ for these FRBs. In this space, $f_\mathrm{depol} = 1$ corresponds to $(L/I)_{500} = (L/I)_{700}$ and points along that line are not undergoing measurable spectral depolarization across the CHIME/FRB band. FRBs below this line are depolarizing, while those above it are experiencing increasing $L/I$ with decreasing frequency. We also compare against the expected distribution of $(L/I)_{500}$ and $(L/I)_{700}$ if the observed FRB population was undergoing RM scattering due to multi-path propagation, following Equation \ref{eq:depol}, with $\sigma_\mathrm{RM} \in [0.1, 10]~\mathrm{rad}~\mathrm{m}^{-2}$ and assuming an intrinsic $L/I = 1.0$ at the time of emission. Here, it is important to note that we are only sensitive to a limited range of $\sigma_\mathrm{RM}$ values with our data. Setting a minimum difference of $|(L/I)_{700} - (L/I)_{500}| \geq 0.05$ and requiring that at least $(L/I)_{700} \geq 0.2$, such that it is possible to detect a polarized signal, we would be able to detect $0.5 \lesssim \sigma_\mathrm{RM} \lesssim 5.0~\mathrm{rad}~\mathrm{m}^{-2}$. In the bottom panel of Figure \ref{fig:depol}, we display the histogram of $f_\mathrm{depol}$ values. Most FRBs in this subset are consistent with $f_\mathrm{depol} = 1$, meaning that their $L/I$ is constant over the CHIME/FRB band. Out of the 23 sources, 21 have less than a $20$\% change in their linear polarization fraction between $500$ and $700$~MHz (i.e., they have $0.8 < f_\mathrm{depol} < 1.2$). There are two notable outliers: (i) FRB 20190217A with $f_\mathrm{depol} = 0.8 \pm 0.1$ and (ii) FRB 20181214C with $f_\mathrm{depol} = 1.8 \pm 0.2$. 

\begin{figure*}[ht!]
\begin{center}
    \includegraphics[width=0.32\textwidth]{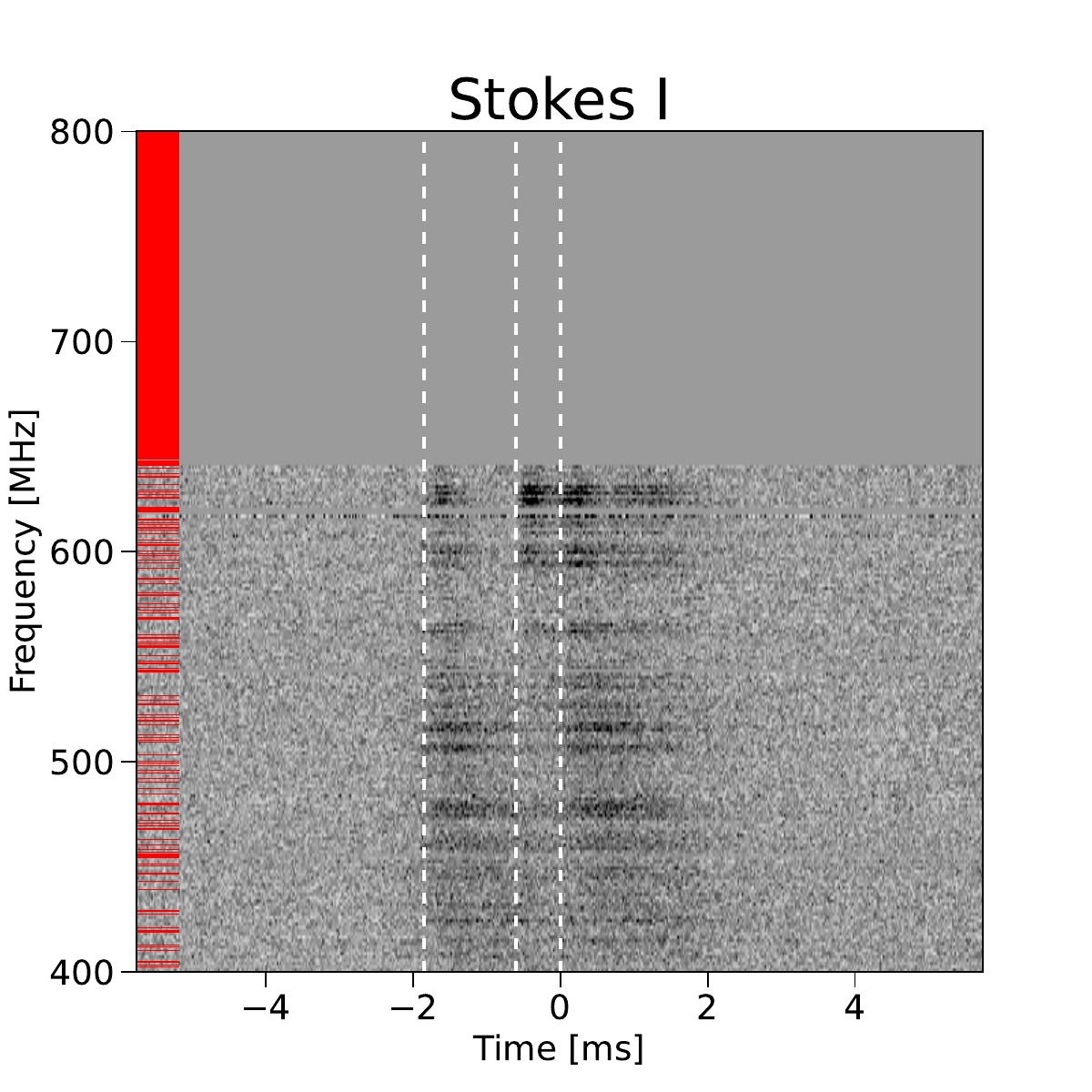}
    \includegraphics[width=0.32\textwidth]{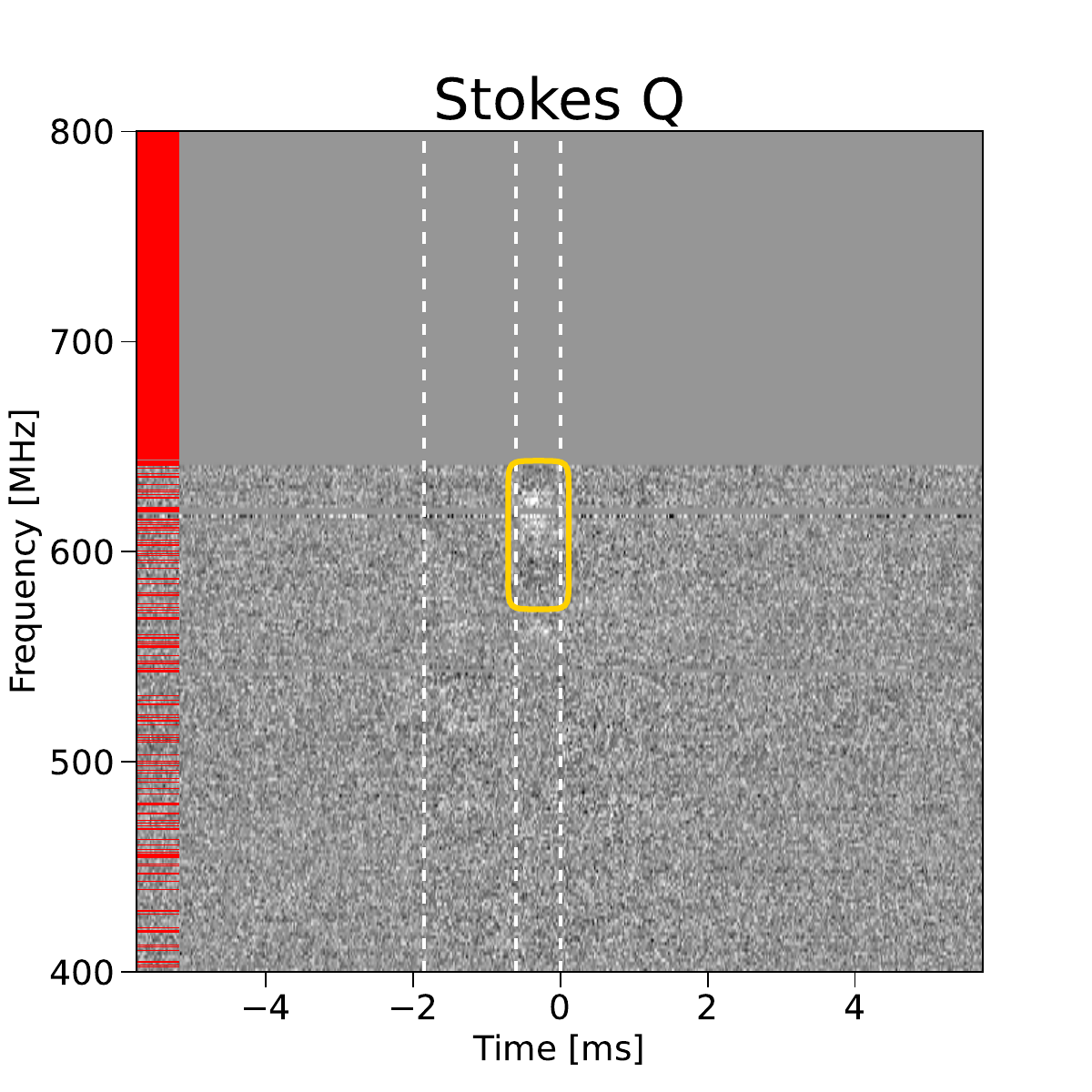}
    \includegraphics[width=0.32\textwidth]{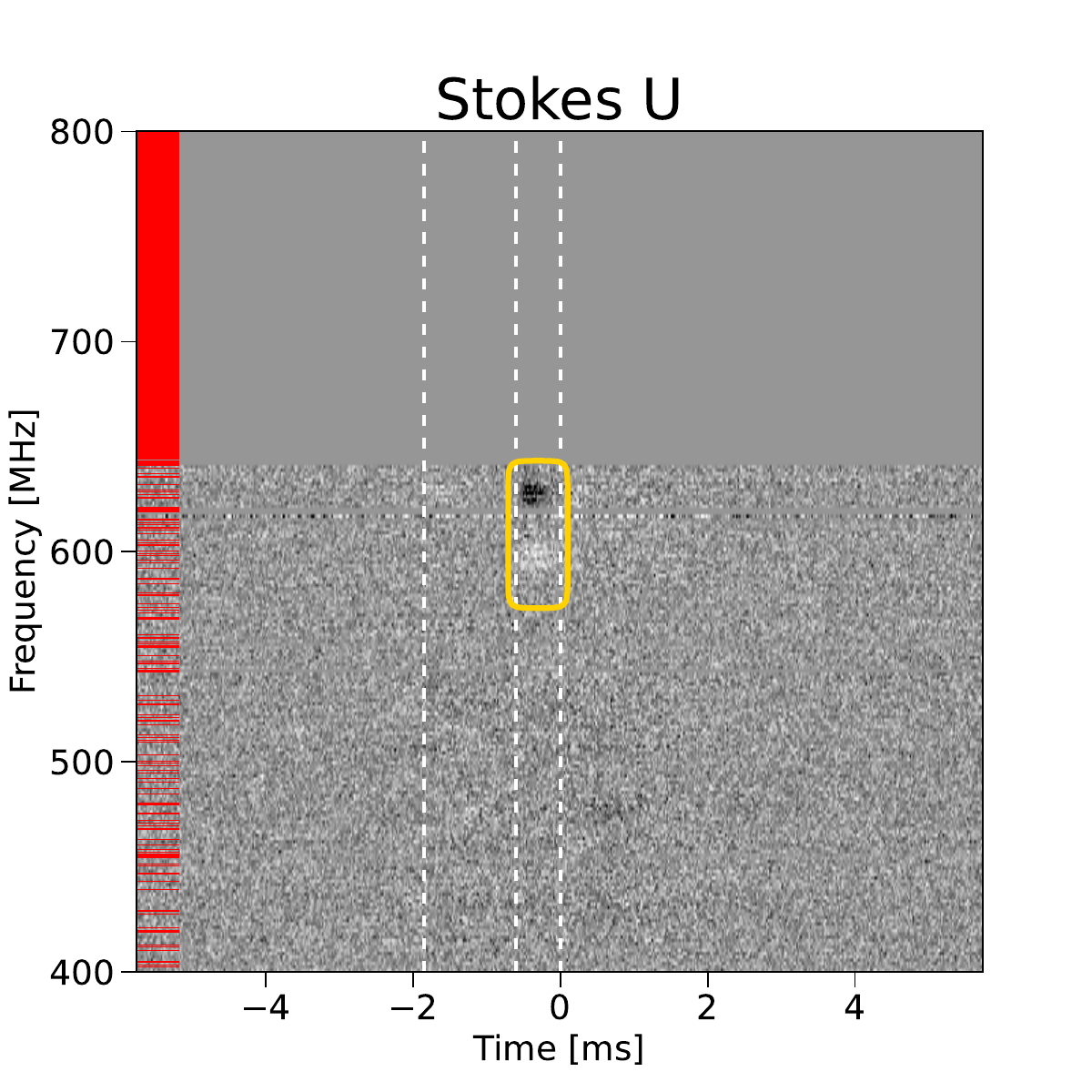}
    \caption{Stokes $I$ (left), $Q$ (middle), and $U$ (right) waterfalls for FRB 20190227A which is composed of at least three components. Dashed white lines are used to highlight the approximate start of each burst component. The narrowband component with measured linear polarization is highlighted in yellow. The red streaks to the left of each waterfall plot highlight frequency channels that are masked out due to radio frequency interference. Note that the data have been downsampled by a factor of $n_\mathrm{down} = 8$ and thus the resolution of the flagged radio frequency interference channels and Stokes images do not match exactly. The first and third component are unpolarized with $L/I \leq 0.219$ and $L/I \leq 0.190$, respectively, while the second component is polarized with $L/I = 0.50 \pm 0.01$ and has $\mathrm{RM} = 62.9 \pm 0.9~\mathrm{rad}~\mathrm{m}^{-2}.$}
    \label{fig:possible_depol}
\end{center}
\end{figure*}

One FRB, for which we only have data between $400-650$~MHz, that was labelled unpolarized (FRB 20190227A) has unique polarimetric properties that may be indicative of depolarization. The leftmost panel of Figure \ref{fig:possible_depol} shows the Stokes $I$ dynamic spectra of FRB 20190227A, which is composed of at least three components over $\sim 4$~ms, though it is ambiguous whether the third component is comprised of one or two subcomponents. Two of these components span $\sim 400 - 650$~MHz, while the other component is only visible over $\sim 600 - 650$~MHz in the baseband data. This narrowband component displays some faint emission in Stokes $Q$ (middle panel of Figure \ref{fig:possible_depol}) and in Stokes $U$ (rightmost panel of Figure \ref{fig:possible_depol}), while the other components appear unpolarized. Considering only the narrowband component, we find a polarized detection with $L/I = 0.50 \pm 0.01$ and $\mathrm{RM} = 62.9 \pm 0.9~\mathrm{rad}~\mathrm{m}^{-2}$. Over the two broadband, unpolarized components we place upper limits on their $L/I$ of $\leq 0.219$ and $\leq 0.190$, respectively. We note that in the first unpolarized component, there is a $\sim 2-3 \sigma$ peak in the FDF at the same RM as the polarized component. The lack of emission below $\sim$600\,MHz makes it difficult to constrain the scattering timescale of the polarized component but the burst widths of the first two components appear comparable at $\gtrsim 600$~MHz. Looking at the same FRB in the intensity data from the first CHIME/FRB catalog \citep{2021ApJS..257...59C} shows that the burst emission extends up to $\sim 800$~MHz, suggesting that the bursts may be part of a downward-drifting envelope that extends from higher frequencies. However, without high time resolution baseband data over the entire CHIME/FRB band, we cannot determine the precise morphology of the sub-bursts at $> 650$~MHz.

\subsubsection{Comparing with FRBs at different observing frequencies} \label{sec:results_depol_freq}
\begin{figure}[ht]
    \includegraphics[width=0.47\textwidth]{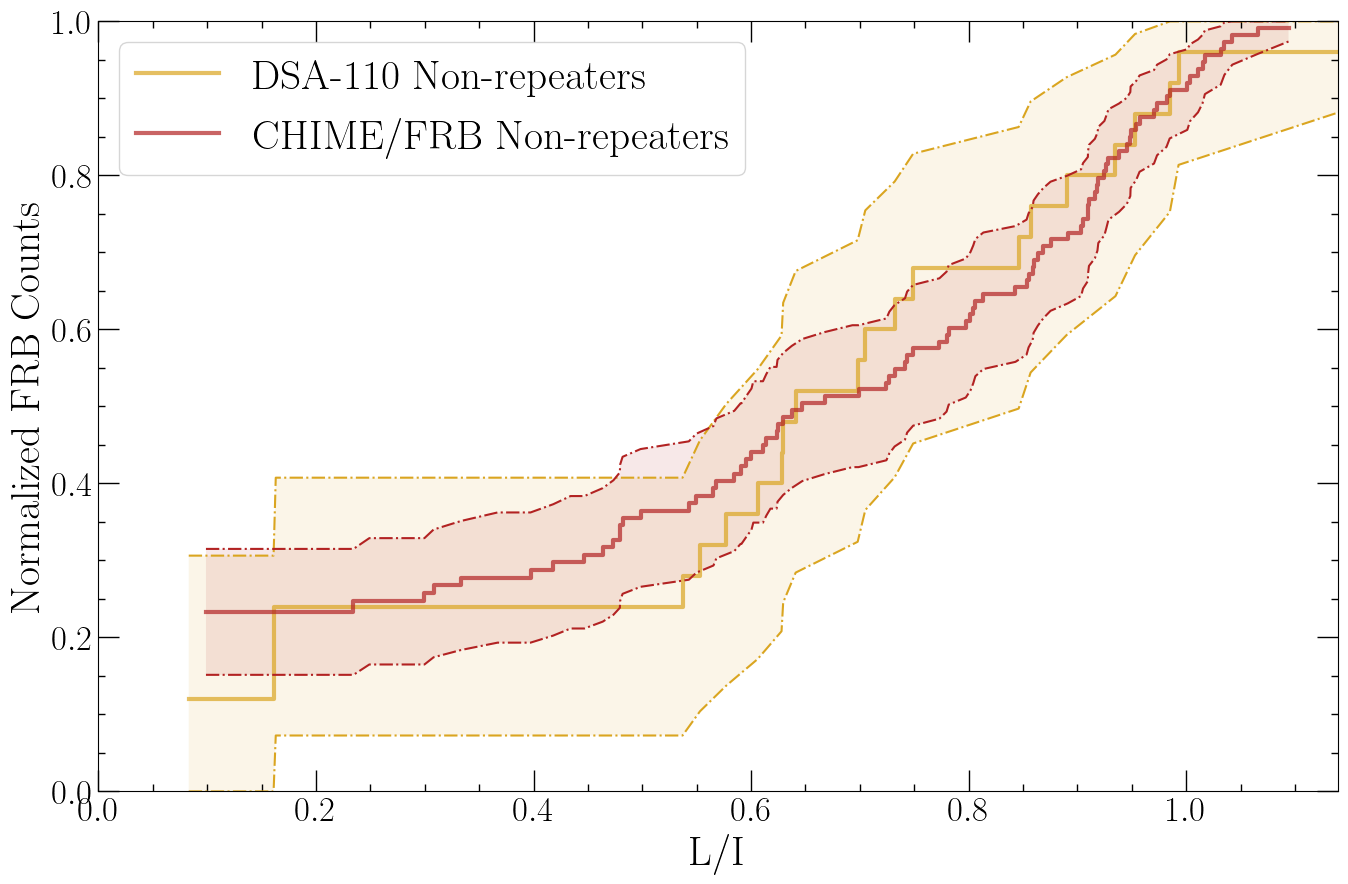}
    \caption{$L/I$ empirical CDFs, as computed using the Kaplan-Meier method, of the non-repeating FRB population with significant RM detections as observed at $400-800$~MHz using CHIME/FRB in this work (red line; 84 polarized FRBs and 29 unpolarized FRBs) and at $1.28-1.53$~GHz with the DSA-110 \citep[gold line; 20 polarized FRBs and 5 unpolarized FRBs;][]{2023arXiv230806813S}. The $95$\% CIs in each respective CDF is plotted as a shaded red and golden region.}
    \label{fig:chime_dsa_comp}
\end{figure}

Having not found any strong depolarization within the CHIME/FRB band, we endeavor to expand our search to a broader frequency range by comparing our results to published FRB populations observed with other instruments. Recently, \cite{2023arXiv230806813S} report polarization properties for $25$ non-repeating FRBs, $20$ of which have a significant RM detection. Their observations were conducted using the 110-antenna Deep Synoptic Array (DSA-110) between $1.28$ and $1.53$~GHz, with a center observing frequency of $\sim 1.4$~GHz. We note that their set of FRB sources is completely distinct from the sample presented in this work (i.e., there are no co-detected sources between the two sets of data). In Figure \ref{fig:chime_dsa_comp}, we compare the $L/I$ CDFs for our CHIME/FRB non-repeating sample with that of \cite{2023arXiv230806813S} (both computed using the Kaplan-Meier method) and find that the $L/I$ distribution observed with the DSA-110 is in close agreement with the one seen in our data. Applying AD, KS, Peto \& Peto and log-rank tests to the two distribution shows that they likely arise from the same underlying distribution ($p_\mathrm{AD} \geq 0.25$, $p_\mathrm{KS} = 0.83$, $p_\mathrm{PP} = 0.96$ and $p_\mathrm{LR} = 0.77$). The similarity between the two populations suggests that there is no clear systematic depolarization between $400$~MHz and $1.53$~GHz. These results are also consistent with five non-repeating FRBs detected by the Australian Square Kilometer Array Pathfinder, with a central observing frequency of $1.27$~GHz and a bandwidth of $336$~MHz, that show no evidence of spectral depolarization \citep{2024MNRAS.527.4285U}. The consistent distribution of $L/I$ across $400$~MHz and $1.53$~GHz also suggests that $L/I$ is not very sensitive to redshift effects. However, it is possible that with larger samples, a dichotomy might one day emerge.

\subsubsection{Beam Depolarization} \label{sec:results_beam_depol}
One method by which FRBs may incur frequency-dependent depolarization is via multi-path propagation through inhomogeneous magneto-ionic media. For most of our sample, we are limited to one band-averaged $L/I$ measurement across the emitting frequency range of each FRB. Therefore, the only way to uniformly compute $\sigma_\mathrm{RM}$ for our sample would be to impose that all FRBs are emitted with $L/I = 1.0$ and their measured $L/I$ is solely the result of beam depolarization. This process would provide poorly constrained upper limits on $\sigma_\mathrm{RM}$ at best and is much better suited for FRBs that are observed over a large range of frequencies \citep[e.g.,][who used observations of repeating FRBs with observing frequencies ranging from $\sim 0.1 - 5.0$~GHz]{2022Sci...375.1266F}. If indeed FRBs are completely linearly polarized at the time they are emitted and the spread in the $L/I$ distribution is caused by beam depolarization, we should still observe a negative correlation between $L/I$ and $|\mathrm{RM}_\mathrm{EG}|$. That is, FRBs whose emission propagates through more dense and strongly magnetized environments should undergo a higher amount of frequency-dependent depolarization from multi-path propagation.

\begin{figure}[ht]
    \includegraphics[width=0.47\textwidth]{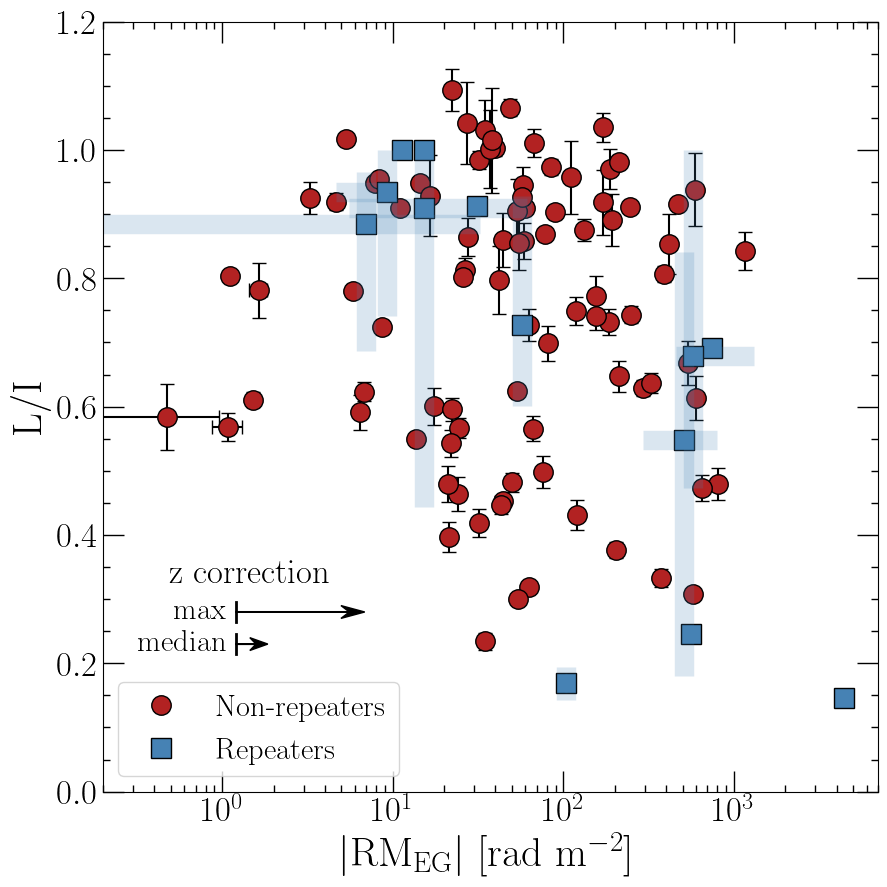}
    \caption{Linear polarization fraction $L/I$ for both the repeating (blue) and non-repeating (red) sample of FRBs plotted as a function of their $|\mathrm{RM}_\mathrm{EG}|$. Following the same setup as Figure \ref{fig:dm_rm}, the measurement uncertainties on the non-repeating sample are denoted by black lines while the intrinsic variation among repeater bursts, which are much larger than the respective measurement uncertainties, are presented as blue shaded regions. Spearman's rank correlation coefficient suggests a marginal monotonic correlation for only the repeating FRBs but no significant monotonic correlation for the non-repeating FRBs or in the combined data set. Based on Figures \ref{fig:depol} and \ref{fig:chime_dsa_comp} we do not expect $L/I$ to be very sensitive to redshift effects. The median and maximum expected shift between $\left|\mathrm{RM}_\mathrm{EG}\right|$ and $\left|\mathrm{RM}_\mathrm{host}\right|$ due to redshift corrections (based on our distribution of $z$ from Section \ref{sec:results_z_corr}) are presented in the lower left corner. Overall, a $z$ correction would not significantly change the $L/I$--$|\mathrm{RM}_\mathrm{EG}|$ correlation we observe.}
    \label{fig:rm_li}
\end{figure}

To test this scenario, we plot $L/I$ as a function of $|\mathrm{RM}_\mathrm{EG}|$ in Figure \ref{fig:rm_li}. Again, we apply Spearman's rank correlation coefficient on the $L/I - |\mathrm{RM}_\mathrm{EG}|$ relation for only the non-repeaters, only the repeaters, and the combined sample, respectively. We find no evidence for a monotonic relationship between $L/I$ and $|\mathrm{RM}_\mathrm{EG}|$ in the non-repeater only and combined data sets. We find a marginally significant negative monotonic correlation between $L/I$ and $|\mathrm{RM}_\mathrm{EG}|$ in the repeater only data set ($p_\mathrm{SR} = 0.002$). Note, however, that the repeater data set is only comprised of 13 sources and therefore we do not draw strong conclusions from this correlation. The corresponding test statistic and $p$-values are reported in Table \ref{tb:spearmanr_tests}. For the non-repeating FRBs, this suggests that there is not a strong relationship between the density and/or magnetic field strength of the FRB local environment and the level of observed linear polarization. In Figure \ref{fig:rm_li}, we are not able to account for the redshift effects on each FRB, which could smear out a potential correlation between $L/I$ and $|\mathrm{RM}_\mathrm{EG}|$. However, based on our previous results that show a consistent distribution of $L/I$ between $400$~MHz and $1.53$~GHz (e.g., see Figures \ref{fig:depol} and \ref{fig:chime_dsa_comp}), we do not expect $L/I$ to be very sensitive to redshift effects. Further, we plot the median and maximum shift in $|\mathrm{RM}_\mathrm{EG}|$ due to redshift corrections (based on our distribution of $z$ from Section \ref{sec:results_z_corr}) in the lower left part of the plot. As a result, the location of most FRBs in this plot would not be drastically affected by a $z$ correction and, thus, we argue that the $L/I$--$|\mathrm{RM}_\mathrm{EG}|$ correlation would not significantly change even if we were able to accurately derive $\left|\mathrm{RM}_\mathrm{host}\right|$ for each FRB.

\subsection{Comparing with FRB burst rates} \label{sec:results_rr}
In this Section, we test whether the observed $L/I$, $|\mathrm{RM}_\mathrm{EG}|$, or $\left|\beta\right|$ values are correlated with the observed FRB burst rates \citep[the bursts rates are obtained from][however, note that not all FRBs in our sample have burst rate estimates available]{2023ApJ...947...83C}. We apply Spearman's rank correlation coefficient to each of the three pairs of correlations (burst rate--$L/I$, burst rate--$|\mathrm{RM}_\mathrm{EG}|$, and burst rate--$\left|\beta\right|$), first for only the non-repeaters, then for only the repeaters, and finally for the combined repeater plus non-repeater data set. The Spearman's rank correlation coefficients, $S_\mathrm{SR}$, and associated $p$-values, $p_\mathrm{SR}$, are reported in Table \ref{tb:spearmanr_tests}. For the combined data set we furthermore calculate the Kendall $\tau$ correlation coefficient in the three tests where left-censored data are present. We find no evidence for a monotonic correlation between the observed burst rate and any of the three polarization properties (observed $L/I$, $|\mathrm{RM}_\mathrm{EG}|$, and $\left|\beta\right|$) in the non-repeater only data set, in the repeater only data set, or in the combined data set. Again, as we are not able to correct for redshift effects, the expected $(1+z)$ burst rate scaling could smear out possible correlations. However, the median shift in burst rate is only a factor of $\sim 1.24$. Following the same argument as in Section \ref{sec:results_beam_depol}, we do not believe a redshift correction would significantly change the results of our correlation tests.

\begin{table}[ht]
\setlength{\tabcolsep}{7pt}
\centering
\caption{Summary of Spearman rank and Kendall $\tau$ correlation coefficients and $p$-values testing whether a monotonic relationship exists between: $\mathrm{log}_{10}\left(|\mathrm{RM}_\mathrm{EG}|\right)$ and $\mathrm{log}_{10}\left(\mathrm{DM}_\mathrm{EG}\right)$, $|\mathrm{RM}_\mathrm{EG}|$ and $L/I$, burst rate and $L/I$, burst rate and $|\mathrm{RM}_\mathrm{EG}|$, and burst rate and $\left|\beta\right|$ for only the non-repeaters, only the repeaters, and the combined repeater plus non-repeater data set. The Kendall $\tau$ method is only used when data sets are partially left-censored, which is the case only for the last three correlation tests. The $p$-values that are $< 0.05$ are presented in bold face.} \label{tb:spearmanr_tests}
\begin{tabular}{ccc} 
\hline
\hline
Parameters & $S_\mathrm{SR}$ & $p_\mathrm{SR}$\\
\hline
\multicolumn{3}{c}{{\textbf{Non-repeaters only}}} \\
\hline
$\mathrm{log}_{10}\left(|\mathrm{RM}_\mathrm{EG}|\right)$ -- $\mathrm{log}_{10}\left(\mathrm{DM}_\mathrm{EG}\right)$ & $0.287$ & \textbf{0.006}\\
$|\mathrm{RM}_\mathrm{EG}|$ -- $L/I$ & $-0.052$ & $0.626$\\
Burst rate -- $L/I$ & $-0.086$ & $0.492$\\
Burst rate -- $|\mathrm{RM}_\mathrm{EG}|$ & $-0.214$ & $0.085$\\
Burst rate -- $\left|\beta\right|$ & $-0.095$ & $0.447$\\
\hline
\multicolumn{3}{c}{{\textbf{Repeaters only}}} \\
\hline
$\mathrm{log}_{10}\left(|\mathrm{RM}_\mathrm{EG}|\right)$ -- $\mathrm{log}_{10}\left(\mathrm{DM}_\mathrm{EG}\right)$ & $0.154$ & $0.616$\\
$|\mathrm{RM}_\mathrm{EG}|$ -- $L/I$ & $-0.770$ & \textbf{0.002}\\
Burst rate -- $L/I$ & $0.127$ & $0.709$\\
Burst rate -- $|\mathrm{RM}_\mathrm{EG}|$ & $0.218$ & $0.519$\\
Burst rate -- $\left|\beta\right|$ & $0.127$ & $0.709$\\
\hline
\multicolumn{3}{c}{{\textbf{Non-repeaters and repeaters}}} \\
\hline
$\mathrm{log}_{10}\left(|\mathrm{RM}_\mathrm{EG}|\right)$ -- $\mathrm{log}_{10}\left(\mathrm{DM}_\mathrm{EG}\right)$ & $0.252$ & \textbf{0.010}\\
$|\mathrm{RM}_\mathrm{EG}|$ -- $L/I$ & $-0.189$ & $0.057$\\
Burst rate -- $L/I$ & $-0.157$ & $0.173$\\
Burst rate -- $|\mathrm{RM}_\mathrm{EG}|$ & $-0.077$ & $0.506$\\
Burst rate -- $\left|\beta\right|$ & $-0.056$ & $0.630$\\
\hline
& $S_\mathrm{K\tau}$ & $p_\mathrm{K\tau}$ \\
\hline
Burst rate -- $L/I$ & $-0.002$ & $0.973$\\
Burst rate -- $|\mathrm{RM}_\mathrm{EG}|$ & $0.052$ & $0.499$\\
Burst rate -- $\left|\beta\right|$ & $0.080$ &$0.296$\\
\hline
\end{tabular}
\end{table}

\section{Discussion} \label{sec:discussion}
\subsection{Typical polarization properties of FRBs} \label{sec:discussion_typical_properties}
Our sample of $128$ non-repeating FRBs, $118$ of which have polarization information ($89$ polarized bursts and $29$ unpolarized bursts), increases the total number of FRB sources with polarization properties by a factor of $\sim 3$. In the following subsections, we discuss the typical polarization properties of our sample and compare them to our understanding of FRB polarimetry prior to this work (see Section \ref{sec:intro} for a summary).

\subsubsection{Linear polarization}
In the $400-800$~MHz band, we find that the mean and median observed levels of linear polarization are $63.3$\% and $64.7$\%, respectively, after accounting for unpolarized upper limits (see Figure \ref{fig:all_hist_cdf}). The distribution of $L/I$ is skewed towards $L/I = 1.0$, with 14\% of polarized FRBs being consistent with $100$\% linear polarization. There is, however, a significant spread in the $L/I$ distribution with 16\% of FRBs being less than $50$\% linearly polarized. Furthermore, 29 FRBs did not reach the $6 \sigma$ polarization detection threshold and for these bursts we place upper limits on $L/I$. Most of the unpolarized bursts have linear polarization upper limit constraints less than $30$\%, with the lowest constraints putting FRB 20190115B and FRB 20190619A at less than $10$\% linearly polarized. This suggests either that there is a range of possible linear polarization levels intrinsic to the FRB emission mechanism or that many FRBs in our sample have been depolarized. We discuss the latter possibility further in Section \ref{sec:discussion_depol}, where we consider the implications of the lack of evidence for depolarization in our data..

\subsubsection{Magneto-ionic environment} \label{sec:discussion_env}
We find that FRBs in our sample typically have a moderate $|\mathrm{RM}_\mathrm{EG}|$ with a mean of $135.4~\mathrm{rad}~\mathrm{m}^{-2}$ and a median of $53.8~\mathrm{rad}~\mathrm{m}^{-2}$. Further, in Section \ref{sec:results_z_corr}, we found that the distribution of $|\mathrm{RM}_\mathrm{host}|$ is likely shifted a few tens of $\mathrm{rad}~\mathrm{m}^{-2}$ higher based on expected redshift corrections. The majority of our sample have $|\mathrm{RM}_\mathrm{EG}|$ in agreement with the median expected $|\mathrm{RM}_\mathrm{EG}|$ contribution from an FRB that is embedded randomly within a MW like host galaxy \citep[on the order of $\sim 10^{0} - 10^{1}~\mathrm{rad}~\mathrm{m}^{-2}$][]{2019MNRAS.488.4220H}, or one originating near a star forming region in the host \citep[on the order of $\sim 10^{1} - 10^{2}~\mathrm{rad}~\mathrm{m}^{-2}$][]{2019MNRAS.488.4220H}, at $z \sim 0.24$ (the median expected redshift of our sample, as derived in Section \ref{sec:results_z_corr}). This result is in agreement with the preponderance of FRB host galaxies being star forming, with the majority of these hosts tracing the star forming main sequence for galaxies \citep{2023ApJ...954...80G} and with \citet{2023arXiv231010018B} who showed that all hosts of localized nearby FRBs are spirals. The peak of the Galactic pulsar $|\mathrm{RM}|$ distribution is approximately $\sim 70~\mathrm{rad}~\mathrm{m}^{-2}$, though the standard deviation is quite large as the distribution extends to $\gtrsim 10^3~\mathrm{rad}~\mathrm{m}^{-2}$ \citep[based on the $1494$ pulsars with RM measurements in the Australia Telescope National Facility pulsar catalog V1.7.1][]{2005AJ....129.1993M}\footnote{The Galactic pulsar catalog is hosted online at \href{https://www.atnf.csiro.au/research/pulsar/psrcat/}{https://www.atnf.csiro.au/research/pulsar/psrcat/}}. The peak of both our $|\mathrm{RM}_\mathrm{EG}|$ and $|\mathrm{RM}_\mathrm{host}|$ distributions are approximately consistent with that of the Galactic pulsar distribution. While Galactic pulsar LoS do not probe the entire extent of the Galaxy, agreement between the pulsar and FRB populations increases our confidence that FRB local environments may be similar to those found in external, MW-like galaxies.

The highest $|\mathrm{RM}_\mathrm{EG}|$ in our sample is $1161~\mathrm{rad}~\mathrm{m}^{-2}$ for FRB 20190214C. It is the lone source with $|\mathrm{RM}_\mathrm{EG}| \geq 10^3~\mathrm{rad}~\mathrm{m}^{-2}$, while only seven other sources exceed $|\mathrm{RM}_\mathrm{EG}| > 500~\mathrm{rad}~\mathrm{m}^{-2}$. We know that $|\mathrm{RM}_\mathrm{EG}| \leq |\mathrm{RM}_\mathrm{host}|$ and so, even without a redshift correction, this subset of FRBs is consistent with the expectations for a source embedded within a dense surrounding environment \citep[e.g., a supernova remnant, which is expected to have RM contributions up to $\sim 10^3~\mathrm{rad}~\mathrm{m}^{-2}$;][]{2018ApJ...861..150P}. It is possible that some other FRBs in our sample with lower $|\mathrm{RM}_\mathrm{EG}|$ also end up having $|\mathrm{RM}_\mathrm{host}|$ of this magnitude, but we cannot be certain without their redshifts. Within our Galaxy, we have seen some pulsars with similarly large $|\mathrm{RM}|$s up to $10^3 - 10^4~\mathrm{rad}~\mathrm{m}^{-2}$ originating near the dense Galactic center \citep[e.g., pulsars PSRs J1746$-$2849, J1746$-$2850, J1746$-$2856, and J1745$-$2912;][]{2023MNRAS.524.2966A}. Therefore, at least a few non-repeating FRBs originate in extremely dense and/or highly magnetized environments, though we cannot ascertain the specific properties of the environments from the $|\mathrm{RM}_\mathrm{EG}|$ measurements alone.

On the other hand, 14 of our 89 polarized FRBs have $|\mathrm{RM}_\mathrm{EG}| < 10~\mathrm{rad}~\mathrm{m}^{-2}$ and, hence, are candidates for FRBs originating in clean magneto-ionic environments. One of the reasons these are only candidates is that a redshift correction could place their $|\mathrm{RM}_\mathrm{host}|$ values substantially higher. There has been some evidence to suggest that the circumgalactic medium of galaxies could contribute only a few $\mathrm{rad}~\mathrm{m}^{-2}$ to the RM of background sources. In particular, \cite{2023A&A...670L..23H} find an excess RM of $3.7~\mathrm{rad}~\mathrm{m}^{-2}$ for background polarized sources that have an impact parameter of $< 100$~kpc with nearby inclined galaxies. It is possible that FRBs inhabiting environments far away from the Galactic disk (e.g., the circumgalactic medium and/or globular clusters) would therefore have very small $|\mathrm{RM}_\mathrm{EG}|$. However, even given a redshift for the FRB host galaxy, our ability to confirm the existence of an FRB in a clean magneto-ionic environment (i.e., low density and/or weakly magnetized with $|\mathrm{RM}_\mathrm{EG}| \sim 0~\mathrm{rad}~\mathrm{m}^{-2}$) is limited by the accuracy of the current foreground MW RM map \citep{2022A&A...657A..43H}. This map has a resolution of $46.9~\mathrm{arcmin}^{2}$ and relies on an RM catalog with a density of only $\sim 1~\mathrm{source}~\mathrm{deg}^{-2}$ \citep[primarily from the National Radio Astronomy Observatory Very Large Array Sky Survey][]{2009ApJ...702.1230T}. Even at high Galactic latitudes, the MW RM contribution to extragalactic RMs varies by up to $\sim 10~\mathrm{rad}~\mathrm{m}^{-2}$ over degree angular scales \citep{2015A&A...575A.118O, 2022A&A...657A..43H}. To accurately compute the MW RM contribution towards FRBs, we therefore require a grid of RMs from background radio galaxies with a much higher sky density. This will only become available for large portions of the sky with upcoming radio surveys such as the Polarisation Sky Survey of the Universe's Magnetism \citep[POSSUM;][]{2010AAS...21547013G}. Otherwise, dedicated follow-up observations around FRB sky positions are needed to construct sufficiently high density grids of RMs to determine whether these FRBs originate in low density and/or weakly magnetized environments.

We find only a marginally significant $p_\mathrm{SR}$ between $\mathrm{log}_{10}\left(|\mathrm{RM}_\mathrm{EG}|\right)$ and $\mathrm{log}_{10}\left(\mathrm{DM}_\mathrm{EG}\right)$, and our results are in agreement with the same correlation seen by \cite{2023ApJ...954..179M} and \cite{2023ApJ...957L...8S} using FRBs with identified host galaxies. The marginal correlation may suggest that while a fraction of $\mathrm{DM}_\mathrm{EG}$ originates from the same local environment as $\mathrm{RM}_\mathrm{EG}$, a significant amount of the $\mathrm{DM}_\mathrm{EG}$ is accumulated from other sources (e.g., the full extent of the host galaxy and the IGM). Interestingly, the correlation is stronger in non-repeaters and in the combined repeaters plus non-repeaters data sets than in only the repeaters. This may be due to the smaller sample size of the repeating population, namely that we do not have as many low $|\mathrm{RM}_\mathrm{EG}|$ or high $\mathrm{DM}_\mathrm{EG}$ repeaters. Physically, this may imply that repeating FRBs are embedded in environments that host stronger magnetic fields than non-repeaters and, therefore, impart RMs that are disproportionately larger than the DM contributed through that same medium. The marginal dichotomy seen in $\left| \beta \right|$ between repeaters and non-repeaters lends some credence to this idea and is discussed further in Section \ref{sec:discussion_rep_nonrep}.

We find typical $\left| \beta \right|$ values of order $\sim 0.1 - 1~\mu\mathrm{G}$ with values for some non-repeaters as low as $\sim 10^{-2} - 10^{-3}~\mu\mathrm{G}$. While it is possible for some FRBs to exist in clean environments \citep[e.g., FRB 20200120E which is located in a globular cluster;][]{2022Natur.602..585K} or for some FRBs to have magnetic fields that are oriented predominantly in the plane of the sky rather than along the LoS, we do not expect typical magnetic field strengths of $\lesssim 1~\mu\mathrm{G}$ for an FRB population that has, so far, been localized primarily to spiral galaxies \citep[e.g.,][]{2023arXiv231010018B}. For reference, \cite{2001SSRv...99..243B} finds that the average total magnetic field strength in the MW disk ranges from $\sim 4~\mu\mathrm{G}$ (at 16~kpc from the Galactic center) to $\sim 10~\mu\mathrm{G}$ (at 3~kpc from the Galactic center). Meanwhile, upper limits for the LoS magnetic field strength of IGM filaments have been placed at $\sim 21~\mathrm{nG}$ using a specific FRB LoS \citep{2016Sci...354.1249R} and at $\sim 30~\mathrm{nG}$ from a cross-correlation between the RM of radio galaxies and large scale structure \citep{2021MNRAS.503.2913A}. Therefore, many of the values of $\left| \beta \right|$ derived in this work fall below the average total magnetic field strength in MW-like galaxies. However, there are still a handful of repeating and non-repeating FRBs with $\left| \beta \right| \gtrsim 1~\mu\mathrm{G}$ that are roughly consistent with the average total magnetic field strength in the outskirts of MW-like galaxies (assuming they have, on average, comparable magnetic field strengths in the plane of the sky direction as well). 

Applying a statistical redshift correction to our polarized non-repeating FRBs, we found a mean $\left|\left<B_{\parallel,\mathrm{host}} \right>\right|$ of $4.62~\mu\mathrm{G}$. As this is only the LoS component of the host magnetic field strength, the total magnetic field strength is likely larger, which makes this broadly consistent with the range of total magnetic field strengths in the MW disk \citep{2001SSRv...99..243B}. There is still a small fraction of the $\left|\left<B_{\parallel,\mathrm{host}} \right>\right|$ distribution in Figure \ref{fig:rm_b_z_corr} with values $\lesssim 10^{-1}~\mu\mathrm{G}$. This low $\left|\left<B_{\parallel,\mathrm{host}} \right>\right|$ tail can be explained by FRB LoS in which the magnetic field is largely in the plane of the sky or a population of FRBs that originate in environments that are much less magnetized than those typically found in disks of MW-like galaxies.

\subsection{Comparing repeaters and non-repeaters} \label{sec:discussion_rep_nonrep}
There has been evidence showing a dichotomy in the observed properties of repeating and non-repeating FRBs, namely in regards to their burst duration and emitting bandwidth \citep{2021ApJ...923....1P, 2023ApJ...947...83C}, and to a less significant extent, in their $\mathrm{DM}_\mathrm{EG}$ distributions \citep{2023ApJ...947...83C}. This may be a selection effect, as it is easier to detect fainter repetitions from closer (i.e., on average lower DM) repeating FRB sources \citep[see also][]{2021A&A...647A..30G,2023PASA...40...57J}. On average, repeating FRBs have larger burst widths, narrower emitting bandwidths, and lower $\mathrm{DM}_\mathrm{EG}$ than non-repeaters. In this work, we compare 13 repeating and 89 non-repeating polarized FRBs observed with CHIME/FRB to ascertain whether this observed dichotomy between the two classes of FRBs extends to their polarization properties. 

Using the Peto \& Peto and Log-Rank tests, we find no difference in the $L/I$ CDFs (accounting for the 29 $L/I$ upper limits for non-repeaters) between non-repeating and repeating FRBs. However, note that we have not incorporated all repeating FRB bursts that are unpolarized into the $L/I$ CDF for repeaters. Though we do not suspect that this effect will dramatically change the outcome of our result as most repeaters in this sample consist of only $1-2$ bursts with available baseband data. So, we do not appear to be missing a large fraction of unpolarized bursts for most of these repeaters.

Notably, the non-repeater distribution has a stronger tail towards lower $|\mathrm{RM}_\mathrm{EG}|$, whereas only two repeating FRBs have $|\mathrm{RM}_\mathrm{EG}| < 10~\mathrm{rad}~\mathrm{m}^{-2}$. This may hint towards the possibility that non-repeating FRBs more frequently occupy clean environments than repeating FRBs, but proving that this is the case would require a larger sample of repeating and non-repeating FRBs or a more accurate foreground MW RM map than is currently available. We observe no non-repeating FRBs with extreme $|\mathrm{RM}|$s as have been seen in some prolific repeaters \citep[FRB 20121102 and FRB 20190520B;][]{2018Natur.553..182M, 2023Sci...380..599A}. In part, this may be due to intra-channel depolarization in CHIME/FRB (a significant change in the polarization angle within a single frequency channel), which causes a $\sim 50$\% depolarization at 600~MHz for $|\mathrm{RM}| \sim 5000~\mathrm{rad}~\mathrm{m}^{-2}$. However, the highest $|\mathrm{RM}|$ FRB detected by CHIME/FRB to date is FRB 20190417A with $|\mathrm{RM}| = 4429.8~\mathrm{rad}~\mathrm{m}^{-2}$ \citep{2023ApJ...951...82M}. The lack of very high $|\mathrm{RM}|$ detections for non-repeaters suggests that the population does not reside in the same extremely dense and highly magnetic environments as some repeaters or that they undergo severe beam depolarization and become unpolarized in the CHIME/FRB band. We also cannot rule out that the lack of high $|\mathrm{RM}|$ detections with CHIME/FRB is exacerbated by the instrumental bias from intra-channel depolarization. Overall, there is significant overlap between the $|\mathrm{RM}_\mathrm{EG}|$ distributions of repeating and non-repeating FRBs, and we see that both populations are broadly consistent with being drawn from the same underlying distribution. 

We do see marginal evidence for a dichotomy in the $\left| \beta \right|$ distributions of repeating and non-repeating FRBs at the $98.3$\% (with the AD test) and $95.8$\% (with the KS test) levels, with the non-repeaters having lower typical $\left| \beta \right|$ than their repeating counterparts. This may hint towards the idea that repeating FRBs, on average, originate in more magnetized environments, which would be consistent with the lack of very high $|\mathrm{RM}_\mathrm{EG}|$ non-repeaters in our sample. However, we caution against the over interpretation of this result for a few reasons. 

We emphasize that $\left| \beta \right|$ is only a lower limit on the $\left|\left<B_{\parallel,\mathrm{host}}\right>\right|$ and a significant difference in the distribution of $\left| \beta \right|$ does not guarantee that such a divergence exists in the true $\left|\left<B_{\parallel,\mathrm{host}}\right>\right|$. The implicit assumption behind Equation \ref{eq:B} is that the RM and DM are imparted from the same medium. Devoid of any redshift information of the FRB host galaxies, our values of $\mathrm{DM}_\mathrm{EG}$ still bear a substantial contribution from the IGM while $\mathrm{RM}_\mathrm{EG}$, on average, traces the FRB host galaxy and local environment contributions. This means that $\left| \beta \right|$ is sensitive to the distance to the FRB source, with closer sources having a higher $\left| \beta \right|$. To better conceptualize this, consider two FRBs at redshifts $z_a=0.3$ and $z_b=0.5$, respectively, that are otherwise identical (assume for both FRBs that $\mathrm{DM}_\mathrm{host} = 100~\mathrm{pc}~\mathrm{cm}^{-3}$ and $\mathrm{RM}_\mathrm{host} = 100~\mathrm{rad}~\mathrm{m}^{-2}$; implying that both have $\left|\left<B_{\parallel,\mathrm{host}}\right>\right| \sim 1.232~\mu\mathrm{G}$). Approximating the IGM contribution to the DM as $\mathrm{DM}_\mathrm{IGM} \sim 1000z~\mathrm{pc}~\mathrm{cm}^{-3}$ based on the observed DM-$z$ relation \citep{2020Natur.581..391M} gives $\mathrm{DM}_\mathrm{IGM,a} = 300~\mathrm{pc}~\mathrm{cm}^{-3}$ and $\mathrm{DM}_\mathrm{IGM,b} = 500~\mathrm{pc}~\mathrm{cm}^{-3}$. Scaling the observed $\mathrm{DM}_\mathrm{host}$ and $\mathrm{RM}_\mathrm{host}$ with redshift produces $\mathrm{DM}_\mathrm{host} (1+z_a)^{-1} = 77~\mathrm{pc}~\mathrm{cm}^{-3}$, $\mathrm{DM}_\mathrm{host} (1+z_b)^{-1} = 67~\mathrm{pc}~\mathrm{cm}^{-3}$, $\mathrm{RM}_\mathrm{host} (1+z_a)^{-2} = 59~\mathrm{rad}~\mathrm{m}^{-2}$, and $\mathrm{RM}_\mathrm{host} (1+z_b)^{-2} = 44~\mathrm{rad}~\mathrm{m}^{-2}$. Therefore, the $\left| \beta \right|$ we would derive for these two FRBs would be a factor of $\sim 2$ different, solely due to the difference in their redshifts.

With this in mind, the apparent dichotomy in repeating and non-repeating FRB $\left| \beta \right|$ should not be interpreted as definite evidence for two populations with distinct magneto-ionic environments. Our results, instead, point to this scenario as being a possibility and the precise local magneto-ionic environments of repeating and non-repeating FRBs warrant further exploration with a large sample of localized FRBs with associated host galaxies and redshifts, as will be available with the CHIME/FRB Outriggers in the near future \citep{2022AJ....163...48M}.

Another subtle point that arises from applying tests that compare the CDFs of repeaters and non-repeaters is that the results do not incorporate implicit variance in the underlying populations which could lead to an unknown uncertainty in the $p$-values. This is easy to imagine in the case of repeaters, where a single repeating source may have a spread in their measured properties (such as $L/I$, $\mathrm{DM}_\mathrm{EG}$, $|\mathrm{RM}_\mathrm{EG}|$, and $\left| \beta \right|$) that is not captured by the median CDF across all repeaters. If non-repeaters are (as of yet) single detections of repeating sources, then we are only drawing one sample from a wider distribution in these properties and that possible underlying variance is also unaccounted for in the $p$-values. However, with large sample sizes such as our non-repeating FRB sample, we average over the source-to-source variances which would reduce the uncertainty on the $p$-values. We are still limited by the number of CHIME/FRB repeaters, especially in the case of repeaters with only a few bursts, as we do not have a solid grasp on the variances in their polarization properties. In this regard, we will further explore any potential dichotomies in FRB polarimetry using a larger sample of CHIME/FRB repeaters in upcoming work by Ng et al. (in preparation). This work will more than double the sample size of CHIME/FRB repeaters with polarization properties.

\subsection{Depolarization} \label{sec:discussion_depol}
\subsubsection{\texorpdfstring{$L/I$}{L/I} variations across 400-800~MHz}
If the large spread in $L/I$ values for the FRB population is indicative of widespread depolarization due to multi-path propagation, we would expect to see a systematic difference in measured $L/I$ distribution at different observing frequencies. However, within the CHIME/FRB band, we see depolarization at the $\sim 20$\% level in only one source, FRB 20190217A. There also exists one outlier, FRB 20181214C, for which $(L/I)_{500}$ is $\sim 80$\% larger than $(L/I)_{700}$. This source, however, is very faint and may suffer more from instrumental polarization than many of the other FRBs. Similar analysis on a much larger sample of FRBs is required to determine whether FRB 20181214C is only an outlier or if there truly exists a population of FRBs in which there is an increase in $L/I$ with decreasing frequency. Aside from these two sources, all other FRBs from this subset have $0.8 < f_\mathrm{depol} < 1.2$. This indicates that most FRBs in the CHIME/FRB frequency band are not undergoing a precipitous decrease in $L/I$, as might be expected from Equation \ref{eq:depol} or a power-law dependence for $L/I$ with frequency. Further, as shown in Figure \ref{fig:rm_li}, we find no evidence for a negative monotonic correlation between $|\mathrm{RM}_\mathrm{EG}|$ and $L/I$ for the non-repeaters, as would be expected if the observed distribution of $L/I$ were due to depolarization from multi-path propagation.

\subsubsection{Comparing with an FRB population at 1.4~GHz}
Extending beyond the CHIME/FRB population, \cite{2023arXiv230806813S} recently presented a sample of 25 FRBs with polarization properties observed in the range $1.28-1.53$~GHz. Comparing our sample (which is at $400-800$~MHz) with that of \cite{2023arXiv230806813S}, we find that the $L/I$ distributions are not statistically different and the mean $L/I$ are in close agreement. Furthermore, the fraction of unpolarized events in both samples are approximately in agreement; $\sim 25$\% of our sample does not have a significant $L$ detection while this number is $\sim 20$\% for the DSA-110 sample (although they apply a higher significance threshold in their RM search, $9\sigma$, compared to the $6\sigma$ threshold used in this work). This comparison, together with the predominantly constant $L/I$ seen in the $23$ FRBs within the CHIME/FRB band, supports the conclusion that widespread depolarization is not present in the observed FRB population. It is possible that most FRBs do not undergo a rapid decline in their $L/I$ between $1.53$~GHz and $400$~MHz and instead this typically occurs at lower or higher frequencies. We should also remember that the population observed in this work and that by \cite{2023arXiv230806813S} are two distinct sets of FRB sources (i.e., there are no co-detections of the same sources). Therefore, while a consistent mean $L/I$ between the two populations points to a lack of depolarization on a population level, it is certainly possible that any individual source could undergo frequency-dependent depolarization between $400$~MHz and $1.53$~GHz. As the DSA-110 and CHIME/FRB sky coverage overlap significantly, a particularly interesting avenue for future work would be to search for frequency-dependent depolarization specifically in non-repeating FRBs co-detected by both instruments.

On the other hand, \cite{2022Sci...375.1266F} claim that a handful of repeating FRBs may be undergoing spectral depolarization over frequency ranges spanning a few hundred MHz to a few GHz (though more observations are needed across this wide frequency range to definitively conclude the existence of depolarization in these FRBs). As derived in Section \ref{sec:results_depol_frac}, CHIME/FRB is only sensitive to a small range of $\sigma_\mathrm{RM}$ values within its frequency band. Hence, we require large spectro-polarimetric FRB samples at higher observing frequencies than $1.53$~GHz and at lower observing frequencies than $400$~MHz to better discern the prevalence of frequency-dependent depolarization across the entire FRB population. Further, most of the repeating FRBs for which \cite{2022Sci...375.1266F} fit a $\sigma_\mathrm{RM}$ value have $|$RM$|$s $\gtrsim 500~\mathrm{rad}~\mathrm{m}^{-2}$, values which are significantly higher than the typical $|\mathrm{RM}_\mathrm{EG}|$ that we find for non-repeaters in our sample. It is possible that only FRBs embedded in very dense and/or strongly magnetized environments have a sufficiently large $\sigma_\mathrm{RM}$ to clearly show frequency-dependent depolarization at the observing frequencies of CHIME/FRB and other FRB surveys. 

\subsubsection{A case study of FRB 20190227A}
While inspecting the unpolarized/depolarized FRBs we came across FRB 20190227A, shown in Figure~\ref{fig:possible_depol}. This FRBs is comprised of two broadband ($400-650$\,MHz) linearly unpolarized  ($\lesssim 20$\%) components and one narrowband ($600-650$\,MHz; but note we have no baseband data $> 650$\,MHz) component that is linearly polarized at the $\sim 50$\% level with $\mathrm{RM} = 62.9 \pm 0.9~\mathrm{rad}~\mathrm{m}^{-2}$. These components are each separated by $\sim 2~\mathrm{ms}$ and, while the two broadband, unpolarized components show scattering tails, the narrowband, polarized component does not seem to extend to low enough frequencies to determine if it is scatter-broadened in the same way. 

There is a precedent for $L/I$ changes from subburst to subburst in the FRB population: for FRB 20181112A, which consists of a total of four components, there is a visible decrease in $L/I$ of $\sim 0.2$\ from the first component to the third component \citep{2020ApJ...891L..38C}. The large spread in the $L/I$ distribution of our non-repeating sample, in conjunction with the lack of population-wide depolarization, suggests that FRBs are emitted at a range of intrinsic $L/I$ values and that not all FRBs are $100$\% linearly polarized. It is possible that the $\gtrsim 30$\% change in linear polarization between components in FRB 20190227A is explained by a change in the angle between our LoS and the center of an emitting beam. For example, \cite{2019Sci...365.1013D} find that $L/I$ indeed decreases with increasing distance from the beam center for radio emission from pulsar PSR J1906+0746.

Alternatively, we could attempt to explain the change in $L/I$ between components by a change in the circumburst media through which the FRB propagates. If the decrease in $L/I$ of $\gtrsim 0.3$ between the unpolarized and polarized components is due to propagation effects, the circumburst medium must have changed significantly on a $\Delta t \sim 2~\mathrm{ms}$ timescale. The length scale over which such a change could occur is $c \Delta t \lesssim 600~\mathrm{km}$ (ignoring any relativistic effects). The relative velocity of the emitting source to the observer, $v_\mathrm{source}$, is likely $\ll c$ so, assuming $v_\mathrm{source} = 500~\mathrm{km}~\mathrm{s}^{-1}$ \citep[which is relatively high for typical pulsar velocities;][]{2017A&A...608A..57V}, the length scales for the evolving circumburst medium would only be $\sim 1~\mathrm{km}$. While variations in scattering timescales have been observed for other sources such as FRB 20190520B \citep[over $2.9$~minutes;][]{2023MNRAS.519..821O} and in the Crab pulsar \citep[over $15$~days;][]{2018MNRAS.479.4216M}, these are both over timescales that are many orders of magnitude longer than what would be required in the case of FRB 20190227A. Predicated upon this and the extremely small length scales required in this scenario, we believe it is unlikely that the decrease in $L/I$ seen between components of FRB 20190227A is caused by propagation effects in the circumburst medium. Instead, it is more likely that the subburst to subburst variation in $L/I$ originates from the FRB emission mechanism.

\subsection{Comparison to Galactic pulsars and magnetars} \label{sec:discussion_pulsars_magnetars}
Since at least some FRBs are likely produced by neutron stars \citep[for arguments pointing towards magnetar origins of FRBs, see][]{2020Natur.587...59B,2020Natur.587...54C, 2022Natur.607..256C,2024arXiv240209304M}, it is interesting to compare the polarized properties of FRBs with those of Galactic neutron stars, even though the latter are less luminous in the radio band by many orders of magnitude. Whereas all FRBs are detected directly, pulsars are often studied by means of their average or stacked pulse profiles over many rotational periods; the most direct comparison would be between FRBs and pulsar single pulses.

Radio emission from canonical pulsars is understood to be produced by charged particles being accelerated near the magnetic poles of the stars, typically leading to one observed pulse per rotational phase of the neutron star (or two, if the radio beams are broad and the beam angle with respect to the LoS is favorable or if the magnetic axis is offset towards $90~\mathrm{deg}$ with respect to the rotational axis). The radio pulses are typically moderately linearly polarized. About $60$\% of pulsar PA curves are reproducible by a ``rotating vector model,'' \citep{2023MNRAS.520.4801J} where, depending on impact parameter and magnetic inclination angle, the observed pulses exhibit a continuous swing in PA over the pulse but can also show a more flat curve \citep{1969ApL.....3..225R}.

In Section~\ref{sec:results_rvm}, we fit the RVM to a non-repeating FRB subsample and found fits that generally match the quality of that obtained from pulsars. Unlike pulsars, the non-repeating FRB sample studied here have unknown periods, thus making the RVM under-constrained with equally plausible fits for a wide range of assumed periods and geometries \citep{2024arXiv240209304M}. The overwhelming preponderance of flat PA curves of this sample is markedly different from that of the pulsar population. If FRBs do originate within the dipolar magnetospheres of rotating neutron stars, the flatness of the PA curves indicates a preference for RVM models for which the magnetic and rotation axes are preferentially aligned ($\alpha=0^{\circ}$) or anti-aligned ($\alpha=180^{\circ}$) to a greater degree than that seen in the Galactic pulsar sample. Counter-intuitively, this clustering is seemingly at odds with a young progenitor scenario in which the rotation and magnetic axes would be expected to be more randomly oriented \citep[e.g.,][]{1996ApJ...458..265G}, due to insufficient timescales for alignment ($\gtrsim 1$ Myr) via electromagnetic torques \citep[e.g.,][]{2010MNRAS.402.1317Y}, however, evidence for this alignment is lacking \citep{2006ApJ...643..332F}. Alternatively, the PA curves displayed by FRBs may not be suitably interpreted via the RVM, and alternative scenarios involving more complex magnetic field configurations \citep[e.g., multipole models][]{2023ApJ...958...78Q} and/or emission regions far from the neutron star surface cannot be ruled out. Indeed, the recent observation of high luminosity radio bursts occurring at random rotational phases in Galactic FRB source SGR 1935+2154 \citep{2023SciA....9F6198Z} is inconsistent with the simple pulsar-like RVM framework explored here.

One interesting avenue to pursue in future work would be to classify the PA curves observed in the repeating FRB population and determine whether they also exhibit mostly constant PA profiles or if they have more variable PA profiles.

\subsubsection{Galactic analogs for FRB polarization properties}
The Crab pulsar is one of the brightest and most well-studied pulsars, and demonstrates the variety of radio emission that \emph{one} neutron star can produce. Together with other pulsars that have $> 100$\,kG magnetic fields near the light cylinder, it does not adhere to the canonical pulsar model and shows at least seven distinct radio emission components over one rotation that are likely produced at different sites near the surface and in the outer magnetosphere \citep{2016JPlPh..82c6302E,2019MNRAS.483.1731L}. While the Crab's main pulse is only weakly polarized, individual nanoshots within the main pulse can be strongly polarized with drastic polarization changes from nanoshot to nanoshot \citep{2003Natur.422..141H}. The high-frequency interpulse and two high-frequency components, on the other hand, are 50\%--100\% linearly polarized \citep{1999ApJ...522.1046M, 2016ApJ...833...47H}. Components do not show PA swings within a single component, but the PA changes from component to component, similar to what is seen for FRBs with multiple components and variable PAs (bottom panel of Fig.~\ref{fig:pa_examples}). Other young pulsars show similar PA jumps from component to component \citep{2008MNRAS.391.1210W}.

Radio bursts from magnetically-powered magnetars are highly linearly polarized, with polarization fractions 60\%--100\%, and display both constant and variable PA profiles over the bursts \citep[e.g.,][]{2017ARA&A..55..261K, 2021MNRAS.502..127L}. These PAs are most similar to the single component constant and variable PA categories of FRBs in this work. Magnetar bursts arrive over a wide range of rotational phases, which is typically interpreted as the emission being produced higher in the neutron star's magnetosphere, instead of in a narrow beam near the magnetic poles.

Sensitive observations with the FAST telescope of the Galactic magnetar SGR 1935+2154 provide evidence for two emission modes existing at the same time in one object \citep{2023SciA....9F6198Z}: low-luminosity pulsar-like pulses were found to occur in a narrow ($\sim 5$\%) region of the rotational phase, whereas FRB-like bursts that are eight orders of magnitude brighter than the pulsar-like emission, occur at random phases. The folded pulse profiles are 12\%--65\% linearly polarized and show no signs of PA swing across the integrated pulse.

Qualitatively, the variety in PA profiles and $L/I$ values observed for FRBs seems to be produced by SGR 1935+2154 and the Crab pulsar, so it is not necessary to invoke novel populations of sources to explain different polarization properties seen in the FRB population, but it may be necessary to invoke different emission mechanisms. For a more quantitative result, \citet{2023arXiv230806813S} compared the polarised properties of various classes of pulsars to FRBs and found that $L/I$ values of FRBs seem to match best the polarization properties of young pulsars, with characteristic ages $< 10^5$ years, but are on average higher than that of the overall population of pulsars. So while FRBs are much more luminous than the radio emission from Galactic neutron stars, we find analogs to the polarization properties of the FRB population among the most energetic neutron stars known in the Milky Way.

\subsection{Using FRB DMs and RMs to estimate redshifts} \label{sec:discussion_z_lowlim}
\subsubsection{Method and assumptions}
FRBs with available polarimetry may provide a unique opportunity to constrain redshifts for sources without a clear host galaxy association. In this Section, we take 20 FRBs (15 non-repeaters and 5 repeaters) from the literature that have DMs, RMs, and confirmed host galaxies with a measured redshift, $z$, and attempt to independently derive upper and lower limits on $z$ using only their respective DMs and RMs. The TNS names for all 20 FRBs are listed in Figure \ref{fig:z_lower_est} and their DM, RM, and literature references are given in Table \ref{tb:lit_pol}.

We start by deriving an upper limit on $z$ using the $\mathrm{DM}_\mathrm{obs}$ of each FRB. In deriving the upper limit, we want to be conservative and attribute as much of the FRB DM budget as is reasonably possible to the $\mathrm{DM}_\mathrm{IGM}(z)$ component. So, from the total observed DM of each FRB, we subtract a conservative $0.5$ times the estimated MW DM contribution along their respective LoS, according to the YMW16 model. We assume that the remaining DM is all accumulated through the IGM (i.e., assuming $\mathrm{DM}_\mathrm{halo} = \mathrm{DM}_\mathrm{host} = 0~\mathrm{pc}~\mathrm{cm}^{-3}$). Using the Macquart relation, we derive an upper limit for the FRB redshift as:
\begin{equation}
z_\mathrm{upper} \sim \frac{\mathrm{DM}_\mathrm{obs} - 0.5(\mathrm{DM}_\mathrm{disk})}{1000}\label{eq:z_upper} \,.
\end{equation}
Variations of this technique have been broadly used to get $z$ upper limits for FRBs in the past. The novel aspect that we propose is to then incorporate the FRB RMs to obtain a $z$ lower limit.

To derive a lower limit, we need values for both $\mathrm{DM}_\mathrm{obs}$ and $\mathrm{RM}_\mathrm{obs}$ and impose some assumptions on the observed parallel magnetic field strength of the host galaxy. In this case, we want a lower limit and so we are more generous in attributing DM contributions to the MW. We subtract $1.5$ times the estimated YMW16 MW DM contribution along their respective LoS and also subtract $\mathrm{DM}_\mathrm{halo} = 30~\mathrm{pc}~\mathrm{cm}^{-3}$. For the $\mathrm{RM}_\mathrm{obs}$, we subtract the expected MW RM using \cite{2022A&A...657A..43H}, as was done in Section \ref{sec:rm_corr} to get $\left| \mathrm{RM}_\mathrm{EG} \right|$. We make an assumption on the minimum LoS averaged magnetic field strength in the FRB host galaxy such that:
\begin{equation}
\left|\left<B_{\parallel,\mathrm{host}}\right>\right| = 1.232\frac{\left|\mathrm{RM}_\mathrm{host}\right|}{\mathrm{DM}_\mathrm{host}} \geq 1~\mu\mathrm{G}\label{eq:b_req} \,.
\end{equation}
While we choose a minimum of $1~\mu\mathrm{G}$ in this work, this threshold can be adjusted. Using Equation \ref{eq:b_req}, we derive an upper limit on $\mathrm{DM}_\mathrm{host}$ as:
\begin{equation}
\mathrm{DM}_\mathrm{host} \leq 1.232 \frac{\left|\mathrm{RM}_\mathrm{EG}\right|}{1~\mu\mathrm{G}}~\mathrm{pc}~\mathrm{cm}^{-3}\label{eq:dm_host_req} \,.
\end{equation}
Subtracting the $\mathrm{DM}_\mathrm{host}$ upper limit from the remaining DM budget and assigning the remaining DM to the IGM component, we use the Macquart relation to derive a lower limit on $z$ as:
\begin{equation}
z_\mathrm{lower} \sim \frac{\mathrm{DM}_\mathrm{obs} - 1.5(\mathrm{DM}_\mathrm{disk}) - \mathrm{DM}_\mathrm{halo} - \left(1.232 \frac{\left|\mathrm{RM}_\mathrm{EG}\right|}{1~\mu\mathrm{G}}\right)}{1000}\label{eq:z_lower} \,.
\end{equation}

\subsubsection{Results and interpretation of failure modes}
\begin{figure*}[ht!]
\begin{center}
    \includegraphics[width=0.99\textwidth]{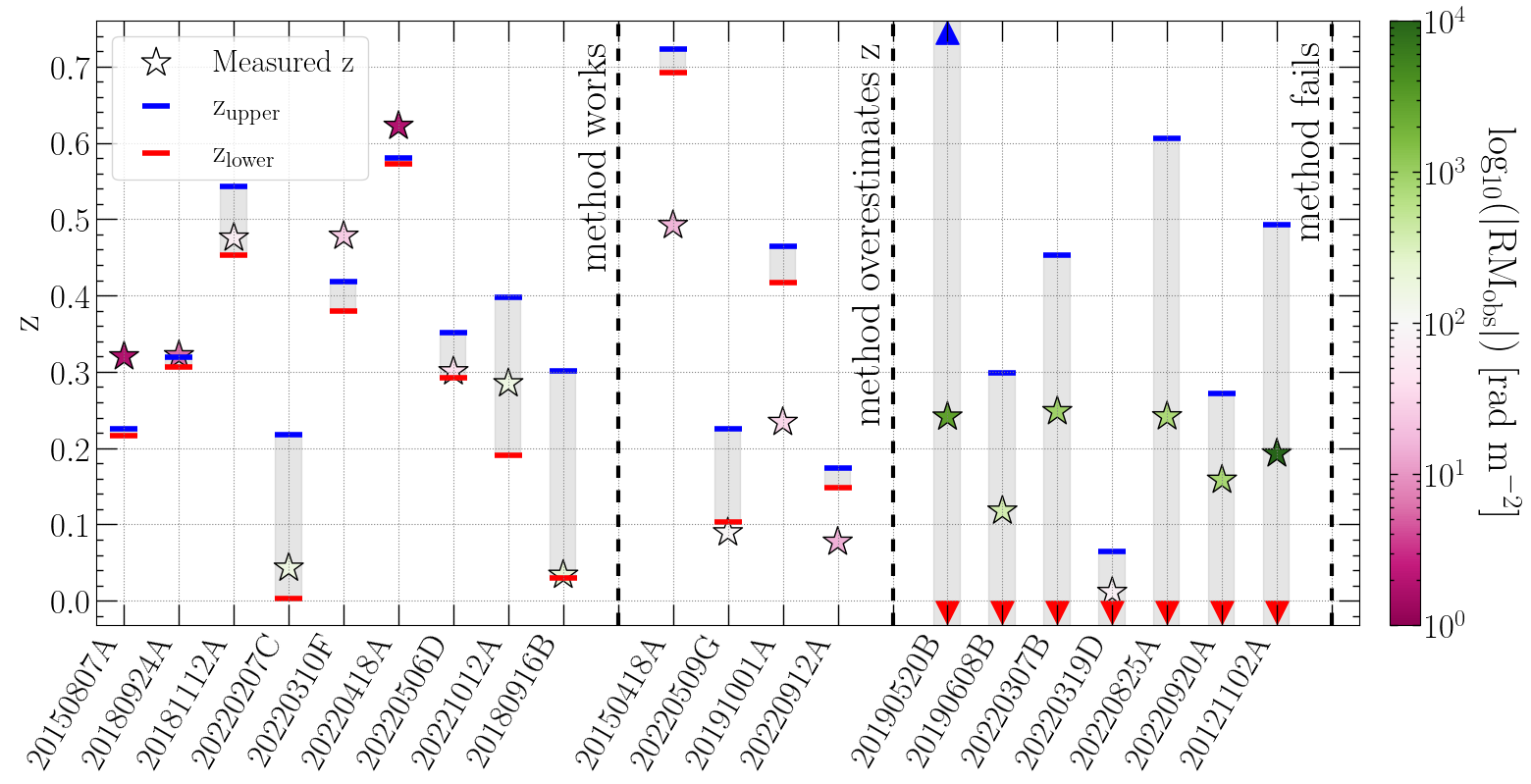}
    \caption{An experiment to test how well our method of deriving a $z$ lower limit works on a sample of twenty FRBs that have measured RM and are associated with host galaxies that have known $z$. The $z$ of the twenty FRBs are plotted as stars and are colored by their respective measured $\mathrm{log}_{10} \left( \left| \mathrm{RM}_\mathrm{obs} \right| \right)$ (see Table \ref{tb:lit_pol} for details and references). The x-axis lists the names of each FRB. Independently derived redshift upper (blue bars) and lower (red bars) limits are overplotted for each FRB with a shaded grey region encompassing the redshift range in between. In cases where the redshift limits are above or below the plotted range, triangular markers are used as indicators instead of bars. The plot is broken up into three sections by dashed black lines: (i) ``method works'', where the derived $z$ lower limits are reasonable compared to the measured $z$; (ii) `method overestimates'', where the $z$ lower limits higher than the independently measured $z$; and (iii) ``method fails'', where the $z$ lower limits are negative.}
    \label{fig:z_lower_est}
\end{center}
\end{figure*}

In Figure \ref{fig:z_lower_est}, we compare the constraints from this method for 20 FRBs that have been associated with host galaxies for which a redshift is known. The main purpose of this plot is to underscore the scenarios in which the method of deriving $z$ lower limits, $z_\mathrm{lower}$, does or does not work well. In almost all cases, the $z$ upper limit, $z_\mathrm{upper}$, falls above the measured $z$ of the FRBs. For the few cases where it does not, we suspect that the LoS may be through a less dense region of the IGM such that the true $\mathrm{DM}_\mathrm{IGM}-z$ relation is shallower than $\mathrm{DM}_\mathrm{IGM} \sim 1000 \times z$. 

The $z_\mathrm{lower}$ provides a reasonable estimate (i.e., between $z=0$ and the measured $z$ of the FRB host galaxy) for $9/20$ FRBs, as shown in the left part of the plot that is labelled ``method works''. Further, the $z_\mathrm{lower}$ method returns a limit that is closer to the independently measured $z$ than the $z_\mathrm{upper}$ estimated from the observed DM alone (i.e., $z - z_\mathrm{lower} < z_\mathrm{upper} - z$) for $5/9$ of these FRBs. For FRBs 20181112A, 20220207C, 20220506D, and 20180916B we find that $z - z_\mathrm{lower} \leq 0.04$, indicating that the $z_\mathrm{lower}$ method provides a fairly accurate estimate of the measured $z$.

For $4/20$ FRBs, we find that the $z_\mathrm{lower}$ values overestimate the measured $z$. All four of these FRB have $\left| \mathrm{RM} \right| < 10^2~\mathrm{rad}~\mathrm{m}^{-2}$ and, therefore, the $\mathrm{DM}_\mathrm{host}$ required to achieve a $1~\mu\mathrm{G}$ strength LoS magnetic field is small and we could be overestimating the $\mathrm{DM}_\mathrm{IGM}$ component in these cases. Physically, this could be due to the magnetic field being largely in the plane of the sky or encountering field reversals along the LoS that cause the $\left| \mathrm{RM}_\mathrm{host} \right|$ to be small compared to the $\mathrm{DM}_\mathrm{host}$ accumulated while propagating through the host galaxy. Alternatively, these FRBs may have accumulated DM in excess of the mean $\mathrm{DM}_\mathrm{IGM}-z$ relation by propagating through dense regions of the IGM or by intersecting structures, such as galaxy halos or clusters. This would lead to the $z_\mathrm{lower}$ values derived using Equation \ref{eq:z_lower} overestimating the true $z$ of the FRB host galaxy. These FRBs are plotted in the middle section of Figure \ref{fig:z_lower_est}, labelled ``method overestimates $z$''.

For the last $7/20$ FRBs, the $z_\mathrm{lower}$ values were found to be less than $0$. A negative $z$ is, of course, unphysical in this context but we discuss the cause of this failure mode as it informs the conditions under which we can apply the method outlined in the previous section. These FRBs fall on the high end of the $\left| \mathrm{RM} \right|$ distribution with six out of seven having $\left| \mathrm{RM} \right| > 10^2~\mathrm{rad}~\mathrm{m}^{-2}$. This group of FRBs also contains prolific repeaters that have very large RM amplitudes (e.g., FRB 20121102A and FRB 20190520B). The cause of this failure mode is that when the $\left| \mathrm{RM} \right|$ of a source is very high, we require a proportionally high $\mathrm{DM}_\mathrm{host}$ to satisfy the assumed $1~\mu\mathrm{G}$ LoS magnetic field strength. The required $\mathrm{DM}_\mathrm{host}$ can exceed the measured DM in some cases, leading to a negative $z_\mathrm{lower}$ estimate. Based on this, we caution that this technique of deriving a $z$ lower limit fails in cases where there is a large $\left| \mathrm{RM} \right|$ contribution from the immediate environment around the FRB (e.g., a supernova remnant or wind nebula) as the LoS magnetic field strength in these regions likely far exceeds the $1~\mu\mathrm{G}$ threshold. The FRBs in this category are on the right section of Figure \ref{fig:z_lower_est}, labelled ``method fails''. Adjusting the $\left|\left<B_{\parallel,\mathrm{host}}\right>\right|$ threshold in Equation \ref{eq:b_req} to $4~\mu\mathrm{G}$ resolves the unphysical $z_\mathrm{lower}$ (i.e., $z_\mathrm{lower} > 0$) for $6/7$ of these FRBs. The remaining outlier is FRB 20121102A, which has inferred magnetic field strengths on the order of $\sim 1~\mathrm{mG}$ \citep{2018Natur.553..182M}. 

Overall, we have shown that the combination of FRB DMs and RMs can be used to derive informative $z$ upper limits and lower limits, assuming a minimum LoS magnetic field strength in the host galaxy, for some FRBs where we have low to moderate $\left| \mathrm{RM} \right|$ (i.e., $\sim 10^1 - 10^3~\mathrm{rad}~\mathrm{m}^{-2}$). While there are clear failure modes, we see that the method works for $45$\% of FRBs in our tests and could prove to be useful, especially on a population level, when there is no clear host galaxy association or if there is no photometric or spectroscopic redshift available for a host galaxy candidate. In the future, other source or burst properties, such as burst morphology, could be investigated as an additional way of identifying FRBs for which this method is or is not well suited.

\section{Conclusions} \label{sec:conclusions}
We have conducted polarimetric analyses for the 128 non-repeating fast radio bursts (FRBs) from the first baseband catalog of the Canadian Hydrogen Intensity Mapping Experiment FRB collaboration, more than doubling the total number of FRB sources with measured polarization properties to date. Of the 128 FRBs, 89 have a significant ($> 6\sigma$) linearly polarized detections and their polarized spectra are well-fit by the CHIME/FRB polarization pipeline. For these FRBs, we analyze their observed time and frequency averaged linear polarization fractions $L/I$, rotation measures (RMs), dispersion measures (DMs), polarization position angle (PA) profiles, and frequency-dependent depolarization between $400-800$~MHz, and further derive lower limits on the magnetic field strength parallel to the line of sight in the FRB host environment, $\left| \beta \right|$. We apply a statistical redshift correction to derive a distribution of rest frame rotation measure amplitudes and average magnetic field strengths parallel to the line of sight arising from the FRB host galaxies. Another 29 FRBs do not have significant RM detections and for these events we place upper limits on their linear polarization fractions based on our $6\sigma$ detection threshold. Our polarimetric sample is compared to published CHIME/FRB repeaters and FRB populations measured at other frequencies. The key conclusions of this work are as follows:
\begin{enumerate}
    \item The median observed $L/I$ of our non-repeating sample is $0.647$ and the distribution has a large spread, suggesting that the common assertion that FRBs are typically $\sim 100$\% linearly polarized is not always accurate. The median extragalactic component of the observed RM, $|\mathrm{RM}_\mathrm{EG}|$, of our non-repeater sample is $53.8~\mathrm{rad}~\mathrm{m}^{-2}$. Applying a statistical redshift correction, we get a median rest frame RM from the host galaxy of $|\mathrm{RM}_\mathrm{host}| = 75.9~\mathrm{rad}~\mathrm{m}^{-2}$. The observed $|\mathrm{RM}_\mathrm{EG}|$ and rest frame $|\mathrm{RM}_\mathrm{host}|$ distributions are mostly consistent with RM expectations from FRBs embedded in Milky Way-like galaxies or near star forming regions in their host galaxy. There are 14 sources with $|\mathrm{RM}_\mathrm{EG}| < 10~\mathrm{rad}~\mathrm{m}^{-2}$, which are candidates for FRBs originating from low density and/or weakly magnetized environments (or could possibly have high redshift host galaxies) but their verification requires a more accurate accounting of the Milky Way RM contribution than is currently available.
    \item We find no evidence for a dichotomy in the observed $L/I$ (including the unpolarized bursts) or $|\mathrm{RM}_\mathrm{EG}|$ distributions between repeaters and non-repeaters. Note that, for the $L/I$ statistical tests, we are not able to fully account for individual repeater bursts that may have been unpolarized but do not believe this significantly effects our results. We do, however, see marginal evidence for non-repeaters and repeaters having different $\left| \beta \right|$ distributions, suggesting that repeaters may originate from more dense and/or more highly magnetized environments, but we also discuss biases that could make this conclusion false. 
    \item We define four archetypes for PA behavior based on a reduced $\chi^2_\nu$ fit to a constant PA profile and the existence of multiple components within the burst: (i) single component with constant PA (containing 57\% of the polarized non-repeating FRBs), (ii) single component with variable PA (10\%), (iii) multiple components with constant PA (22\%), and (iv) multiple components with variable PA (11\%). Radio bursts from magnetars, Crab pulses, and radio emission from some young pulsars provide useful Galactic analogs that have high typical $L/I$ and can produce constant PAs. Specifically, some Crab pulse components and young pulsars show similar component-to-component PA jumps as seen for FRBs that fall under the ``multiple components with variable PA'' archetype.
    \item In a subset of 23 bursts with $400$~MHz emitting bandwidths, 21 have a depolarization ratio between $700$~MHz and $500$~MHz of $0.8 < f_\mathrm{depol} < 1.2$ and do not show any clear signs of frequency-dependent changes in $L/I$. Only one FRB shows depolarization at the $20$\% level and one FRB shows a $80$\% increase in $L/I$ over the CHIME/FRB band. Across all $89$ polarized non-repeating FRBs in our sample, we do not find evidence for a negative monotonic correlation between $|\mathrm{RM}_\mathrm{EG}|$ and $L/I$, as might be expected from depolarization stemming from multi-path propagation. We also find similar $L/I$ distributions between non-repeating FRBs observed at $400-800$~MHz and those at $1.4$~GHz, further supporting the lack of frequency-dependent depolarization seen in the CHIME/FRB band. In all, our findings suggest that there is no observable frequency-dependent depolarization on the population level for non-repeating FRBs and the observed scatter in their $L/I$ distribution is likely intrinsic to the FRB emission mechanism.
    \item We derive a technique for obtaining lower limits on FRB redshifts by using their observed DMs and RMs and making an assumption about the lower limit of the host galaxy line of sight magnetic field strength. Testing on a sample of FRBs with known redshifts, the technique appears to work best in cases of moderate $\left| \mathrm{RM}_\mathrm{obs} \right|$ values but has difficulty with FRBs that have either extremely low or high $\left| \mathrm{RM}_\mathrm{obs} \right|$ values. In cases where independent redshift information is not available, this technique can provide some loose constraints on the FRB host redshift. However, it should be applied cautiously as peculiar lines of sight (e.g., extremely low or high $\left| \mathrm{RM}_\mathrm{obs} \right|$ values, particularly under or over-dense regions of the intergalactic medium, intervening galaxies and/or clusters) are not always well constrained. 
\end{enumerate}

While our work expands the known FRB sample with polarized properties by a factor of $\sim 3$, there are still many promising avenues for future work to better understand FRB polarization. First, we require a higher resolution Milky Way RM map to place tighter constraints on the FRB host RM contribution. Doing so would enable us to more carefully examine the subset of FRBs that appear to be stemming from clean magneto-ionic environments that may be similar to, for example, FRB 20200120E, which is located within a globular cluster \citep{2022Natur.602..585K}. It would also be interesting to see whether the preponderance of of constant PA profiles observed in non-repeating FRBs extends to the repeating FRB population or if repeaters exhibit more variable PA profiles. Another exciting prospect is studying the spectro-polarimetry of potential co-detections between the CHIME/FRB and the Deep Synoptic Array. Doing so would provide data spanning a range of frequencies between $400$~MHz and $1.53$~GHz for individual non-repeating FRBs and allow for much greater leverage in frequency when searching for spectral depolarization. Additionally, the CHIME/FRB Outriggers \citep[][]{2022AJ....163...48M, 2024arXiv240207898L} are expected to provide host galaxy associations and redshift information for many FRBs in the near future, enabling precise studies of magnetic fields in FRB host galaxies and the relation between RM and redshift.

\section*{Acknowledgements}
We acknowledge that CHIME is located on the traditional, ancestral, and unceded territory of the Syilx/Okanagan people. We are grateful to the staff of the Dominion Radio Astrophysical Observatory, which is operated by the National Research Council of Canada. CHIME is funded by a grant from the Canada Foundation for Innovation (CFI) 2012 Leading Edge Fund (Project 31170) and by contributions from the provinces of British Columbia, Québec and Ontario. The CHIME/FRB Project is funded by a grant from the CFI 2015 Innovation Fund (Project 33213) and by contributions from the provinces of British Columbia and Québec, and by the Dunlap Institute for Astronomy and Astrophysics at the University of Toronto. Additional support was provided by the Canadian Institute for Advanced Research (CIFAR), McGill University and the Trottier Space Institute at McGill thanks to the Trottier Family Foundation, and the University of British Columbia. 
The baseband recording system for CHIME/FRB is funded in part by a CFI John R. Evans Leaders Fund award to IHS.
The University of Toronto operates on the traditional land of the Huron-Wendat, the Seneca, and most recently, the Mississaugas of the Credit River; we are grateful to have the opportunity to work on this land. 
FRB research at UBC is supported by an NSERC Discovery Grant and by CIFAR. 
A.P. is funded by the NSERC Canada Graduate Scholarshops--Doctoral program.
Z.P. was a Dunlap Fellow and is supported by an NWO Veni fellowship (VI.Veni.222.295).
B.M.G. acknowledges the support of the Natural Sciences and Engineering Research Council of Canada (NSERC) through grant RGPIN-2022-03163, and of the Canada Research Chairs program.
M.B. is a McWilliams fellow and an International Astronomical Union Gruber fellow. M.B. also receives support from the McWilliams seed grant.
A.M.C. is funded by an NSERC Doctoral Postgraduate Scholarship.
A.P.C. is a Vanier Canada Graduate Scholar.
V.M.K. holds the Lorne Trottier Chair in Astrophysics \& Cosmology, a Distinguished James McGill Professorship, and receives support from an NSERC Discovery grant (RGPIN 228738-13), from an R. Howard Webster Foundation Fellowship from CIFAR, and from the FRQNT CRAQ.
C.L. is supported by NASA through the NASA Hubble Fellowship grant HST-HF2-51536.001-A awarded by the Space Telescope Science Institute, which is operated by the Association of Universities for Research in Astronomy, Inc., under NASA contract NAS5-26555.
K.W.M. holds the Adam J. Burgasser Chair in Astrophysics and is supported by NSF grants (2008031, 2018490).
K.N. is an MIT Kavli Fellow. 
A.B.P. is a Banting Fellow, a McGill Space Institute~(MSI) Fellow, and a Fonds de Recherche du Quebec -- Nature et Technologies (FRQNT) postdoctoral fellow.
K.S. is supported by the NSF Graduate Research Fellowship Program.

\facilities{CHIME}

\software{Astropy \citep{2013A&A...558A..33A, 2018AJ....156..123A, 2022ApJ...935..167A},
Matplotlib \citep{Hunter:2007}, 
NADA \citep{NADApackage},
Nested Sampling \citep{2004AIPC..735..395S},
NumPy \citep{harris2020array},
PyGEDM \citep{2021PASA...38...38P},
RM-CLEAN \citep{2009A&A...503..409H},
RM-synthesis \citep{2005A&A...441.1217B}, 
RM-tools \citep{2020ascl.soft05003P},
SciPy \citep{2020SciPy-NMeth},
Seaborn \citep{Waskom2021}.}

\appendix

\section{Polarization properties of FRBs from literature} \label{app:A}
\begin{figure*}[ht!]
\begin{center}
    \includegraphics[width=0.99\textwidth]{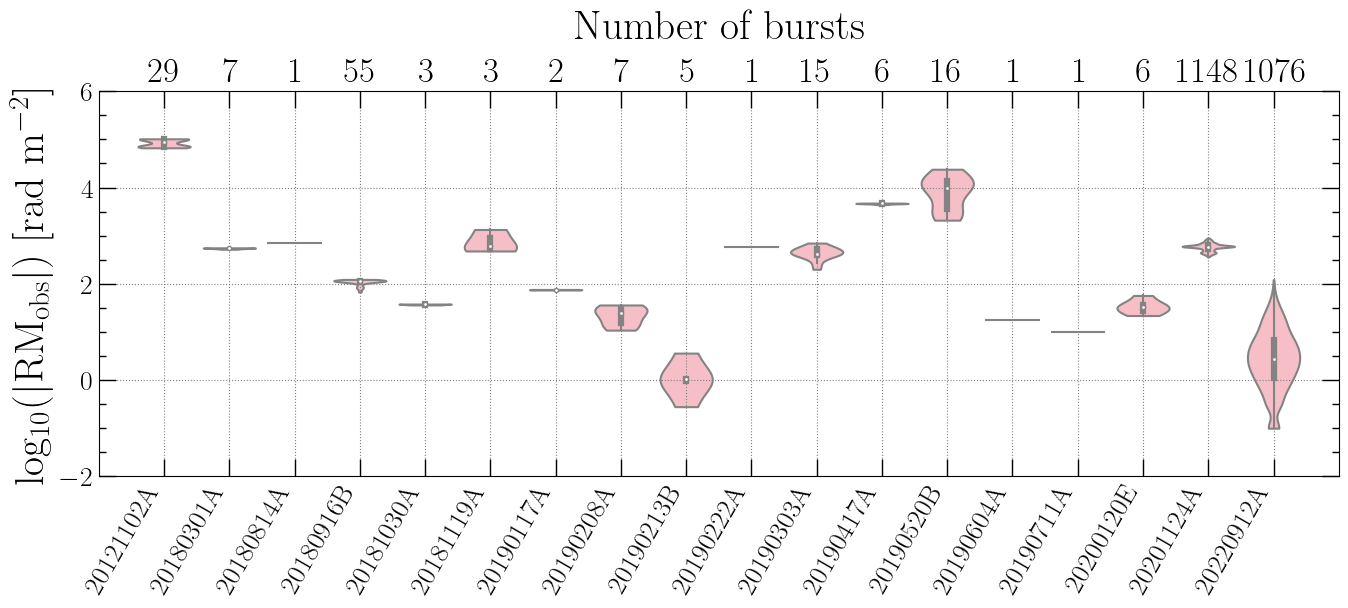}
    \caption{Violin plot of the observed rotation measure magnitude for the 18 repeating FRBs currently published in the literature (see Table \ref{tb:lit_pol}). KDEs of the $\mathrm{log}_{10}\left(|\mathrm{RM}_\mathrm{obs}|\right)$ distribution for each repeater are shown in pink and truncated at the minimum and maximum observed values. All of the KDEs have been normalized such that the width of each violin plot marker is equal. A white dot highlights the median $\mathrm{log}_{10}\left(|\mathrm{RM}_\mathrm{obs}|\right)$ of each repeater with a grey box representing the interquartile range and vertical grey lines showing the full observed range of $\mathrm{log}_{10}\left(|\mathrm{RM}_\mathrm{obs}|\right)$. The number of polarized bursts for each repeater is given at the top of the plot.}
    \label{fig:lit_rm_violin}
\end{center}
\end{figure*}

\begin{figure*}[ht!]
\begin{center}
    \includegraphics[width=0.99\textwidth]{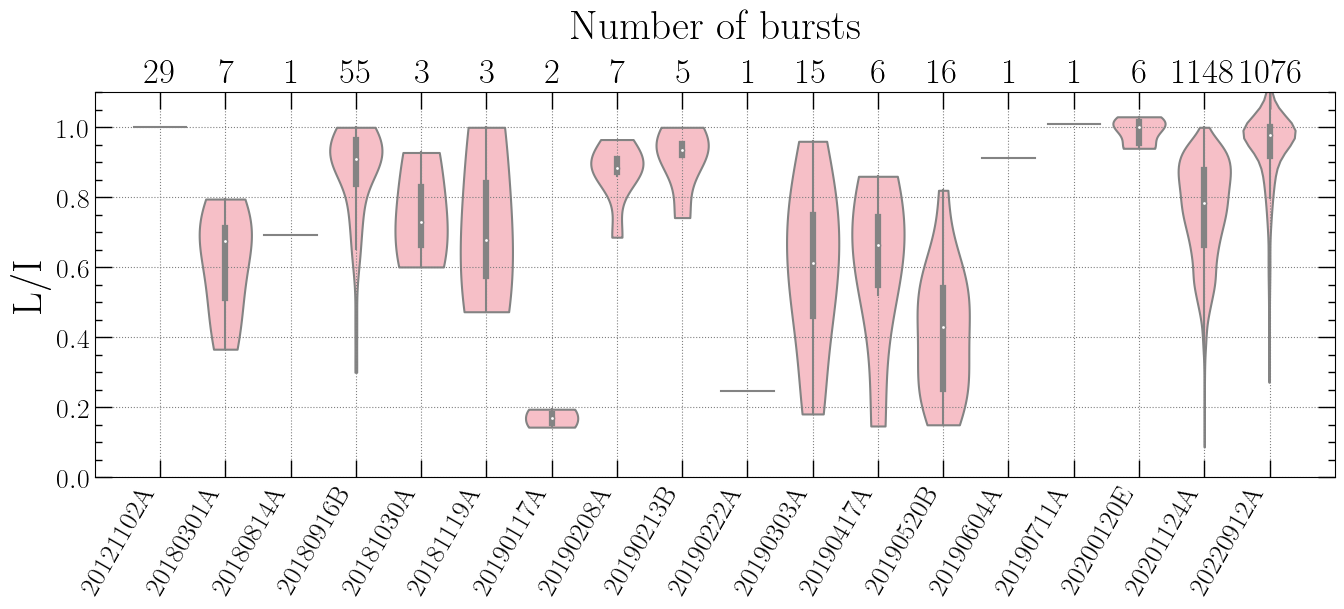}
    \caption{Analogous to Figure \ref{fig:lit_rm_violin} for the observed $L/I$ distribution for the 18 repeating FRBs from Table \ref{tb:lit_pol}.}
    \label{fig:lit_li_violin}
\end{center}
\end{figure*}

In Table \ref{tb:lit_pol}, we summarize the properties of the published polarized FRB sample in literature prior to this work. In Figure \ref{fig:lit_rm_violin} and \ref{fig:lit_li_violin}, we plot the $\mathrm{log}_{10}\left(|\mathrm{RM}|\right)$ and $L/I$, respectively, of each repeater from Table \ref{tb:lit_pol} as a violin plot to better visualize the spread in observed polarimetry of repeaters. The number of bursts included for each repeater is provided as a second x-axis at the top of each plot. 

\newpage
\setlength{\tabcolsep}{5pt}
\setlength{\LTcapwidth}{1.0\textwidth}
\begin{center}
\begin{longtable}{ccccccc}
\caption{Summary of published FRB polarization properties prior to this work. The FRB TNS name is presented in column 1. Column 2 provides the observing frequency $\nu_\mathrm{obs}$ in GHz and column 3 lists the total number of bursts with polarization properties for that source. Column 4 provides the measured DM range and column 5 provides the RM range for each FRB source. Columns 6 and 7 show the time and frequency averaged linear polarization fraction, $L/I$, and circular polarization fraction, $|V|/I$, respectively. Note that in the case of repeaters, a range encompassing the 95\% confidence interval for the respective parameter is provided and these FRBs are highlighted in boldface. In some cases, a repeater may only have a few bursts and the 95\% confidence interval (computed assuming a Gaussian distribution) falls outside the range of the data; in these cases we instead report the minimum and maximum observed values for the given parameter and these are italicized. Recall that $L/I$ may be affected by the observing frequency, $\nu_\mathrm{obs}$, so, for some repeaters, a large range of $L/I$ could be due to a wide span in $\nu_\mathrm{obs}$. In the online version of this manuscript, we provide a machine readable version of this table.} \label{tb:lit_pol}\\
\hline
\hline
FRB Name & $\nu_\mathrm{obs}$ & Number of & DM & RM & $L/I$ & $|V|/I$\\
 & (GHz) & bursts & (pc cm$^{-3}$) & (rad m$^{-2}$) &  & \\
\hline
\endfirsthead

\multicolumn{7}{c}
{{\bfseries \tablename\ \thetable{} -- continued from previous page}} \\
\hline
FRB Name & $\nu_\mathrm{obs}$ & Bursts & DM & RM & $L/I$ & $|V|/I$\\
 & (GHz) &  & (pc cm$^{-3}$) & (rad m$^{-2}$) &  & \\
\hline
\endhead

\hline \multicolumn{2}{c}{{Continued on next page}} \\ \hline
\endfoot
\hline
\endlastfoot

\textbf{FRB 20121102A} & $4.5^{1,2}$ & $29$ & $[560.5, 561.4]$ & $[74037, 90998]$ & $1.0$ & $\sim 0.0$\\
\textbf{FRB 20180301A} & $1.3^{3}$ & $7$ & $[516.3, 516.8]$ & $[536.4, 555.9]$ & $[0.50, 0.73]$ & $<0.10$\\
\textbf{FRB 20180814A} & $0.6^{4}$ & $1$ & $189.0$ & $699.8$ & $0.69$ & $<0.20$\\
\textbf{FRB 20180916B} & $0.4^{5}$, $0.1^{6}$, $0.8^{7}$, $0.6^{8}$ & $55$ & $[349.0, 349.2]$ & $[-112.9, -105.6]$ & $[0.83, 0.91]$ & $\sim 0.0$\\
\textbf{FRB 20181030A} & $0.6^{4}$ & $3$ & $103.5$ & \textit{[36.4, 38.0]} & \textit{[0.60, 0.93]} & $<0.20$\\
\textbf{FRB 20181119A} & $0.6^{4}$ & $3$ & \textit{[363.6, 364.3]} & \textit{[481.0, 1344]} & \textit{[0.47, 1.0]} & $<0.20$\\
\textbf{FRB 20190117A} & $0.6^{4}$ & $2$ & \textit{[393.1, 395.8]} & \textit{[74.30, 76.31]} & \textit{[0.14, 0.20]} & $<0.20$\\
\textbf{FRB 20190208A} & $0.6^{4}$ & $7$ & $[579.6, 579.9]$ & $[-14.56, 24.66]$ & $[0.81, 0.94]$ & $<0.20$\\
\textbf{FRB 20190213B} & $0.6^{4}$ & $5$ & \textit{[301.4, 301.7]} & \textit{[-3.630, 1.060]} & \textit{[0.74, 1.0]} & $<0.20$\\
\textbf{FRB 20190222A} & $0.6^{4}$ & $1$ & $459.6$ & $571.2$ & $0.25$ & $<0.20$\\
\textbf{FRB 20190303A} & $1.3^{9}$, $0.6^{4}$ & $15$ & $[221.5, 221.7]$ & $[-516.4, -388.3]$ & $[0.48, 0.73]$ & $<0.20$\\
\textbf{FRB 20190417A} & $1.3^{9}$, $0.6^{4}$ & $6$ & $1378$ & $[4548, 4729]$ & $[0.40, 0.80]$ & $<0.20$\\
\textbf{FRB 20190520B} & $6.0^{9}$, $2.4^{10}$, $5.6^{10}$ & $16$ & $[1201, 1211]$ & $[-6685, 5830]$ & $[0.33, 0.51]$ & $[\sim 0.0, 0.17]$\\
\textbf{FRB 20190604A} & $0.6^{4}$ & $1$ & $552.5$ & $-17.80$ & $0.91$ & $<0.20$\\
\textbf{FRB 20190711A} & $1.3^{11}$ & $1$ & $587.87$ & $10.00$ & $1.0$ & $\sim 0.0$\\
\textbf{FRB 20200120E} & $0.6^{12}$, $1.4^{14}$ & $6$ & $[87.75, 87.79]$ & $[-44.76, -24.97]$ & $[0.96, 1.0]$ & $[\sim 0.0, 0.11]$\\
\textbf{FRB 20201124A} & $1.4^{14}$, $2.4^{15}$, $1.3^{9, {16}}$ & $1148$ & $[413.1, 413.2]$ & $[-594.5, -582.3]$ & $[0.75, 0.77]$ & $[0.08, 0.09]$\\
\textbf{FRB 20220912A} & $0.6^{17}$, $1.4^{18}$, $1.3^{19}$ & $1087$ & $[225.1, 235.1]$ & $[-2.37, 3.33]$ & $[0.94, 0.96]$ & $[0.10, 0.12]$\\
FRB 20110523A & $0.8^{20}$ & $1$ & $623.3$ & $-186.1$ & $0.44$ & $<0.30$\\
FRB 20150418A & $1.4^{21}$ & $1$ & $776.2$ & $36.00$ & $0.09$ & $\sim 0.0$\\
FRB 20150807A & $1.4^{22}$ & $1$ & $266.5$ & $12.00$ & $0.80$ & $\sim 0.0$\\
FRB 20150215A & $1.4^{23}$ & $1$ & $1106$ & $1.600$ & $0.43$ & $0.03$\\
FRB 20160102A & $1.4^{24}$ & $1$ & $2596$ & $-220.6$ & $0.35$ & $0.06$\\
FRB 20171209A & $1.4^{25}$ & $1$ & $1457.4$ & $121.6$ & $1.0$ & $0.0$\\
FRB 20180311A & $1.4^{25}$ & $1$ & $1570.9$ & $4.8$ & $0.75$ & $0.11$\\
FRB 20180714A & $1.4^{25}$ & $1$ & $1467.92$ & $-25.9$ & $0.91$ & $0.05$\\
FRB 20180924A & $1.3^{11}$ & $1$ & $362.2$ & $22.00$ & $0.90$ & $0.13$\\
FRB 20181112A & $1.3^{26}$ & $1$ & $589.3$ & $10.50$ & $0.92$ & $<0.34^\mathrm{a}$\\
FRB 20190102C & $1.3^{11}$ & $1$ & $364.5$ & $-105.0$ & $0.82$ & $0.05$\\
FRB 20190608B & $1.3^{11}$ & $1$ & $340.1$ & $353.0$ & $0.91$ & $0.09$\\
FRB 20190611B & $1.3^{11}$ & $1$ & $332.6$ & $19.00$ & $0.93$ & $0.15$\\
FRB 20191001A & $0.9^{27}$ & $1$ & $507.9$ & $55.5$ & $0.57$ & $0.05$\\
FRB 20191108A & $1.4^{28}$ & $1$ & $588.1$ & $474.0$ & $0.70$ & $<0.10$\\
FRB 20201020A & $1.4^{29}$ & $1$ & $398.59$ & $110$ & $-^\mathrm{b}$ & $-^\mathrm{b}$\\
FRB 20220121B & $1.4^{30}$ & $1$ & $313.421^\mathrm{c}$ & $-4.60$ & $0.71$ & $0.51$\\
FRB 20220204A & $1.4^{30}$ & $1$ & $612.584^\mathrm{c}$ & $-11.0$ & $0.63$ & $0.07$\\
FRB 20220207C & $1.4^{30,31}$ & $1$ & $262.3$ & $162.5$ & $0.75$ & $0.11$\\
FRB 20220208A & $1.4^{30}$ & $1$ & $440.73^\mathrm{c}$ & $-23.3$ & $1.0$ & $0.11$\\
FRB 20220307B & $1.4^{30,31}$ & $1$ & $499.2$ & $-947.2$ & $0.70$ & $0.10$\\
FRB 20220310F & $1.4^{30,31}$ & $1$ & $462.2$ & $11.4$ & $0.61$ & $0.07$\\
FRB 20220319D & $1.4^{30,31}$ & $1$ & $111.0$ & $59.9$ & $0.16$ & $0.03$\\
FRB 20220330D & $1.4^{30}$ & $1$ & $467.788^\mathrm{c}$ & $-122.2$ & $0.73$ & $0.03$\\
FRB 20220418A & $1.4^{30,31}$ & $1$ & $623.5$ & $6.1$ & $0.64$ & $0.03$\\
FRB 20220424E & $1.4^{30}$ & $1$ & $863.932^\mathrm{c}$ & $129.0$ & $0.58$ & $0.25$\\
FRB 20220506D & $1.4^{30,31}$ & $1$ & $396.9$ & $-32.4$ & $0.95$ & $0.31$\\
FRB 20220509G & $1.4^{30,31}$ & $1$ & $269.5$ & $-109.0$ & $0.99$ & $0.05$\\
FRB 20220726A & $1.4^{30}$ & $1$ & $686.232^\mathrm{c}$ & $499.8$ & $0.94$ & $0.04$\\
FRB 20220825A & $1.4^{30,31}$ & $1$ & $651.2$ & $750.2$ & $0.54$ & $0.07$\\
FRB 20220831A & $1.4^{30}$ & $1$ & $1146.14^\mathrm{c}$ & $772.1$ & $0.85$ & $0.02$\\
FRB 20220920A & $1.4^{30,31}$ & $1$ & $315.0$ & $-830.3$ & $0.89$ & $0.15$\\
FRB 20221012A & $1.4^{30,31}$ & $1$ & $442.2$ & $165.7$ & $0.63$ & $0.01$\\
FRB 20221029A & $1.4^{30}$ & $1$ & $1391.75^\mathrm{c}$ & $-155.8$ & $0.86$ & $0.07$\\
FRB 20221101A & $1.4^{30}$ & $1$ & $1475.53^\mathrm{c}$ & $4670$ & $0.55$ & $0.11$\\
FRB 20221101B & $1.4^{30}$ & $1$ & $491.554^\mathrm{c}$ & $-32.2$ & $0.99$ & $0.19$\\
\hline
\hline
\multicolumn{7}{l}{$^1$ \cite{2018Natur.553..182M}; $^2$ \cite{2021ApJ...908L..10H}; $^3$ \cite{2020Natur.586..693L}; $^4$ \cite{2023ApJ...951...82M}; $^5$ \cite{2020ApJ...896L..41C};} \\
\multicolumn{7}{l}{$^6$ \cite{2021ApJ...911L...3P}; $^7$ \cite{2022ApJ...932...98S}; $^8$ \cite{2023ApJ...950...12M}; $^9$ \cite{2022Sci...375.1266F}; $^{10}$ \cite{2023Sci...380..599A};} \\
\multicolumn{7}{l}{$^{11}$ \cite{2020MNRAS.497.3335D}; $^{12}$ \cite{2021ApJ...910L..18B}; $^{13}$ \cite{2022NatAs...6..393N}; $^{14}$ \cite{2021MNRAS.508.5354H}; $^{15}$ \cite{2022MNRAS.512.3400K};} \\
\multicolumn{7}{l}{$^{16}$ \cite{2022Natur.609..685X}; $^{17}$ \cite{2022ATel15679....1M}; $^{18}$ \cite{2023ApJ...949L...3R}; $^{19}$ \cite{2023ApJ...955..142Z};} \\
\multicolumn{7}{l}{$^{20}$ \cite{2015Natur.528..523M}; $^{21}$ \cite{2016Natur.530..453K}; $^{22}$ \cite{2016Sci...354.1249R}; $^{23}$ \cite{2017MNRAS.469.4465P}; $^{24}$ \cite{2018MNRAS.478.2046C};} \\
\multicolumn{7}{l}{$^{25}$ \cite{2019MNRAS.488..868O}; $^{26}$ \cite{2020ApJ...891L..38C}; $^{27}$ \cite{2020ApJ...901L..20B}; $^{28}$ \cite{2020MNRAS.499.4716C};}\\
\multicolumn{7}{l}{$^{29}$ \cite{PM2023}; $^{30}$ \cite{2023arXiv230806813S}; $^{31}$ \cite{2023ApJ...957L...8S}; $^\mathrm{a} |V|/I$ ranges from $\sim 0.0$ to $0.34$ across} \\
\multicolumn{7}{l}{the burst; $^\mathrm{b}$ Polarization calibration was corrupted by radio frequency interference, therefore no $L/I$ or $V/I$ are reported;} \\
\multicolumn{7}{l}{$^\mathrm{c}$ DM obtained from the TNS entry for the FRB. DM uncertainty is not reported.} \\
\end{longtable}
\end{center}

\section{Polarization position angle uncertainty estimates} \label{app:B}
Our estimate of the PA uncertainty follows \cite{2019ApJ...878...92V}:
\begin{equation}
\sigma_\psi = \frac{1}{2} \frac{Q_\mathrm{obs,derot} U_\mathrm{obs,derot}}{Q_\mathrm{obs,derot}^2 + U_\mathrm{obs,derot}^2} \left[ \left( \frac{\sigma_{Q_\mathrm{obs,derot}}}{Q_\mathrm{obs,derot}} \right)^2 \left( \frac{\sigma_{U_\mathrm{obs,derot}}}{U_\mathrm{obs,derot}} \right)^2 \right]^{1/2}\,, \label{eq:pa_err}
\end{equation}
where $\sigma_{Q_\mathrm{obs,derot}}$ and $\sigma_{U_\mathrm{obs,derot}}$ are the measured rms noise in $Q_\mathrm{obs,derot}$ and $U_\mathrm{obs,derot}$, respectively. We note that Equation \ref{eq:pa_err} is an approximation for the true PA measurement uncertainties and becomes increasingly inaccurate for lower S/N in $L$. Setting a minimum threshold on the S/N of $L$ ($\mathrm{S/N}(L)_\mathrm{thresh}$) can help to ensure that we are only presenting PA results for which we can correctly characterize the uncertainty. To determine a sensible value for $\mathrm{S/N}(L)_\mathrm{thresh}$, we construct a simple simulation as follows.

\begin{figure*}[ht!]
    \centering
    \includegraphics[width=0.69\textwidth]{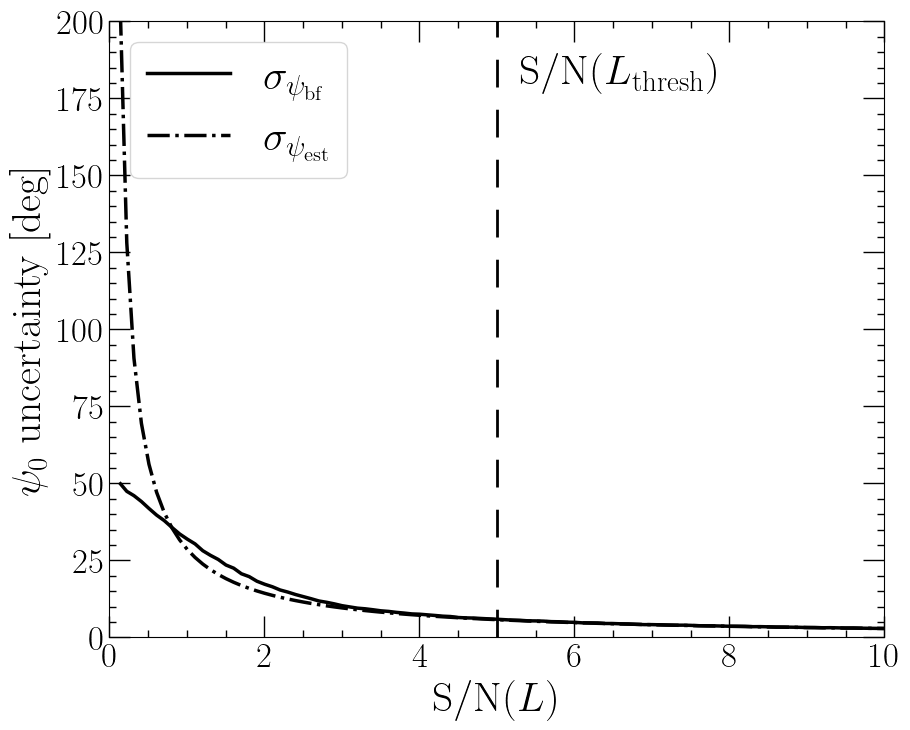}
    \caption{The estimated PA uncertainty using Equation \ref{eq:pa_err}, $\sigma_{\psi,\mathrm{est}}$ (dash-dotted line), compared to the uncertainty derived via a brute force approach (solid line) as a function of the S/N in $L$ in our synthetic polarized measurements. The vertical dashed line indicates $\mathrm{S/N}(L)_\mathrm{thresh} = 5$, which we apply to our FRB data.}
    \label{fig:pa_err}
\end{figure*}

We synthesize a set of polarized signals by taking $10 000$ draws of Stokes $Q_\mathrm{sim}$ and $U_\mathrm{sim}$ from normal distributions with means $\mu_{Q_\mathrm{sim}}$ and $\mu_{U_\mathrm{sim}}$ and standard deviations $\sigma_{Q_\mathrm{sim}}$ and $\sigma_{Q_\mathrm{sim}}$ with arbitrary units:
\begin{equation}
Q_\mathrm{sim} \sim \mathcal{N}(\mu_{Q_\mathrm{sim}}, \sigma_{Q_\mathrm{sim}}=1)\,, \label{eq:Q_sim}
\end{equation}
\begin{equation}
U_\mathrm{sim} \sim \mathcal{N}(\mu_{U_\mathrm{sim}}=0.1, \sigma_{Q_\mathrm{sim}}=1)\,. \label{eq:U_sim}
\end{equation}
Here, we set $\sigma_{Q_\mathrm{sim}} = \sigma_{Q_\mathrm{sim}} = 1$, while progressively increasing $\mu_{Q_\mathrm{sim}}$ in the range $[0.1, 10]$ in intervals of $0.1$. This creates a set of $100$ synthetic polarized measurements (each with $10 000$ simulated points in $Q-U$ space) in which most of the linearly polarized signal arises from Stokes $Q$ and the linearly polarized S/N grows incrementally with increasing $\mu_{Q_\mathrm{sim}}$. For each synthetic polarized measurement, we compute a distribution of the PAs following Equation \ref{eq:pa_derot}, assuming that our simulated $Q_\mathrm{sim}$ and $U_\mathrm{sim}$ have been de-rotated. Then we estimate the PA measurement uncertainty based on the approximation in Equation \ref{eq:pa_err}, $\sigma_{\psi,\mathrm{est}}$, and also via a brute force approach by taking the standard deviation of the resulting PA distribution, $\sigma_{\psi,\mathrm{bf}}$. Figure \ref{fig:pa_err} shows $\sigma_{\psi,\mathrm{est}}$ (dash-dotted line) and $\sigma_{\psi,\mathrm{bf}}$ (solid line) as a function of S/N in $L$. We find that $\sigma_{\psi,\mathrm{est}}$ and $\sigma_{\psi,\mathrm{bf}}$ begin to visibly diverge at a $L$ S/N of $\lesssim 3$, with a mean $|\sigma_{\psi,\mathrm{est}} - \sigma_{\psi,\mathrm{bf}}| \sim 0.1~\mathrm{deg}$ at $L$ S/N of $> 3$. 

Therefore, we set a conservative $\mathrm{S/N}(L)_\mathrm{thresh} = 5$ in our analysis (depicted by the dashed line in Figure \ref{fig:pa_err}) and mask out any points in the PA, $L$, $V$, $L/I$, and $|V|/I$ profiles that fall below this limit. This helps to ensure that the error properties of the PAs that we analyze are well estimated, which is particularly important when considering PA variations intrinsic to the source (e.g., in Section \ref{sec:pa_var}).

\bibliography{ref}{}
\bibliographystyle{aasjournal}

\end{document}